\documentclass[12pt,a4paper]{article}
\usepackage[margin=1in]{geometry}
\usepackage[utf8]{inputenc}
\usepackage{natbib}
\usepackage{hyperref}
\usepackage{verbatim}
\usepackage{amsmath}
\usepackage{amsfonts}
\usepackage{amssymb}
\usepackage{amsthm}
\usepackage{mathtools}
\usepackage{bm}
\usepackage{xparse}
\usepackage{graphicx}
\usepackage{xcolor}
\usepackage{textcase}
\usepackage{url}
\usepackage{lipsum}
\usepackage{appendix}
\usepackage{epstopdf}
\usepackage{setspace}
\usepackage{centernot}
\usepackage{braket}

\usepackage{titlesec}

\usepackage[pagewise]{lineno}

\usepackage{float}

\usepackage{caption}
\usepackage{subcaption}

\titleformat{\section}
  {\normalfont\scshape \filcenter \large}{\thesection}{1em}{}

\titleformat{\subsection}
  {\normalfont\scshape \filcenter}{\thesection}{1em}{}

\usepackage{indentfirst}

\DeclareMathOperator{\Tr}{Tr}

\usepackage{amsthm}

\usepackage{authblk}

\usepackage{dsfont}

\begin{document} 

\title{Quantum Interference and the Limits of Separability}

\author[1, 2]{Sebastian Horvat}
\affil[1]{University of Vienna, Department of Philosophy, Universitätsstraße 7, 1010 Vienna, Austria}
\affil[2]{University of Vienna, Faculty of Physics, Boltzmanngasse 5, 1090 Vienna, Austria}
\date{}                     
\setcounter{Maxaffil}{0}
\renewcommand\Affilfont{\itshape\small}


\maketitle

\begin{abstract}
Quantum theory implies, and empirical evidence confirms, that while particles \textit{can} exhibit wave-like behavior in interferometric experiments, this behavior is so limited as \textit{not} to allow for third- and higher-order interference. The article at hand shows that this possibility-impossibility structure suggests the universal validity of a principle that regulates statistical correlations between spatiotemporally localized events, \textit{independently} of the nature of the objects that may or may not partake in these events. Roughly, and up to some qualifications, the said principle mandates that \textit{any} joint influence of $m$ mutually spacelike separated events on \textit{another} event, be such, that it can be separated by \textit{at least} $\lceil \frac{m}{2} \rceil$ mediating events, and in some cases, by \textit{no more} than $\lceil \frac{m}{2} \rceil$ mediating events. The structure of quantum interference thus teaches us that events can influence each other in a non-separable fashion, but that this non-separability has a certain exactly quantifiable limit.
\end{abstract}

\bigskip

\bigskip

There are occasions in life that present us with something valuable to be learned without the need of stepping outside of our home, by letting what is already there, inside of our home, unfold and exhibit itself in a new light. The web that interweaves quantum phenomena, quantum theorizing and quantum experimentation - and that we are needless to say technologically very much at home with - seems to present us with such an occasion, or this is at least what is perceived both by the interpreter and by the reconstructor of the quantum, by those that believe that there is room for learning something new about our world by reflecting on the achievements delivered to us by our quantum-mechanical predecessors.\footnote{The ``interpreter'' aims at understanding what is really the case in quantum phenomena, what the ``physical world'' is really like, given the empirical validity of quantum theory (see e.g., the various entries in Freire \textit{et al}., 2022). The ``reconstructor'' aims at re-building or reformulating quantum theory, oriented by certain theoretical and practical desiderata, e.g., aiming for a higher transparency of the theory's axioms, simplicity and mathematical consistency (Goyal, 2023, and references therein). Both the interpreter and the reconstructor - the metaphysician and the engineer, so to say - thus see quantum theory as problematic, as needing to be reformed, better understood, rebuilt. Our investigation shares this negative judgment, but instead of engaging in interpretation or reconstruction, it takes one particular and relatively opaque aspect of the quantum and it aims to squeeze out, to extract, a new - and hopefully relatively less opaque - principle therefrom.} However that may be, it is also the working conviction shared by the article that will hereby follow, wherein some very basic and well known features of quantum interferometric experiments will be shown to contain such valuable world-lessons. More particularly, it will be shown that the \textit{possibility} of \textit{second}-order particle-interference and the corresponding \textit{impossibility} of \textit{third}-order particle-interference suggest the universal validity of a certain principle that constrains the separability of the statistical correlations occurring between adequately spatiotemporally situated events. What in the first instance appear to be curious conditions in which point-like objects exhibit wave-like behavior will thereby be tentatively suggested to imply the validity of a curious principle that regulates event occurrences in a wide range of physical phenomena, if not in any physical phenomenon whatsoever. 

A self-enclosed physical principle that stands on its own feet ought to ideally allow for a straight up \textit{logico-mathematical} exposition that makes the principle understandable prior to its potential subsequent justification and exemplification. It is however far more pedagogical, at least when making first contact with the principle, to follow its \textit{genetic} exposition, the one that confronts us with various examples, asks us to view them under a certain angle, and thereby induces us into seeing these particulars as signs or instances of the sought universal. We will naturally choose the latter mode of presentation, thereby tracing the following genetic line. In section 1 we will review the double- and multiple-slit experiments, together with the aforementioned possibility and impossibility of respectively second- and third-order particle-interference. In section 2 we will turn to a recently introduced generalization of these experiments, and point out that they also exhibit an analogous possibility-impossibility structure. In section 3 we will introduce general interference phenomena, that further generalize the latter experiments, and that are definable exclusively by reference to localized events and statistical correlations holding between them. In section 4 we will make several observations about these phenomena, based on a selection of examples and on some analytically proven statements. This will finally motivate the introduction of our principle in sections 5 and 6, where we will also critically discuss its significance and its relation to other physical principles.


\section*{I. The Double-slit and Multiple-slit Experiments}

\begin{figure}
\centering
\includegraphics[width=0.5\linewidth]{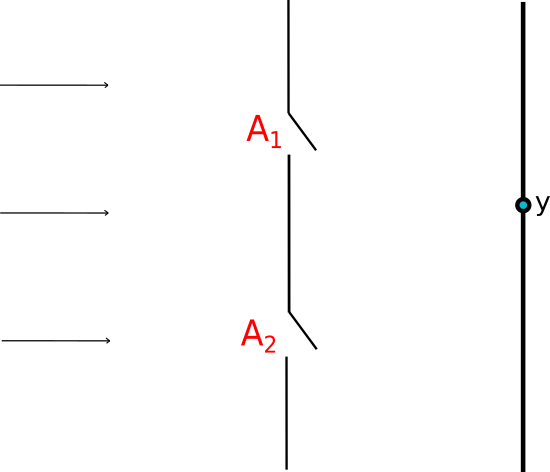}
\caption{The double-slit experiment.}

\label{fig1}
\end{figure}

We will start by reviewing the double-slit experiment, one of the paradigm examples of the wave-like behavior of particles. Stated schematically, the double-slit experiment consists in a particle being sent on a plate pierced by two slits, each of which can be either open or closed - henceforth denoted with 1 or 0. If the particle passes through the plate, it is subsequently detected at a screen that lies behind the plate - see Figure \ref{fig1}. More precisely, the experiment can be partitioned into three temporal steps:\\
$t_1$: The two slits are set in configurations $a_1$ and $a_2$, where $a_i \in \left\{0,1 \right\}$.\\
$t_2$: A particle is sent on the plate, via a method that is independent of $a_1$ and $a_2$.\\
$t_3$: The particle is either detected somewhere on the screen or is not detected at all.

Let $P(y|a_1a_2)$ be the conditional probability that the particle is detected at location $y$ on the screen, given that the two slits were set in configurations $a_1$ and $a_2$, and let
\begin{equation}\label{eq1}
I_2^{(y)}=P(y|11)-P(y|10)-P(y|01)    
\end{equation}
be the so-called \textit{2nd order interference term}, here relativized to location $y$. As it is well known, experiments and elementary quantum theory both show the possibility of $I_2^{(y)}$ being different from $0$. This is in turn in tension with Newtonian mechanics, according to which a non-vanishing $I_2^{(y)}$ can only be generated by waves, but not by individual particles. This possibility of a particle so-to-say statistically interfering with itself has furthermore been stated to defy any ``classical'' explanation whatsoever, even one that is open to modifications of Newtonian mechanics (Feynman \textit{et al}., 1963). Indeed, $I_2^{(y)}$ is determined to vanish, provided that: (i) the particle can reach the screen by passing either through the first slit or through the second slit, (ii) the particle cannot pass through a closed slit, and (iii) if the particle passes through one slit, then its subsequent movement is independent of the configuration of the other slit.\footnote{Statements (i)-(iii) imply that $I_2^{(y)}=0$ by (arguably) ordinary reasoning, one that complies with the rules of ``classical'' propositional logic. Some have attempted to sever this inference - and to thus maintain an aura of physical ``classicality'' while still concording with empirical evidence - by arguing that the aforementioned rules of reasoning do not apply to the case at hand (Putnam, 1969; Maudlin, 2005; see also Horvat \& Toader, 2025, section 4).}

Whereas assumptions (i) and (ii) codify some basic expectations about how billiard-ball-like objects move - i.e. that they always have a definite location and that they cannot pass through some other material bodies - assumption (iii) expresses the belief that the particle in the experiment is interacting exclusively with the slits. Putting this latter assumption into doubt appears to be far less radical than claiming that the particle can somehow be at multiple locations at once or that it can somehow penetrate through a closed slit without leaving any trace behind. And in fact, it is assumption (iii) that is put into doubt both by the De Broglie-Bohm model of the double-slit experiment and by the recently proposed toy model due to Catani \textit{et al.} (2023a, 2023b): both of these models drop the third assumption by positing a further degree of freedom that can be influenced by the configurations of both slits and that can in turn influence the particle's motion, thereby somewhat deflating the peculiarity of the phenomenon of particle-interference.\footnote{According to the De Broglie-Bohm theory, the pilot wave is what does the said job of mediating between the particle and the distant slit (Goldstein, 2025). In Catani \textit{et al.}'s toy model, this job is instead done by the phase degrees of freedom associated to the spatial modes that can possibly be occupied by the particle.} 

The above two models suggest that the wave-like behavior of quantum particles in the double-slit experiment is not as curious as Feynman and his likes might have contended. However, note that they do so only by speculating on some arguably ad-hoc posits (in the case of Catani \textit{et al.}'s model)\footnote{The ad-hoc posits referred to above are the phase degrees of freedom associated to the spatial modes possibly occupied by the particle. In order to transform this positing into a legitimate and possibly empirically corroborated postulation, one would need to articulate a broader theory, one that is not restricted to one particular class of experiments, but that presents the posited entities in a broader variety of phenomena. The authors presumably recognize this, as they refer to their proposal as merely being a \textit{toy} model. However, the conclusions drawn by the authors - that many features of quantum interference are, in a sense, not as problematic as usually thought - \textit{seem} to be stronger than ones warranted to be drawn from a mere toy model. Indeed, their model \textit{arguably} does not provide a legitimate explanation of quantum interferometric phenomena. Instead, it only correctly reproduces the data extractable from these phenomena, while adhering to the structural constraint of ``generalized non-contextuality'' (see Schmid et al., 2024,  and references therein). While the latter constraint might even be - in addition to mathematico-logical consistency - a \textit{necessary} condition for a model to be a legitimate contender at explaining a certain phenomenon, it can by no means amount to a \textit{sufficient} condition that guarantees that it does adequately explain that phenomenon. This short critical note admittedly requires further elaboration that cannot be pursued here (but see Horvat, 2025, sections 3 and 4.2, for some more thoughts that pull in this direction).} or by taking up on the usual conundra faced by De Broglie-Bohm's theory\footnote{For a review of the main problems faced by De Broglie-Bohm's theory, consult Passon (2025).}. Indeed, despite the existence of these two classical-like models, we will later see that there \textit{is} after all something curious about the possibility of particle-interference in the double-slit experiment, and that neglecting this kernel of curiosity would also lead to the neglecting of the apparent validity of a universal principle suggested by the \textit{impossibility} of \textit{third}-order particle interference. Before reaching that stage, let us first review the said impossibility that appears in experiments with multiple slits.

\begin{figure}
\centering
\includegraphics[width=0.5\linewidth]{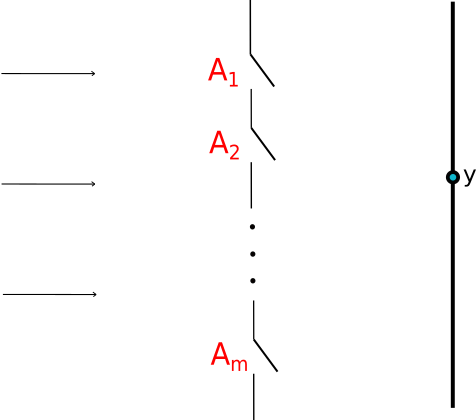}
\caption{The $m$-slit experiment.}

\label{fig22}
\end{figure}

As the name suggests, \textit{multiple-slit experiments} are straightforward generalizations of the double-slit experiment, with the sole difference being that the plate is now pierced by more than two slits, say, by $m$ of them (see Figure \ref{fig22}). 
Letting $P(y|\mathbf{a})$ denote the probability of the particle landing at $y$ conditioned on the slits having been configured as $\mathbf{a}\equiv (a_1,...,a_m)$, the $m$\textit{-th order interference term} reads as: 

\begin{equation}
I_m^{(y)}=\sum_{a_1,...,a_m=0}^1 (-1)^{\sum_{i=1}^m a_i}P(y|\mathbf{a}).   
\end{equation}

As it can be readily verified, the generalized $I_m^{(y)}$ reduces to Eq. \eqref{eq1} for $m=2$, since $P(y|00)=0$ holds, or in other words, since the particle cannot reach the screen if both slits are closed. The reason these experiments are worthy of special consideration is the following: both quantum theory and current experimental evidence show that $I_m^{(y)}=0$, for $m\geq 3$, regardless of the kind of particle used in the experiment and of the method by which it is sent through the slits.\footnote{The properties of multiple-slit experiments were first explored by Sorkin (1994). Besides the experiments that have been conducted thereafter - and that have, as expected, corroborated the predictions of quantum mechanics (Sinha \textit{et al.}, 2010) - several works on these experiments have focused on the study of information-theoretic principles that could, in a sense, explain their aforementioned properties (see Barnum \textit{et al.}, 2017, and references therein).} Whereas the double-slit experiment told us that particle-like objects can behave in a wave-like fashion, multiple-slit experiments are now telling us that this wave-like behavior is so constrained as not to allow for higher-order interference in form of a non-vanishing $I_{m\geq3}^{(y)}$. The interference of particles in multiple-slit experiments thus exhibits a peculiar ``possibility-impossibility structure'' somewhat analogous to the structure exhibited by EPR-like experiments: whereas experiments that violate Bell inequalities are possible, no experiment is possible in which the said violation is so high as to violate \textit{Tsirelson's bound} (according to quantum theory and to current experimental evidence) (Tsirelson, 1980). Similarly, in our case, whereas experiments are possible in which a non-zero $I_{2}^{(y)}$ is generated by a single particle, no such experiment can generate a non-zero $I_{m\geq3}^{(y)}$ (again: according to quantum theory and to current experimental evidence). 

An important difference should however be noted between the possibility-impossibility structure featured in multiple-slit experiments and the one exhibited in EPR-like experiments. The difference is that the possibility discovered by Bell and the impossibility discovered by Tsirelson appear to apply to any phenomenon whatsoever, as long as they feature events that are spatiotemporally and statistically related to each other in a certain fashion. The events in question are thereby left unspecified, constrained neither to be related to specific instruments - such as slits and their configurations - nor to specific objects and their properties - such as particles and their classical or quantum states. As already announced, the principle that we are ultimately after in our investigation here is also supposed to apply universally across phenomena, and to constrain correlations between events, independently of their nature. The way by which we are going to reach such a general principle is by abstracting away from the various instrumental and theoretical specifications of multiple-slit experiments and ultimately reaching a possibility-impossibility structure that is articulable by resources that are almost as minimal and general as the ones needed to articulate the Bell-Tsirelson possibility-impossibility structure.\footnote{The need for the ``almost''-caveat will be clear by the end of the article.} Before reaching this final stage of abstraction, we need to however still recap an intermediate stage that generalizes our possibility-impossibility structure to the broader class of - as we will here call them - \textit{semi-general interference experiments}.

\section*{II. Semi-general Interference Experiments}

\begin{figure}
\centering
\includegraphics[width=0.5\linewidth]{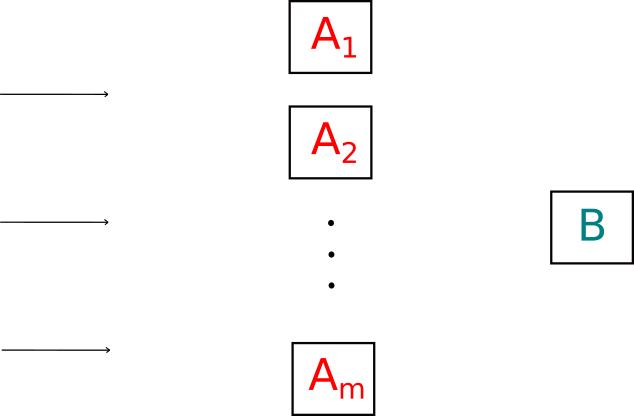}
\caption{The semi-general interference experiment of order $m$.}

\label{fig3}
\end{figure}

The class of semi-general interference experiments contains the wide range of experiments that can be obtained from multiple-slit experiments by replacing their slits and screens by certain other devices and by allowing for more than one particle to be involved therein (Horvat \& Dakić, 2021a). More specifically, in any such experiment, a collection of particles interacts with $m$ devices and, upon potentially reaching another $(m+1)$-st device, an output is elicited in the latter - see Figure \ref{fig3}. All of the said devices are assumed to implement \textit{local} transformations, roughly meaning that a device can change the state of only those particles that pass through it. Additionally, the $m$ intermediate devices - the ones that generalize the slits - are assumed not to be capable of increasing the number of particles, i.e. none of them can turn $l$ incoming particles into $k>l$ outgoing particles. For example, an intermediate device might thus implement a joint unitary operator that acts on the internal degrees of freedom of the particles that pass through it, leaving the spatial degree of freedom unaltered.\footnote{The details regarding the case in which the intermediate devices implement such unitaries will be given in Section IV.II. For details regarding how the devices act on the incoming particles in the general case, see (Horvat \& Dakić, 2021a).}

The experiments can once again be partitioned into three timesteps:\\
$t_1$: The intermediate devices are set in configurations $\mathbf{a}\equiv a_1,...,a_m$, where $a_i \in \left\{0,1 \right\}$.\\
$t_2$: A collection of particles is sent towards the devices, via a method that is independent of $\mathbf{a}$.\\
$t_3$: The $(m+1)$-st device produces an output $b \in \left\{0,1 \right\}$.

Letting $P(b|\mathbf{a})$ be the probability that output $b$ is produced by the final device, given that the intermediate devices have previously been set in configurations $\mathbf{a}$, we will say that
\begin{equation}\label{eq3}
    I_m=\frac{1}{2^m}\sum_{\mathbf{a}}P(\oplus_{i=1}^ma_i|\mathbf{a})-\frac{1}{2}
\end{equation} 
is the \textit{semi-general m-th order interference term}, where $\oplus_{i=1}^ma_i \equiv \left(\sum_{i=1}^ma_i \right) \text{mod 2}$. 

Semi-general interference experiments obey the following properties (Horvat \& Dakić, 2021a):  
\begin{itemize}
    \item $m$-th order interference can be generated by no less than $m$ \textit{classical} particles: more precisely, there are experiments with $n$ classical particles that generate $I_{m}\neq0$ for $m\leq n$, but any such experiment necessarily obeys $I_{m}=0$, for all $m>n$;
    \item $m$-th order interference can be generated by no less than $\lceil \frac{m}{2} \rceil$ \textit{quantum} particles: more precisely, there are experiments with $n$ quantum particles that generate $I_{m}\neq0$ for $m\leq 2n$, but any such experiment necessarily obeys $I_{m}=0$ for $m> 2n$.
\end{itemize}

These properties generalize the possibility-impossibility structure from the previous section along two dimensions. First, the semi-generalized order of interference turns out to be additive under composition of particles, be that they are classical or quantum: 3rd-order interference is thus for instance possible, but only by the involvement of at least two quantum or three classical particles; more generally, $m$-th order interference can be produced by no less than $m$ classical or no less than $\lceil m/2 \rceil$ quantum particles. Second, this possibility-impossibility structure holds for a wide range of experiments that include devices operating on the internal degrees of freedom of possibly entangled particles: in particular, the impossibility of $n$ quantum particles to produce more than $2n$-th order interference holds regardless of the dimensionality of their internal degrees of freedom. 

Note that semi-general interference experiments and the possibility-impossibility structure exhibited therein do amount to generalizations of ordinary multiple-slit experiments and their previously mentioned possibility-impossibility structure. The latter indeed amount to a special case of the former, in which (i) only one particle is involved, (ii) the intermediate devices are instantiated by slits, and (iii) the final device is instantiated by the detection of the particle at some location $y$ on a screen.\footnote{$I_m$ accordingly reduces to $I^{(y)}_m$ via the identification of $P(0|\mathbf{a})$ with $P(y|\mathbf{a})$. Furthermore, note that semi-general interference experiments also generalize other interferometric experiments, such as the Mach-Zehnder experiment (and its generalization with multiple arms), wherein the intermediate devices are realized by phase-shifters and the final device by a particle detector.} The possibility-impossibility structure that we previously pointed out to hold in multiple-slit experiments thus turns out to be a merely contingently isolated instance of a structure that holds within the broader class of semi-general interference experiments. Notice however that this class, albeit less narrow than the class of multiple-slit experiments, is still defined by reference to certain natural objects (the particles and their number) and to certain instruments (that do not increase the number of particles). In what follows we will ascend to a further layer of abstraction, liberating ourselves from particles, transformations and devices - a layer that is intended to apply universally to spatiotemporally localized events, regardless of the natural or artificial objects that they might or might not be related to.

Before proceeding with our task, note that the principle that we are after is going to be formulated in terms of \textit{events}. Since the usage of the term `event' is generally insufficiently clear in discussions on quantum theory, it is advisable to roughly specify how this term is going to be used throughout the paper. An event will here generally be something that occurs, or that is the case, in some region of spacetime. A \textit{spatiotemporally localized} event will be an event that occurs, or that is the case, within a \textit{relatively small} region of spacetime. Since all events that we will be referring to here are spatiotemporally localized, we will refer to all of them more simply as `events', occasionally dropping the qualifier `spatiotemporally localized' or `localized'. Thus, for example, a bomb exploding roughly at some spatiotemporal location denoted by $x \in \mathbb{R}^4$ is an event; my writing of these words here-and-now is also an event, as well as your reading of these words there-and-then. Now, as we will perpetually be repeating throughout the paper, there is currently no consensus on which events can generally be said to occur in arbitrary quantum phenomena. More specifically, there is no consensus on which events can be said to occur in a quantum phenomenon in those spatiotemporal regions that do not feature measurements or preparations of quantum systems, or more generally, in those regions that feature negligible decoherence. This is the reason why - later, upon discussing quantum phenomena - we will only be referring to those events in quantum phenomena whose occurrences we can all agree upon, such as local preparations and measurements of quantum systems: a paradigmatic quantum-mechanical event will thus be a device being in a specific configuration within some relatively well localized spatiotemporal region. Keeping this proviso in mind - which will still be elaborated in the later sections - we are now safe to proceed with our task.

\section*{III. General Interference Phenomena}

We are aiming at a principle that universally regulates relations between event-occurrences and that is inspired by the possibility-impossibility structure exhibited by multiple-slit experiments, and more generally, by semi-general interference experiments. The first step is thus to re-describe semi-general interference experiments in terms of spatiotemporally localized events, and then to tentatively abstract from their previous specifications. As we have seen in the previous section, the quantity of interest in semi-general interference experiments concerns the statistical correlation between the configurations $\mathbf{a}$ of the $m$ intermediate devices and the output $b$ of the $(m+1)$-st device. What we will however now be concerned with are not atemporal configurations of devices, but spatiotemporally localized \textit{events}: in our case, what matters are the configurations $(\mathbf{a},b)$ of the $(m+1)$ devices \textit{only} at the moments in which they interact with the incoming particles. Let us henceforth for simplicity denote these events again with the same symbols $(\mathbf{a},b)$, meaning for instance that $a_i=0$ and $b=1$ denote the events that respectively correspond to the $i$-th device being configured to $0$ at the moment of its interaction with the particles and the $(m+1)$-st device being configured to $1$ at the moment of its interaction with the particles. It is clear by construction that the spatiotemporal locations at which events $\mathbf{a}$ occur are all mutually spacelike separated, whereas event $b$ is bound to take place in the common causal future of events $\mathbf{a}$. The quantity that will therefore be of interest to us is not the correlation between atemporal configurations of certain devices, but the correlation between a collection of mutually spacelike separated events on the one hand, and an event lying in their common causal future, on the other hand. More particularly, the quantity of interest $I_m=\frac{1}{2^m}\sum_{\mathbf{a}}P(\oplus_{i=1}^ma_i|\mathbf{a})-\frac{1}{2}$ will still be formally equivalent to Eq. \eqref{eq3}, but will now concern the correlation between \textit{event} $b$ and the function $\oplus_ia_i$ of \textit{events} $\mathbf{a}$. 

Now we are almost ready to execute the previously announced step of abstraction and generalization: instead of considering only those phenomena in which events $(\mathbf{a},b)$ correspond to configurations of certain devices at times of interaction with certain particles and that generate $I_m \neq 0$, let us tentatively consider \textit{any} phenomenon in which certain mutually spacelike separated events $\mathbf{a}$ and an event $b$ lying in their common causal future are so correlated as to satisfy $I_m \neq 0$. The class of phenomena that we have tentatively just referred to is rather vast, and contains many other phenomena alongside our semi-general interference experiments. In fact, it includes also those phenomena in which the correlation between $\mathbf{a}$ and $b$ is not established by some mediator, but is adjusted by some common cause, such as, for instance, a trivialized multiple-slit experiment in which the particle is sent through the slits, but its subsequent trajectory is manipulated by an agent that knows the values of $\mathbf{a}$. We will naturally want to exclude such pathologically trivial correlations from our category of interest and focus only on those cases where the occurrence of events $\mathbf{a}$ is, in a certain sense, causally relevant for the occurrence of event $b$. In order to do that, however, we will need to pause and tread more carefully by introducing some auxiliary definitions.\footnote{The definitions that will follow will feature some unfamiliar concepts and notations. A summary of the most relevant ones is provided in Appendix 0.}

Again, the category of phenomena that we are after include, roughly speaking, all those phenomena in which certain mutually spacelike separated events $\mathbf{a}$ and an event $b$ lying in their common causal future are so correlated as to satisfy $I_m \neq 0$, but without there being a conspiratorial common cause that trivializes this correlation. We are clearly thereby concerned with examples pertaining to the wide category of \textit{statistical phenomena}, that is, phenomena that can be described by probability distributions over spatiotemporally localized events. More precisely, a statistical phenomenon is a phenomenon that can be described by a \textit{probabilistic-event model}, specified as follows.\footnote{The category of statistical phenomena and their correlative category of probabilistic event models were introduced in (Horvat, 2025), which also contains a discussion of how quantum phenomena are to be incorporated therein.}

\bigskip

\textbf{Definition 1.} Let $\left\{\Omega_x \right\}_{x \in \mathbb{R}^4}$ be a family of sets, whereby the elements of each $\Omega_x$ denote events that can possibly occur at spatiotemporal location $x$. Let $\mathcal{X}$ be the family of all finite subsets of $\mathbb{R}^4$ and, for each $X \in \mathcal{X}$, let $\mathcal{F}_X$ be the $\sigma$-algebra on $\Omega_X$, where $ \Omega_X = \prod\limits_{x \in X}^{} \Omega_x$. A \textit{probabilistic event model} $\mathbf{P}$ is a family of probability distributions $\mathbf{P}=\left\{(\Omega_X,\mathcal{F}_X, P_X) \right\}_{X\in \mathcal{X}}$, such that for all $X' \subset X$, $P_{X'}$ is the corresponding marginal distribution of $P_{X}$.

\bigskip

A probabilistic-event model, or PE-model, assigns to each spatiotemporal location $x \in \mathbb{R}^4$ a set $\Omega_x$ of events that can possibly occur at that location, say, a light bulb being on or off at location $x$. The model also assigns, to any finite list of locations $X$, a probability $P_X(\omega_X)$ that events $\omega_X \in \prod\limits_{x \in X}^{}\Omega_{x}$ occur at their corresponding locations in $X$. PE-models are thus rich enough to be used to describe phenomena whose featured events exhibit statistical regularities: for instance, a regular coin toss can be described by a PE-model that (i) coordinates the coin's fall with an element $x \in \mathbb{R}^4$, (ii) assigns a set $\Omega_x =\left\{h,t \right\}$ whose elements correspond to the two possible events of the coin landing heads or tails, and (iii) specifies the probability distribution as $P_x(\omega_x)=\frac{1}{2}$. More generally, a PE-model is said to adequately describe a phenomenon if the distributions it posits approximately coincide with the statistical distributions exhibited by the phenomenon. To be sure, the assessment of the relation of adequacy between a given PE-model and a given statistical phenomenon may be highly non-trivial, especially so in the case of quantum phenomena, for which, as mentioned beforehand, we still lack a way of assigning event-occurrences to those spatiotemporal regions that feature negligible decoherence - such as those regions in which no measurements or preparations are occurring. 

Note that one and the same phenomenon may naturally be described by different PE-models corresponding to different ``resolutions'', or degrees of attention to detail. Getting back to the regular coin toss, both of the following PE-models describe it adequately: (i) one that asserts that a coin lands at location $x$ with probability 1, and also (ii) another one that asserts that a coin lands \textit{heads} at location $x$ with probability $\frac{1}{2}$. The latter model describes the same phenomenon in more detail, or in other words, it amounts to a refinement of the former model. Let us accordingly introduce more generally the \textit{refinement}-relation that can hold or not hold between any pair of PE-models.

\bigskip

\textbf{Definition 2.} Let $\mathbf{P}=\left\{(\Omega_X,\mathcal{F}_X, P_X) \right\}_{X\in \mathcal{X}}$ and $\mathbf{P'}=\left\{(\Omega'_X,\mathcal{F}'_X, P'_X) \right\}_{X\in \mathcal{X}}$ be two PE-models. We say that $\mathbf{P'}$ is a \textit{refinement} of $\mathbf{P}$ if and only if there exists a family of functions $\left\{f_x:\Omega'_x \rightarrow \Omega_x \right\}_{x \in \mathbb{R}^4}$, such that $P_X(\omega_X)=P'_X (f^{-1}(\omega_X))$, where $f(\omega_{x_1},...,\omega_{x_n})\equiv(f_{x_1}(\omega_{x_1}),...,f_{x_n}(\omega_{x_n}))$.

\bigskip

Functions $f_x$ in the above definitions can be understood as coarse-graining relations between sets $\Omega'_x$ and $\Omega_x$, so that the latter set offers a coarser description of the possible events occurring at $x$ than the former set. Before proceeding, let us on the way introduce some usual special-relativistic notation that will come of use both now and later in the paper. First, symbols $x,y,z$ will henceforth always be used to denote elements of $\mathbb{R}^4$. Second, $x \prec y$ holds if there is a future-oriented time-like or null-like curve from $x$ to $y$, and $x \neq y$; $x \sim y$ holds if $x$ and $y$ are spacelike separated. Finally, the causal past and the causal future of some region $X \subset \mathbb{R}^4$ will respectively be denoted by $\mathcal{C}^{(-)}_X \equiv \left\{y \in \mathbb{R}^4| \exists x \in X: y \prec x \right\}$ and $\mathcal{C}^{(+)}_X \equiv \left\{y \in \mathbb{R}^4| \exists x \in X: x \prec y \right\}$. With all of these definitions in hand, we are now ready to proceed with our task and to introduce the category of \textit{general interference phenomena}, that, as previously announced, provides the final generalization of semi-general interference experiments. 


\textbf{Definition 3.} A statistical phenomenon is a \textit{general interference phenomenon of order m} if there exist $X=(x_1,...,x_m)$ and $y$, whereby $x_i \sim x_j$ and $y \in \bigcap_{i=1}^m \mathcal{C}_{x_i}^{(+)}$, such that there is a PE-model $\mathbf{P}=\left\{(\Omega_X,\mathcal{F}_X, P_X) \right\}_{X\in \mathcal{X}}$ that adequately describes the phenomenon and satisfies the following conditions: 

\begin{enumerate}
    
    \item $\Omega_{x_i}\cong\Omega_y\cong\left\{0,1 \right\}$
    \item $\frac{1}{2^m}\sum_{\omega_X}P_y(\oplus_{i=1}^m\omega_{x_i}|\omega_X)-\frac{1}{2} \neq 0$
    \item For all $z \in \mathbb{R}^4$, and for any $\mathbf{P'}=\left\{(\Omega'_X,\mathcal{F}'_X, P'_X) \right\}_{X\in \mathcal{X}}$ that refines $\mathbf{P}$ and that adequately describes the phenomenon, and such that $\Omega'_{x_i}=\Omega_{x_i}$, $\Omega'_{y}=\Omega_{y}$: 
    \begin{equation*}
        \left[P'_y(\omega'_y|\omega'_X\omega'_{z}) \neq  P'_y(\omega'_y|\omega'_X) \right] \quad \longrightarrow \quad \left[P'_X(\omega'_X|\omega'_z)=P'_X(\omega'_X) \right]
    \end{equation*}
\end{enumerate}

\bigskip

\begin{figure}
\centering
\includegraphics[width=0.65\linewidth]{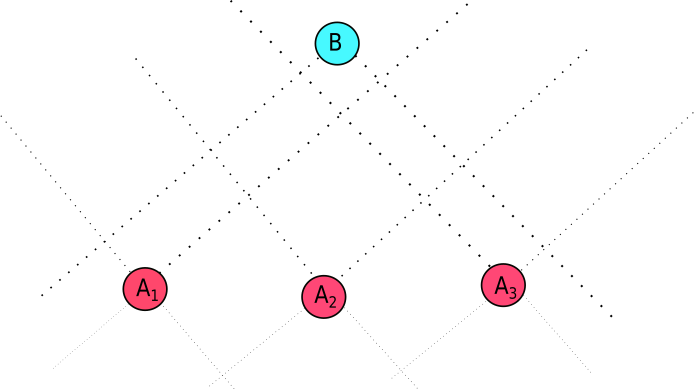}
\caption{Spatiotemporal diagram of a general interference phenomenon of order 3. The dashed lines represent the future or past lightcones of their pertaining locations. Events $(a_1,a_2,a_3)$ (occurring at mutually spacelike separated locations $(x_1,x_2,x_3)$) are so correlated with event $b$ (that occurs at future location $y$) as to generate $I_3 \neq 0$.}

\label{fig4}
\end{figure}

Let us unpack the above definition by explaining how each of its components relates to our previously made remarks about semi-general interference experiments - see Fig. \ref{fig4}. First, note that a general interference phenomenon of order $m$ - or $\text{GIP}_{\text{m}}$ for short - is a statistical phenomenon identified by reference to a collection of $m$ mutually spacelike separated events occurring at $X=(x_1,...,x_m)$ and an event occurring somewhere in their causal future, that is, at some $y \in \bigcap_{i=1}^m \mathcal{C}_{x_i}^{(+)}$. The definition proceeds by referring to the existence of a certain PE-model $\mathbf{P}$ that, according to point 1, assigns dichotomic random events at the said locations, whose event spaces are thereby isomorphic to $\left\{0,1\right\}$. Note that the definition is so far trivial, as in any statistical phenomenon whatsoever certain events can be identified that occur at locations mutually related as above, and that can be described or encoded in terms of bits: for example, even when confronted with a location to which we would canonically ascribe a continuum of possible events, we can nevertheless re-describe the events at that location, in a coarse-grained fashion, in terms of two possible events corresponding to some bi-partition of the said continuum. 

The first non-trivial determination of a $\text{GIP}_{\text{m}}$ appears at point 2 in the definition, which says that the given events at $(X,y)$ need to be, according to model $\mathbf{P}$, so correlated as to satisfy $I_m \neq 0$. As already said before, this condition is a straightforward abstraction of the condition of $m$-th order interference in semi-general interference experiments. Point 3 finally assures the said correlation between events at $X$ and the event at $y$ not to be trivially generated by some common cause at $z$: \textit{if} an event at $z$ has an influence on the event at $y$ without mediation of events at $X$, \textit{then} it cannot have an influence on the events at $X$.\footnote{Let me explain why this constraint needs to be implemented in terms of a \textit{conditional} statement. Namely, any interference experiment will exhibit many past events that \textit{do} influence the events at $X$, as the events at $X$ do not spark out of nowhere: e.g., they might depend on what some experimenters have chosen to do in the past. What is important is just that these causing events do not also have an influence on the event at $y$ without mediation by the events at $X$: the said experimenters are allowed to intervene on what happens at $X$, but are not allowed to directly intervene on what happens at $y$. 

Furthermore, the conditional can contrapositively be expressed as follows: any event at any location $z$ that has an influence on the events at $X$ needs to be such that it does not have an influence on the event at $y$ without mediation of events at $X$. Again, there might be many events that do directly influence the event at $y$ without mediation of the events at $X$: e.g., in a semi-general interference experiment, the probability of the final outcome at $y$ depends not only on the particles that are being measured, but also on the setting of the measurement device that is being used. There is thus an event which does influence the outcome probability without mediation of events at $X$. However, this event does not have any influence on the events at $X$: the settings of the measuring device is to be chosen independently from the events at $X$.} Note that this latter condition needs to be imposed on all refinements $\mathbf{P'}$ of the original model $\mathbf{P}$: indeed, even if there were a common cause at some location $z$, the original model might have been chosen so as not to be refined enough to capture it, which is why we are forced to refer to the class of all refinements of model $\mathbf{P}$.

Le me emphasize that the definition above could admittedly be simplified, but only at the expanse of caution and precision. Our definition is phrased, so-to-say, in the \textit{semantic key}: it defines a class of phenomena, not in terms of their own determinations, but in terms of the determinations of their descriptions, or here, in terms of the properties of their PE-models. One might instead, as it is more usual in empirical science, tentatively switch to the \textit{ontological key}, and speak directly of the supposed determinations of the phenomena themselves. Such a definition would define the $\text{GIP}_{\text{m}}$-class in terms of phenomena that feature dichotomic events at locations $(X,y)$, that are so correlated as to generate $I_m\neq 0$, and whose correlation is not established by a common cause at some location $z$. The danger with this path derives from the fact we mentioned beforehand, which is that we do not (yet) have a complete principled way of coordinating PE-models and quantum phenomena: whereas we certainly all agree on what a PE-model is supposed to assert regarding events that correspond to relatively macroscopic happenings, such as measurements and preparations, we do not (yet) have clear criteria that could apply to the remaining locations, e.g. to those locations lying in between a preparation and a measurement of a photon manipulated in a laboratory. We thus choose, for the time being, to remain in the semantic key, though hoping that a future will come where we will be safe to modulate to the ontological key, and thereby simplify our definitions.

The double-slit experiment, be that the interference pattern is generated by an electromagnetic wave or by an electron, provides a cardinal example of a $\text{GIP}_{\text{2}}$, whereas both classical and quantum semi-general interference experiments of the $m$-th order clearly provide examples of $\text{GIP}_{\text{m}}$-s. But there are certainly also other phenomena that fall under the category of $\text{GIP}_{\text{m}}$-s, besides our previously introduced semi-general interference experiments. One example are a modification of the latter experiments, whereby the intermediate devices are capable of increasing the number of particles: for instance, a machine may be so set up as, depending on its configuration, either to emit or not to emit a photon, without this photon having been previously sent into the machine. Another example might be one in which the correlation between the events at $X$ and at $y$ is established by some operations on a quantum field. And of course, imagination and future research might reveal further examples of $\text{GIP}_{\text{m}}$-s that currently do not, or even \textit{cannot}, come to mind to us.

\section*{IV. Reaching for the Principle}

Let us briefly recap our genetic road, in order to see where we stand. The cardinal stimulus for our investigation was the encounter with the possibility-impossibility structure exhibited by semi-general interference experiments: for any such experiment, $m$-th order interference is achievable with no less than $\lceil m/2 \rceil$ particles. Our aim was then to ideally turn this structure into a principle that universally constrains correlations between events. We thus first took the category of semi-general interference experiments, abstracted away from many of its specifications, and arrived to the wider category of $\text{GIP}_{\text{m}}$-s, definable exclusively in terms of statistical correlations between events. Our goal is now to finally turn the above possibility-impossibility structure into a universal principle that regulates event-occurrences. In other words, we are aiming to take inspiration from the restriction that ``every semi-general interference experiment of order $m$ features at least $\lceil m/2 \rceil$ particles'' and discover the ideally universal restriction that ``every $\text{GIP}_{\text{m}}$ satisfies XYZ'', where ``XYZ'' is to ideally speak universally of events, without zooming into the details of their nature or their artifice.

Recall again that there is an important obstacle in our goal, which is that we do not, as of yet, have a coherent and agreed upon way of assigning event-occurrences to arbitrary regions in quantum phenomena. 
A resolution of this problem, if possible at all, is part of a reaching for a resolution of the broader ontological problems faced by our current grasp of quantum phenomena. And once such a resolution is reached, if at all, the principle that we are after might turn out to be phrasable in ontological terms, that is, in terms of events that occur in our $\text{GIP}_{\text{m}}$-s, without any further qualification. Here we are however in no position to do that, which is why we will take an alternative road that will base itself only on those events in quantum phenomena on whose occurrences we can all agree upon, such as local laboratory preparations and measurements of quantum systems. More precisely, the events that we will be referring to in what follows will consist in certain devices being in some configurations in some relatively well localized spatiotemporal regions, be it that the said devices are used to prepare quantum systems, to act unitarily on them, or to measure them: for example, a photon-detector indicating, roughly within spatiotemporal region $x$, that ``one photon has been detected'', is an event.

Note, however, that if we are to focus only on the said unproblematically assignable kind of events, then it appears that we do not have any means of articulating - merely in terms of correlations between events - the restriction that a $\text{GIP}_{\text{m}}$ features no more than a certain number of particles: indeed, it appears that it is exactly those regions that lie \textit{between} $X$ and $y$ that we should say something about, if we are to generalize the restriction that only a certain amount of particles passes through that region. We will address this issue by the following means: instead of focusing only on the correlations between events occurring in the $\text{GIP}_{\text{m}}$ of interest, we will also thematize the correlations between events occurring in \textit{other physically possible} phenomena, that are somehow related to the original $\text{GIP}_{\text{m}}$. Before fleshing this out, let us offer some preliminary intuition, both on what this strategy is to look like and why it is expected to lead us to a principle with the sought characteristics. We will do so by surveying three examples.

\subsection*{IV.I. Examples}

\textbf{Example 1.} Let us consider the following example of a $\text{GIP}_{\text{2}}$, call it $\mathcal{E}$. In experiment $\mathcal{E}$, two parties that are located at a large distance from each other, $\text{Alice}_{\text{1}}$ and $\text{Alice}_{\text{2}}$, each prepare a photon either in state $\ket{\uparrow}$ or in state $\ket{\downarrow}$, that is, they align the photon's spin either parallelly or anti-parallelly to a certain axis, call it $z$. The two preparations are so coordinated as to occur at spacelike separation. The parties then send their pertaining photons to another distant location, where Bob measures each of the two photons in the $z$-basis and outputs, say on a computer screen, the sum-modulo-two $b \in \left\{ 0,1\right\}$ of his two measurement outcomes. It is obvious that, in each run of the experiment, the output $b$ will be equal to $a_1 \oplus a_2$, where $a_1,a_2 \in \left\{ 0,1\right\}$ are the events that correspond to $\text{Alice}_{\text{1}}$'s and $\text{Alice}_{\text{2}}$'s preparations. It is also clear that this experiment is a $\text{GIP}_{\text{2}}$: the two preparations $a_1$ and $a_2$ occur at spacelike separation, and they are so correlated with the future measurement outcome $b$ as to generate $I_2=\frac{1}{2}\neq 0$; furthermore, there is no conspiratorial common cause that coordinates the two preparations with the measurement outcome. 

Now consider \textit{another} $\text{GIP}_{\text{2}}$, call it $\mathcal{E}^*$, which again features $\text{Alice}_{\text{1}}$ and $\text{Alice}_{\text{2}}$ doing the very same things as in the original experiment $\mathcal{E}$, and again sending their photons towards Bob. The difference is however that the photons are now intercepted by two additional parties lying on the $\text{Alice}_{\text{1}}$-Bob line and the $\text{Alice}_{\text{2}}$-Bob line respectively, call them $\text{Bob}_{\text{1}}$ and $\text{Bob}_{\text{2}}$. $\text{Bob}_{\text{1}}$ and $\text{Bob}_{\text{2}}$ each measure their respective photons in the $z$-basis and then send them towards Bob, who once again measures the photons in the $z$-basis and outputs the sum of the measurement outcomes. Let us denote with $(b_1,b_2)$ the events that correspond to the measurements performed by $(\text{Bob}_{\text{1}}, \text{Bob}_{\text{2}})$ and let us refer to them as \textit{intermediate events} - `intermediate' in the sense that they occur temporally between the events $(a_1,a_2)$ and $b$. It is again obvious that $b$ will in each run be equal to $a_1 \oplus a_2$, since the intermediate events $(b_1,b_2)$ are in each run perfectly correlated to $(a_1,a_2)$. The correlation between $(a_1,a_2)$ and $b$ in $\mathcal{E}^*$ is therefore equal to the correlation between these same events in the original experiment $\mathcal{E}$. Moreover, the intermediate events $(b_1,b_2)$ ``mediate'' the correlation between $(a_1,a_2)$ and $b$, in the sense that their distribution $P^*$ satisfies 

\begin{equation}
    P^*(b|a_1a_2)=\sum_{b_1,b_2}P^*(b|b_1b_2)P^*(b_1b_2|a_1a_2),
\end{equation}
and will as such also be referred to as \textit{mediating events}. Thus, the correlation between $(a_1,a_2)$ and $b$ \textit{could} be mediated by mediating events (as exhibited in $\mathcal{E}^*$), but just happens not to be so mediated in the original experiment $\mathcal{E}$. 

\bigskip

\textbf{Example 2.} The prior example, though involving quantum particles and thus presenting trouble for a straight up assignment of events to the intermediate regions traveled by the photons, is nevertheless, in a sense, almost classical. A very similar experiment could indeed be executed by the use of macroscopic objects, like billiard balls of various colors or sheets with various inscriptions thereon, and these experiments would again feature the possibility of the relevant correlation between $(a_1,a_2)$ and $b$ to be mediated by some intermediate events $(b_1,b_2)$: in fact, these fully classical experiments would already themselves feature such mediating events, without the need of even referring to other physically realizable experiments. Let us now contrast these ``quantum-but-almost-classical'' and classical phenomena on the one hand with, so to say, a ``fully-quantum'' phenomenon on the other hand. The quantum phenomenon that we will consider will again involve three parties $\text{Alice}_{\text{1}}$, $\text{Alice}_{\text{2}}$ and Bob lying on distant vertices of a triangle, and the former two aiming to communicate their inputs to Bob. The difference is that $\text{Alice}_{\text{1}}$ and $\text{Alice}_{\text{2}}$ now share only \textit{one} photon in a superposition of their two locations.\footnote{The experiment presented here is the simplest instance within a class of experiments that has recently been extensively analyzed in the literature, and wherein a plurality of mutually separated parties aim to communicate with a common receiver by the use of \textit{one }shared quantum particle (Horvat, 2019; Horvat \& Dakić, 2021b; Zhang et al., 2022; Chen \textit{et al.}, 2024; Maisriml \textit{et al.}, 2026).} The photon's initial quantum state is thus $\frac{1}{\sqrt{2}}(\ket{1}+\ket{2})$, where $\ket{i}$ is the state of the photon being localized at $\text{Alice}_{\text{i}}$'s location. The two parties each act on the photon by either leaving it intact or by applying a local $\pi$-phase-rotation, thereby transforming the initial state into 

\begin{equation}
    \frac{1}{\sqrt{2}}(e^{i\pi a_1}\ket{1}+e^{i\pi a_2}\ket{2}),
\end{equation}
where $a_1,a_2 \in \left\{ 0,1 \right\}$. Finally, when the photon reaches Bob, it undergoes a projective measurement on states $\ket{\pm}\equiv\frac{1}{\sqrt{2}}(\ket{1}\pm\ket{2})$.\footnote{Let us add a cautionary note here. Bob's measurement is described above in terms of projectors on states $\frac{1}{\sqrt{2}}(\ket{1}\pm\ket{2})$, which might give the impression of the measurement being non-local, thereby casting doubt on the possibility of assigning a localized event to it. This is however merely an artefact of the simplified description of our quantum-optical setup. Bob's measurement is surely spatiotemporally localized, since it amounts to a local interaction between a beam-splitter and a photon: a more accurate quantum-mechanical model could indeed track the unitary evolution of the photon all the way up to the beam-splitter, and then describe in detail their interaction, leading to a measurement outcome. Such details would however amount to unnecessary complications, which is why we choose to work with our simplified models. One is only to keep in mind that all measurements referred to throughout the paper are local, in the sense that they occur within relatively small regions of spacetime.}

It is simple to verify that Bob's measurement outcome $b$ is once again correlated with $(a_1,a_2)$ in such a way that $I_2 = \frac{1}{2} \neq 0$. The relevant difference between this example and the previous ones however lies in the (im)possibility of realizing the correlation between $(a_1,a_2)$ and $b$ by some mediators $(b_1,b_2)$. Namely, note that the only local measurements that do \textit{not }change the number of photons in our experiment are those that effectively detect whether the photon is present at that location or not, thus amounting to a position measurement. The state of the photon, after undergoing such measurement, is either $\ket{1}$ or $\ket{2}$, with 50\% probability, and is in particular independent of $a_1$ and $a_2$, thereby erasing the correlation between these two events and any future outcome of measurements on the photon. Therefore, if the correlation between our events is established by one delocalized photon, then this correlation \textit{cannot} be mediated by events that arise from number-preserving operations. On the other hand, there \textit{is} a physically possible experiment in which \textit{non}-number-preserving operations generate events that preserve and mediate the correlation between $(a_1,a_2)$ and $b$: in fact, one can easily check - see Appendix 1 for details - that the said correlation is preserved by projective measurements on states $\left\{\ket{b_1}_1 \otimes \ket{b_2}_2 \right\}_{b_i =0,1}$, which, in the second-quantization notation, are given by 
\begin{equation}
    \ket{x}_i=\frac{1}{\sqrt{2}}(\mathds{1} +(-1)^xc^{\dagger}_i)\ket{0},
\end{equation}
with $\ket{0}$ being the vacuum state and $c^{\dagger}_i$ the creation operator that ``creates'' one photon at $\text{Alice}_{\text{i}}$'s location. 

Unlike in the case of the previous (almost-)classical examples, our quantum experiment is therefore such that the correlation of interest cannot be mediated by intermediate events, unless these arise out of non-number-preserving operations. In other words, the number of photons incoming towards event $b$ is not equal to one, at least not deterministically so: indeed, it is easy to check that the said number of photons is 1 only with probability $\frac{1}{2}$, and 0 or 2 each with probability $\frac{1}{4}$ (see Appendix 1). Denoting with $X\equiv(x_1,x_2)$ and $y$ the locations at which $(a_1,a_2)$ and $b$ occur, our quantum example is therefore such that there is no physically possible experiment that mediates and preserves the correlation between $(a_1,a_2)$ and $b$ via intermediate events $(b_1,b_2)$, \textit{and} such that the incoming photon number in $\mathcal{C}_{y}^{(-)}$ is probabilistically equal to the incoming photon number in $\mathcal{C}_{X}^{(-)}$. More precisely, there is no such physically possible experiment for which the probability distributions of the total photon number on every spacelike hypersurface $S_X \subset \mathcal{C}_{X}^{(-)}$ are equal to the probability distributions of the total photon number on every spacelike hypersurface $S_y \subset \mathcal{C}_{y}^{(-)}$.

Let us note one important detail concerning the simple quantum example above. The detail is that the invoked non-number-preserving operations amount to projective measurements on states whose particle number is not sharply defined, in that they involve superpositions of the vacuum state and a single-photon state. While such operations are physically possible in the case of photons, and more broadly, in the case of uncharged bosons, the \textit{parity and charge superselection rules} - rules that are deeply ingrained in current theoretical physics and that concord with empirical evidence - prohibit such operations in the case of charged particles and in the case of fermions (see e.g., Wightman, 1995). This implies that, for a straightforward modification of our example in which the photon is replaced by an electron (a charged fermion), there is no experiment in which the correlation of interest is mediated by events arising out of non-number-preserving measurements, such as the ones above. However, as explained in Appendix 1, the correlation of interest \textit{can} nevertheless be mediated and preserved, but only via intermediate events arising out of joint number-preserving operations on the electron and on one \textit{additional} auxiliary electron, previously inserted into the experiment.\footnote{For more details on such ways of circumventing superselection rules, see Aharonov \& Vaidman (2000) and Horvat \textit{et al.} (2020). } Therefore, the incoming electron number in $\mathcal{C}_{y}^{(-)}$ (which amounts to \textit{two} electrons) once again differs from the incoming electron number in $\mathcal{C}_{X}^{(-)}$ (which amounts to \textit{one} electron): more precisely, whereas each spacelike hypersurface $S_X \in \mathcal{C}_{X}^{(-)}$ contains one electron, there are spacelike hypersurfaces $S_y \in \mathcal{C}_{y}^{(-)}$ that contain two electrons.

\bigskip

\textbf{Example 3.} Let us survey one more example before proceeding with the reaching for our principle. We will consider a $\text{GIP}_{\text{2n}}$ that effectively amounts to $n$ replicas of the previously analyzed quantum $\text{GIP}_{\text{2}}$. More precisely, $2n$ parties - $\text{Alice}_{\text{1}}$,...,$\text{Alice}_{\text{2n}}$ - share $n$ particles, such that their joint state is given by

\begin{equation}\label{random 3}
    \frac{1}{2^{n/2}}(\ket{1}_1+\ket{2}_1)\otimes(\ket{3}_2+\ket{4}_2)\otimes . . . \otimes (\ket{2n-1}_n+\ket{2n}_n),
\end{equation}
where $\ket{i}_k$ corresponds to the $k$-th particle being at $\text{Alice}_{\text{i}}$'s location. Each party applies a local $\pi$-phase-rotation or leaves their particle intact, thereby transforming the state into 
\begin{equation}
    \frac{1}{2^{n/2}}(e^{i\pi a_1}\ket{1}_1+e^{i\pi a_2}\ket{2}_1)\otimes(e^{i\pi a_3}\ket{3}_2+e^{i\pi a_4}\ket{4}_2)\otimes . . . \otimes (e^{i\pi a_{2n-1}}\ket{2n-1}_n+e^{i\pi a_{2n}}\ket{2n}_n).
\end{equation}

The particles are then sent to Bob, who measures each particle $i$ in basis $\left\{\ket{+}_i,\ket{-}_i \right\}$, where $\ket{\pm}_i\equiv\frac{1}{\sqrt{2}}(\ket{2i-1}\pm \ket{2i})$, and finally outputs the sum-modulo-2 of the outcomes of the $n$ measurements. It is simple to inspect that this straighforward generalization of our previous example generates $I_{2n}=\frac{1}{2} \neq 0$, thus amounting to a $\text{GIP}_{\text{2n}}$. Once again, it is impossible to preserve and mediate this correlation by $2n$ intermediate events, without a (probabilistic) change in the particle number. Furthermore, this same impossibility holds, for the very same reasons, for any number $k>n$ of intermediate events. It is however clear that there is a physically possible experiment in which the correlation is mediated by $n$ events, without any change in particle number: in fact, the original experiment itself already alludes to this, as Bob's measurement involves $n$ separate measurements on each of the particles, that might as well have been executed prior to reaching his location. Indeed, indicating with $(b_1,...,b_n)$ the events corresponding to the intermediate events arising out of projective measurements on $\left\{\ket{+}_i,\ket{-}_i \right\}$, it holds that $b_i=a_{2i-1}\oplus a_{2i}$, which then obviously preserves and mediates the correlation between $\mathbf{a}$ and $b$.

\subsection*{IV.II. Analytic Evidence}

Note that the instances we surveyed above amounted to $\text{GIP}_{\text{m}}$-s that generate the logically maximum possible interference term $I_{m}=\frac{1}{2}$. We will henceforth denote all such maximal-interference-generating phenomena as \textit{maximal} $\text{GIP}_{\text{m}}$-s. Let us now finally take in all of the examples above and tentatively hypothesize that their features generalize as follows: that while (i) there are maximal $\text{GIP}_{\text{m}}$-s generated by $\lceil m/2 \rceil$ particles that \textit{cannot} allow for the relevant correlation to be mediated by more than $\lceil m/2 \rceil$ intermediate mutually spacelike separated events without a change in particle number, it is nevertheless the case that (ii) any maximal $\text{GIP}_{\text{m}}$ generated by $\lceil m/2 \rceil$ particles \textit{can} be mediated by $\lceil m/2 \rceil$ mutually spacelike separated events, without a change in particle number. Given that a large class of $\text{GIP}_{\text{m}}$-s that current physics acquaints us with are all generated with no less than $\lceil m/2 \rceil$ particles, \textit{if} claim (ii) were to hold, we would be warranted to tentatively put forward the following conjecture: while there are instances of $\text{GIP}_{\text{m}}$-s whose correlation cannot be mediated by more than $\lceil m/2 \rceil$ events without a change in particle number, \textit{any} maximal $\text{GIP}_{\text{m}}$ \textit{whatsoever} allows for its correlation to be mediated by \textit{at least} $\lceil m/2 \rceil$ mutually spacelike separated events, without a change in particle number. Before reaching this moment of conjecture, we are however to dwell more on claim (ii). In fact, whereas the validity of (i) has already been established by our concrete examples above, the justification of claim (ii) - the justification of a purportedly universal statement - requires far more work, in that it has been established merely in the very particular case sketched above. What we will now see is that it \textit{does} turn out to hold for practically all semi-general interference experiments, and therefore for a significantly large subclass of $\text{GIP}_{\text{m}}$-s (the ``practically''-caveat will be expanded on below).

Recall that a semi-general interference experiment of order $m$ consists in a certain number of particles being acted upon by $m$ spatially separated devices and subsequently undergoing a binary measurement. More precisely, any such experiment that can be modelled within non-relativistic quantum theory can be described as follows. Supposing that the experiment involves $n$ particles whose internal degrees of freedom have dimensionality $d$, an arbitrary initial state of the particles is described by $\rho \in \mathcal{L}((\mathbb{C}^m\otimes \mathbb{C}^d)^{\otimes n})$, where $\mathbb{C}^m$ and $\mathbb{C}^d$ are the spaces assigned to the spatial and internal degrees of freedom of each of the $n$ particles. The intermediate devices are furthermore so restricted as to implement local transformations that do not increase the number of particles. That is, for each list of configurations $\mathbf{a}\equiv (a_1...a_m)$, if the $n$ particles are sent through devices $\mathbf{i}\equiv (i_1,...,i_n)$, the output state can depend only on configurations $\mathbf{a}_{\mathbf{i}}\equiv (a_{i_1}...a_{i_n})$, and cannot contain more than $n$ particles. While a general such transformation could be quantum-mechanically described by some number-non-preserving completely positive map (CP-map), here we will restrict ourselves to those cases that can be described in terms of number-preserving unitary operators. For each list of configurations $\mathbf{a}$, the joint action of the intermediate devices will thus be assumed to implement a unitary operator $U^{(\mathbf{a})}$, whereby locality imposes the following decomposition:
\begin{equation}
    U^{(\mathbf{a})}=\sum_{\mathbf{i}} \ket{\mathbf{i}}\bra{\mathbf{i}} \otimes U_{\mathbf{i}}^{(\mathbf{a}_{\mathbf{i}})}.
\end{equation}

In the above equation, each spatial state $\ket{\mathbf{i}}=\ket{i_1...i_n} \in (\mathbb{C}^m)^{\otimes n}$ corresponds to the particles traveling through devices $(i_1,...,i_n)$, and each $U_{\mathbf{i}}^{(\mathbf{a}_{\mathbf{i}})}$ is an $\mathbf{a}_{\mathbf{i}}$-dependent unitary operator acting upon the particles' internal degrees of freedom. In other words, if the particles are sent in state $\ket{i_1...i_n}$, then they undergo a unitary transformation that depends only on inputs $(a_{i_1}...a_{i_n})$, thereby ensuring the locality of the devices' operations. Furthermore, if the incoming particles have mutually non-overlapping spatial wave-functions - i.e. if they are sent in state $\ket{i_1...i_n}$, with $i_k\neq i_l$, for all `$k,l$' - then locality also implies that the unitaries act as follows: 
\begin{equation}
    U^{(\mathbf{a})} \left(\ket{\mathbf{i}}\otimes\ket{\phi}\right) = \ket{\mathbf{i}}\otimes \left[U_{i_1}^{(a_{i_1})} \otimes . . . \otimes U_{i_n}^{(a_{i_n})}\right] \ket\phi,
\end{equation}
where each $U_{i_l}^{(a_{i_l})}$ is a unitary operator that acts on the internal degree of freedom of the $l$-th particle. Indeed, if particle $l$ is sent deterministically to device $i_l$, then its output state can depend only on $a_{i_l}$. Proceeding with the final stage of the experiment, after passing through the $m$ intermediate unitary devices, the particles finally reach the $(m+1)$-st device, whose action can be described by an arbitrary binary POVM $\Pi=\left\{\Pi_0,\Pi_1 \right\}$, with $\Pi_i \in \mathcal{L}((\mathbb{C}^m\otimes \mathbb{C}^d)^{\otimes n})$.\footnote{As we will explain at the end of this subsection, the assumption that the measurement $\Pi$ acts \textit{only} on the $n$ particles restricts somewhat the range of semi-general interference experiments to which the subsequent discussion will apply.}

A non-relativistic quantum-mechanical semi-general interference experiment can therefore be characterized in terms of a triple $T=\left(\rho, U^{(\mathbf{a})}, \Pi \right)^{n}_d$, where $n$ is the number of particles and $d$ the dimensionality of their internal degrees of freedom. The semi-general interference term generated by such a triple is accordingly given by 

\begin{equation}
    I_m(T)=\frac{1}{2^m} \sum_{\mathbf{a}}P_T(\oplus_i a_i|\mathbf{a})-\frac{1}{2},
\end{equation}
with 
\begin{equation}
    P_T(b|\mathbf{a})=\Tr \left( \Pi_b U^{(\mathbf{a})}\rho U^{(\mathbf{a})\dagger}\right).
\end{equation}

As announced beforehand, our goal is to show that all such experiments allow for at least $\lceil m/2 \rceil$ number-preserving local measurements, whose outcomes mediate and preserve the correlation between $\mathbf{a}$ and $b$. In order to show that, let us first note that all such $\text{GIP}_{\text{2n}}$ that feature $n$ quantum particles satisfy a property that will be of use to us, summarized in the following Lemma.

\bigskip

\textbf{Lemma 1.} Let a semi-general interference experiment of order $2n$ be characterized by triple $T=\left(\rho, U^{(\mathbf{a})}, \Pi \right)_d^{n}$, where $\rho=\sum_{\psi} p_{\psi}\ket{\psi}\bra{\psi}$. If $I_m(T)=\frac{1}{2}$, then for each $\ket{\psi}$:

\begin{equation*}
    U^{(\mathbf{a})}\ket{\psi}=G\sum_{B\in\mathcal{B}_{2n}}e^{i\phi_B}\sqrt{q_B}\left[\frac{1} {\sqrt{2}} \left(\ket{\beta^{(0)}}_B + (-1)^{\sum_la_l}\ket{\beta^{(1)}}_B \right) \right],
\end{equation*}
where $G$ is a unitary operator; $\mathcal{B}_{2n}$ is the set of all bipartitions $B:\left\{1,...,2n \right\} \rightarrow \left\{0,1 \right\}$, such that $|b^{-1}(0)|=|b^{-1}(1)|=n$; $q_B$ is a probability distribution over such bipartitions; $\phi_B$ are real coefficients; and $\ket{\beta^{(0)}}_B$ and $\ket{\beta^{(1)}}_B$ are normalized and mutually orthogonal vectors in $(\mathbb{C}^{2n}\otimes\mathbb{C}^{d})^{\otimes n}$

\smallskip

\textit{Proof}. See Appendix 2.

\bigskip

While the particularities of the above statement are relatively opaque, its crux is rather simple: if maximal interference is to be generated, the state of the $n$ particles, after being unitarily transformed, needs to necessarily be of a certain form. This form is in particular such that the dependence on the devices' configurations $(a_1...a_{2n})$ is encoded in $\pi$-phases attached to certain equally weighted components of the wave function, thus somewhat mirroring the examples we sketched beforehand. It can accordingly be shown that any such experiment allows for its correlation between $\mathbf{a}$ and $b$ to be established by further intermediate measurements, as stated in the following theorem.

\bigskip

\textbf{Theorem 1.} Let a semi-general interference experiment of order $2n$ be characterized by triple $T=\left(\rho, U^{(\mathbf{a})}, \Pi \right)_d^{n}$. If $I_{2n}(T)=\frac{1}{2}$, then there exists another semi-general interference experiment of order $2n$ that is characterized by triple $T'=\left(\rho, U^{(\mathbf{a})}, \Pi' \right)_d^{n}$, such that $I_{2n}(T')=\frac{1}{2}$ and

\begin{equation}
    \Pi'_b=\sum_{B \in \mathcal{B}_n}\sum_{b_1...b_n=0}^1G E_B H_B^{\dagger} \left(M^{(B,1)}_{b_1} \otimes . . . \otimes M^{(B,n)}_{b_n} \right) \Pi^{(B)}_b \left(M^{(B,1)}_{b_1} \otimes . . . \otimes M^{(B,n)}_{b_n} \right) H_B E_B G^{\dagger},
\end{equation}
where $G$ and $H_B$ are unitary operators, whereas $\left\{ E_B\right\}_B$, $\left\{M^{(B,1)}_{b_1} \otimes . . . \otimes M^{(B,n)}_{b_n} \right\}_{b_i}$ and $\left\{\Pi^{(B)}_b\right\}_b$ are projective measurements. In particular, $P_{T'}(b|b_1...b_n)=\delta_{b,\oplus_ib_i}$, $P_{T'}(b_1...b_n|\mathbf{a})=\frac{1}{2^{n-1}} \delta_{\oplus_i b_i, \oplus_j a_j}$, and $P_{T'}(B|\mathbf{a})=P_{T'}(B)=q_B$.


\textit{Proof}. See Appendix 3.

\bigskip

The above theorem states that for any semi-general interference experiment $T$ that lies within our large class, there is another experiment $T'$ that achieves maximal interference with the very same input state $\rho$ and the same unitaries $U^{(\mathbf{a})}$, but with a different final measurement $\Pi'$, whereby this latter measurement has the following structure: (i) the particles first undergo a unitary transformation $G$, a projective measurement $\left\{ E_B\right\}_B$, and a unitary $H_B$ ; (ii) then they undergo $n$ possibly spacelike separated number-preserving measurements $\left\{M^{(B,1)}_{b_1} \otimes . . . \otimes M^{(B,n)}_{b_n} \right\}_{b_i}$, with outcomes $(b_1...b_n)$; and (iii) a final measurement is performed that outputs $b=\oplus_i b_i$. Experiment $T'$ thus features $n$ additional possibly mutually spacelike separated events corresponding to the measurement outcomes $(b_1...b_n)$, that mediate and preserve the correlation between $\mathbf{a}$ and $b$. Let us mark this explicitly in the following corollary.

\bigskip

\textbf{Corollary 1.} Let a semi-general interference experiment of order $2n$ be characterized by triple $T=\left(\rho, U^{(\mathbf{a})}, \Pi \right)_d^{n}$. If $I_{2n}(T)=\frac{1}{2}$, then there exists another semi-general interference experiment of order $2n$ that is characterized by triple $T'=\left(\rho, U^{(\mathbf{a})}, \Pi' \right)_d^{n}$, which features $n$ additional mutually spacelike separated number-preserving measurements, whose outcomes $(b_1,...b_n)$ are so distributed that

\begin{equation}
    P_{T'}(b|\mathbf{a})=\sum_{b_1...b_n} P_{T'}(b|b_1...b_n)P_{T'}(b_1...b_n|\mathbf{a}),
\end{equation}
where $P_{T'}(b|b_1...b_n)=\delta_{b,\oplus_i b_i}$
and $P_{T'}(b_1...b_n|\mathbf{a})=\delta_{\oplus_ib_i,\oplus_ja_j}$.

\bigskip

The above statements all concerned semi-general interference experiments with an even number of intermediate devices $m=2n$. Let us now state an analogous - but slightly less general - theorem that concerns cases with odd $m$.  

\bigskip

\textbf{Theorem 2.} Let a semi-general interference experiment of order $(2n-1)$ be characterized by triple $T=\left(\rho, U^{(\mathbf{a})}, \Pi \right)_d^{n}$, where $\rho=\sum_{\psi}p_{\psi}\ket{\psi}\bra{\psi}$. Suppose that for each $\ket\psi$, there are two vectors $\left\{\ket{\mathbf{k}}_{\psi},\ket{\mathbf{l}}_{\psi} \right\} \subset (\mathbb{C}^{2n-1})^{\otimes n}$, such that
$\braket{\mathbf{i}|\psi}\neq 0$ if and only if $\ket{\mathbf{i}} \in \left\{\ket{\mathbf{k}}_{\psi},\ket{\mathbf{l}}_{\psi} \right\}$. 
If $I_{2n-1}(T)=\frac{1}{2}$, then there exists another semi-general interference experiment of order $(2n-1)$ that is characterized by triple $T'=\left(\rho, U^{(\mathbf{a})}, \Pi' \right)_d^{n}$, such that $I_{2n-1}(T')=\frac{1}{2}$, and which features $n$ additional mutually spacelike separated number-preserving measurements, whose outcomes $(b_1,...b_n)$ are so distributed that

\begin{equation}
    P_{T'}(b|\mathbf{a})=\sum_{b_1...b_n} P_{T'}(b|b_1...b_n)P_{T'}(b_1...b_n|\mathbf{a}),
\end{equation}
where $P_{T'}(b|b_1...b_n)=\delta_{b,\oplus_i b_i}$
and $P_{T'}(b_1...b_n|\mathbf{a})=\delta_{\oplus_ib_i,\oplus_ja_j}$.

\smallskip

\textit{Proof}. See Appendix 4.

\bigskip

The only difference between the odd case and the previously stated even case is that we now appeal to a further restriction, which is that the input state $\rho$ is a mixture of states $\ket\psi$ that each have support on exactly \textit{two} different spatial configurations $\left\{\ket{\mathbf{k}}_{\psi},\ket{\mathbf{l}}_{\psi} \right\} \subset (\mathbb{C}^{2n-1})^{\otimes n}$ of the $n$ particles. While the more general proof remains out of analytic reach, in Appendix 5 we provide some preliminary evidence that the statement does generalize, by showing that it holds for certain experiments with states that have support on \textit{three} different spatial configurations. 

Before proceeding, we ought to point out a feature regarding how \textit{mixed} input states $\rho$ are treated in Theorems 1 and 2, and how the assumptions on which these treatments rely affect our overall discussion. Namely, the intermediate operations that appear in the experiments $T'$ constructed in the proofs of the said theorems are such that \textit{different} operations are applied to \textit{different} components $\ket{\psi}$ of the said input mixed state $\rho$. In other words, the constructed protocols rely on the possibility of either (i) accessing the purifying degrees of freedom, in case the mixture is \textit{im}proper (i.e. arising due to entanglement), or (ii) accessing the classical records present in the preparation of the state, in case the mixture is proper.\footnote{More precisely, let $S$ be initially prepared in state $\rho_S= \sum_{\psi} p_{\psi} \ket{\psi}_S \bra{\psi}$. The proofs of the theorems assume that there is a further system $E$, such that their joint state $\rho_{SE}$ satisfies the following: there is a projective measurement $\left\{ \Pi_{\psi}\right\}_{\psi}$ on system $E$, which is such that, if outcome $\psi$ is obtained, the post-measurement state of $S$ is $\ket{\psi}$. The mathematical possibility of there being such additional system with such $\psi$-selecting measurements is ensured by the GHJW-theorem (Gisin, 1989; Hughston \textit{et al}., 1993). The operations $H_B$ appearing in Theorems 1 and 2 are then assumed to have the following form: $H_B=\sum_{\psi}H_B^{(\psi)} \otimes \Pi_{\psi}$, where each $H_B^{(\psi)}$ acts only on system $S$.} That such accessing is needed can be exemplified on a simple variation of Example 3 from the previous subsection. Let 2 particles be used to generate $I_4 = \frac{1}{2}$, with the help of 4 $\pi$-phase shifters, which we will denote with $(s_1,...,s_4)$. Suppose the two particles are sent in a mixture of (a) one particle going through $(s_1,s_2)$, and the other through $(s_3,s_4)$, and (b) one particle going through $(s_1,s_3)$ and the other through $(s_2,s_4)$. It is clear that the intermediate operations need to implement different measurements for each branch of the mixed state: if the particles are in state (a), then the measurements should interfere path $s_1$ with path $s_2$, and path $s_3$ with path $s_4$; conversely, if the particles are in state (b), then the measurements should interfere $s_1$ with $s_3$, and $s_2$ with $s_4$.

The above assumption is not problematic \textit{per se}: according to quantum theory and experimental evidence, a mixed state $\rho$ does indeed arise either due to entanglement or due to classical mixing, thus seemingly always pointing to an environment $E$ that can at least in principle be physically accessed. Our remark does however highlight another \textit{restriction} of our proofs. Namely, note that a semi-general interference experiment has been characterized by triple $T=(\rho, U^{(\mathbf{a})},\Pi)$, where $\Pi$ is assumed to act \textit{only} on system $S$, the one that was initially in state $\rho$. However, noting that any such system $S$ found in a mixed state is necessarily accompanied by an in principle accessible environment $E$, one can also envisage a different kind of semi-general interference experiments, in which the final measurement $\Pi$ is conducted on the \textit{joint} system comprised by $S$\textit{ and} $E$. This broader class of semi-general interference experiments is thus not directly covered by our proofs. However, in Appendix 6 we show that in case the mixture $\rho$ is \textit{proper}, the proofs \textit{do} trivially apply after all. The case of \textit{im}proper mixtures is on the other hand \textit{not} trivially covered by our proofs, as also explained at the end of the same Appendix: a detailed analysis thereof thus exceeds the present article.

\subsection*{IV.III. Evaluating the Evidence}

Let us summarize the road we have traversed in this section. After presenting Examples 1-3, some of their properties motivated us to tentatively suggest that all maximal $\text{GIP}_{\text{m}}$-s allow for their correlation to be mediated, without a change of the number of particles, by \textit{at least} $\lceil m/2 \rceil$ mutually spacelike separated events, and in some cases, by \textit{no more} than $\lceil m/2 \rceil$ such events. In order to explore whether this is really the case we considered a wide range of $\text{GIP}_{\text{m}}$-s - a large subclass of semi-general interference experiments of order $m$ - and proved several statements that show that the aforementioned claim does hold for the phenomena within this range. The question that however remains is how the evidence presented above bears on the more general claim that we are ultimately after. More precisely, two questions still remain. How does our investigation concerning our subclass of semi-general interference experiments bear on the class of \textit{all} semi-general interference experiments? How does the latter investigation bear on the even more general investigation of phenomena pertaining to the class of all $\text{GIP}_{\text{m}}$-s?

Starting with the first question, note that our subclass of semi-general interference experiments was restricted in three regards: (i) the intermediate devices were constrained to implement unitary number-preserving transformations, (ii) in the case of odd $m$, the input state was required to have support on two (or, in some cases, three) spatial configurations of the particles, and (iii) if the input state $\rho$ is an \textit{im}proper mixture, then the final measurement $\Pi$ is required to act only on the $n$ particles used in the experiment. Let us now go through some heuristic considerations that, while by no means proving, do allude to the possibility of relaxing some of these restrictions. As stated in (i), we constrained ourselves only to unitary number-preserving transformations, whereas more general non-number-preserving CP-maps might in principle also be allowed. For instance, the original experiment with multiple slits does not involve unitary operators, as the intermediate devices are implemented by slits, that can either let the incoming particles go through or not. However, in this case, albeit $m$-th order interference can be generated by $\lceil m/2 \rceil$ particles, it is simple to notice that the interference term cannot be \textit{maximal}, i.e. it holds that $I_{m}<\frac{1}{2}$. More generally, if for some configuration $\mathbf{a}$, there is a non-zero probability of the post-transformation state $\rho^{(\mathbf{a})}$ being the vacuum state, then interference again cannot be maximal. \footnote{Let $\mathbf{a}=(a_1a_2...a_m)$ and $\mathbf{\bar{a}}=((a_1 \oplus 1)a_2...a_m)$ be devices' configurations that differ only in the first input. Since $\sum_ia_i\neq\sum_i\bar{a}_i$, a necessary condition for the interference term to be maximal is for the post-transformation states $\ket\psi_{\mathbf{a}}$ and $\ket\psi_{\mathbf{\bar{a}}}$ to be orthogonal. Suppose that $\ket\psi_{\mathbf{a}}=\ket{0}$, where $\ket0$ is the vacuum state. In other words, if the configurations are set to $\mathbf{a}$, the particles are all blocked from passing through. It then follows that $\ket\psi_{\mathbf{\bar{a}}}=\alpha \ket{0} + \beta \ket{\bar{0}}_{a_1}$, where $\ket{\bar{0}}_{a_1}$ is an $a_1$-dependent state that is orthogonal to $\ket0$, and $\alpha$ is the amplitude corresponding to no particle passing through the first device. It thus follows that $\braket{\psi_{\mathbf{a}}|\psi_{\mathbf{\bar{a}}}}=\alpha\neq 0$, implying $I_m < \frac{1}{2}$.} We thus deem it safe to assume the restriction to number-preserving operations to be unproblematic. Furthermore, as previously shown in (Horvat \& Dakić, 2021a), the impossibility of generating $m$-th order interference with less than $\lceil \frac{m}{2}\rceil$ particles holds also in the case of general number-preserving CP-maps; it thus appears reasonable to expect that our hereby proven theorems also extend accordingly. A proof thereof however exceeds the bounds of our investigation.

Passing over to point (ii), note that the analytic result, in the case of odd $m$, concerns only a restricted class of input states, leaving the remaining cases open to further investigation. While an analytic proof would certainly be desirable here, note that it would be quite surprising if the generalization of Theorem 2 turned out not to hold, given the validity of Theorem 1. The hand-wavy, intuitive, reason for why this is so is that a semi-general interference experiment of order $m$ that features $n$ particles is, so to say, the less classical and the more peculiar in proportion to quantity $\frac{m}{n}$. Namely, even ordinary billiard balls can realize the minimal case of $\frac{m}{n}=1$, whereby one can trivially identify our $n$ mediating events; but only a carefully executed quantum experiment can realize the maximal case with $\frac{m}{n}=2$, which, as shown in Theorem 1, nevertheless allows for the existence of our $n$ mediating events. The cases in between these two extremes - between, so to say, the fully classical and the fully quantum - are presumably expected to also allow for the existence of the $n$ mediating events, at pains of an arguably arbitrary discontinuity. In other words, one would expect that the following property holds: for any $m<2n$, if there exists an experiment of order $m$ that features $n$ particles but that does \textit{not} allow for $n$ mediating events, then there also exists an experiment of order $(m+1)$ that features $n$ particles and that does \textit{not} allow for $n$ mediating events. Or, by contraposition: if all experiments of order $m$ that feature $n$ particles allow for $n$ mediating events, then also all experiments of order $m'<m$ that feature $n$ particles allow for $n$ mediating events. If this statement holds, then Theorem 1 implies the generalization of Theorem 2. Here we are unfortunately not in a position to rigorously justify this inference, so we will stay contented with our hereby stated hand wavy intuition and preliminary evidence, leaving a full analysis for another occasion.

Let us finally say a few words on the last restriction (iii), which concerns the relationship between the final measurement $\Pi$ and the status of the mixed state $\rho$ in which the $n$ particles are initially sent. Our results directly apply if $\Pi$ acts exclusively on the $n$ particles used in the experiment, rather than jointly on the particles together with their environment. But, \textit{if} the initial state $\rho$ is a \textit{proper} mixture, then our results \textit{also} apply in the case $\Pi$ acts jointly on the particles and on the environment. The \textit{only} case that is thus \textit{not} covered by our proofs is the case in which the said mixture is improper \textit{and} $\Pi$ acts jointly on the particles and on the environment. This case needs to thus be fully investigated elsewhere. 

In light of the above, despite the various restrictions thereby indicated, we deem it reasonable to take our analytic results as providing good evidence for taking \textit{all} semi-general interference experiments of order $m$ to allow for $\lceil m/2 \rceil$ mediating events. The question that remains now is how this latter statement bears on our tentative conjecture, which aims to take us from the realm of interference experiments to the broader realm of phenomena included in the category of $\text{GIP}_{\text{m}}$-s. First, note that semi-general interference experiments certainly \textit{are} paradigm examples of $\text{GIP}_{\text{m}}$-s: namely, any $\text{GIP}_{\text{m}}$ needs to be such that events $\mathbf{a}$ and $b$ can be identified, whose correlation satisfies $I_m \neq 0$, and such that this correlation is not established by some common cause, but, paradigmatically, by some objects that travel between locations $\mathbf{x}$ and location $y$. A semi-general interference experiment is a natural exemplification of such a phenomenon, in that certain objects interact at locations $\mathbf{x}$, eliciting events $\mathbf{a}$, and thence travel to location $y$, where they again interact with a further object in order to elicit event $b$. However, as mentioned beforehand, there are certainly other examples of $\text{GIP}_{\text{m}}$-s. One such class consists in experiments analogous to the ones we have just studied, but that feature waves, instead of particles. In the classical case - e.g. in case the relevant correlation is established by a sound wave or by an electromagnetic wave - it is clear that the phenomenon allows for $m$ intermediate events, analogous to the case of $m$ classical particles: indeed, the mediating events $(b_1,...b_m)$ can simply be so chosen as to correspond to the values of the wave at locations $(y_1,...y_m)$, where each $y_i$ lies in the immediate causal future of $x_i$. On the other hand, the quantum-field-theoretic case would require a completely new investigation: in fact, it is not even clear whether there are any physically possible $\text{GIP}_{\text{m}}$-s that can be modelled within quantum-field theory, but that \textit{cannot} be modelled in the non-relativistic fashion that we have been following so far. This class of cases, if it exists at all, thus definitely exceeds our capabilities here and should be further investigated elsewhere.

\section*{V. The Principle}

Our analytic evidence, though certainly imperfect and still to be rigorously complemented, does appear to provide strong support to our general claim that concerns the wide category of $\text{GIP}_{\text{m}}$-s. We are thus finally ready to tentatively, boldly, fallibly, conjecture that our general claim \textit{does} hold, and to formulate it more rigorously and abstractly in terms of a general principle. In other words, we want to transform the still somewhat vague observation that ``all maximal $\text{GIP}_{\text{m}}$-s seem to allow for their correlation to be mediated, without a change in particle number, by at least $\lceil m/2 \rceil$ events, and in some cases, by no more than $\lceil m/2 \rceil$ events'' into a rigorously formulated principle that regulates correlations between spatiotemporally localized event-occurrences. Let us accordingly rephrase more precisely each of the elements appearing in our claim, starting with an explication of what it is for a phenomenon to ``allow'' for the occurrence of our mediating events. Recall that the mediating events in question do not need to partake in the phenomenon under consideration, but can also occur in \textit{another} phenomenon, which is sufficiently similar to the original one. In fact, Theorems 1 and 2 state that any semi-general interference experiment $T$ that generates maximal interference is such that there is another $T'$, which features our desired mediating events: in particular, $T$ and $T'$ have in common both the incoming state $\rho$ and the unitaries $U^{(\mathbf{a})}$. Now, note that this commonality of $(\rho,U^{(\mathbf{a})})$ can be re-described in a more abstract manner that rids itself of the use of quantum-mechanical vocabulary: $T$ and $T'$ need to be so related to each other that they agree on what happens in the causal past of $X=(x_1,...,x_m)$, that is, in $\mathcal{C}^{(-)}_{X}$. More precisely, for any PE-model $P$ that adequately describes phenomenon $T$, there needs to exist a PE-model $P'$ that adequately describes $T'$, in such a way that the two models agree on the distribution of events in any arbitrarily selected region within the causal past of $X$: succinctly, $P_Z=P_Z'$, for all $Z \subset \mathcal{C}^{(-)}_{X}$. 

Another component of our proposal that needs to be precisified is the supposed ``mediation'' of the correlation between $\mathbf{a}$ and $b$ on behalf of events $\mathbf{b}\equiv (b_1...b_n)$. Indeed, at pains of trivialization, certain further conditions need to still be imposed, besides the already thematized condition that $P(b|\mathbf{a})=\sum_{\mathbf{b}}P(b|\mathbf{b})P(\mathbf{b}|\mathbf{a})$. One such condition is that there should not exist any other event $b'$ that occurs in the causal past of $b$ and which is so correlated with $\mathbf{a}$ as to generate maximal interference.\footnote{If this was not the case, then even Example 2 above - our paradigmatic quantum $\text{GIP}_{\text{2}}$ - would allow for $m=2$ mediating events. Indeed, a possible experiment could then be constructed in which: (i) the particle is interfered eliciting event $b'$ that generates maximal interference, (ii) the particle is then sent either to the left or to the right side depending on the value of $b'$, (iii) local detectors then check whether the particle passed through the left or through the right side, eliciting spacelike separated events $(b_1,b_2)$ and (iv) after bouncing off a mirror, the particle reaches another detector that measures whether the particle went to the left or to the right side, eliciting event $b$. In this example, the correlation between $b$ and $(a_1,a_2)$ is locally mediated by events $(b_1,b_2)$. However, this mediation is trivial, in that it relies on maximal interference already having been achieved previously at $b'$.} Furthermore, each event $b_i$ can depend on at most $\lceil m/n \rceil$ events among $(a_1...a_m)$: more precisely, $P(b_i|\mathbf{a})=P(b_i|S_i(\mathbf{a}))$, where $S_i(\mathbf{a})\subset (a_1...a_m)$ and $|S_i(\mathbf{a})|\leq \lceil m/n \rceil$. Each $b_i$ can thus be thought of as mediating the correlation between $b$ and at most $\lceil m/n \rceil$ different $a_i$-s, which was for instance the case in Example 3 above, where each $b_i$ mediated the correlation between $(a_{2i-1},a_{2i})$ and $b$.\footnote{In particular, we do not want to allow for \textit{one} event $b_i$ to mediate the correlation between $b$ and \textit{all} events $\mathbf{a}$: indeed, if we allowed for this, the other events $b_j \neq b_i$ would be redundant, effectively amounting to a mediation exclusively on behalf of the one selected event $b_i$.}

The final component that we still need to address is the reference to the preservation of the number of particles. Recall that our principle would trivialize without \textit{some} such reference, in that even paradigmatic particle-interference experiments allow for mediating events, \textit{if} additional particles of the same type are inserted in the experiment. The way we articulated this constraint beforehand, under Example 2, was by comparing the particle number at spacelike hypersurfaces in $\mathcal{C}^{(-)}_X$ with the particle number at spacelike hypersurfaces in $\mathcal{C}^{(-)}_y$. Remember now that our goal is to ideally reach a principle that is \textit{universal} and that thus regulates event-occurrences \textit{independently} of the objects that may or may not feature therein. While this ideal goal has hereby not been reached, in that we are still referring to the particle number, note that we can come closer to it by recognizing the following. 

All of the evidence that we surveyed in the previous section is only sensitive to relative \textit{differences} in the particle-number between various spacelike hypersurfaces, making the absolute number of particles involved in the phenomena of no relevance. Now, note that two collections of particles of the same type that differ in the number of their constituents necessarily also differ in some of their other properties: e.g. a collection consisting only of 1 electron contains less charge and less mass than a collection consisting of 2 electrons. More generally, and more precisely, let $C$ and $C'$ be two collections of particles of the same type, let $n_C$ and $n_{C'}$ be the number of their respective constituents, and let $\mathcal{Q}_C=\left(Q^{(i)}_C \right)_i$ and $\mathcal{Q}_{C'}=\left(Q^{(i)}_{C'} \right)_i$ be lists containing all of the \textit{non-dynamical quantities} associated to the two collections (e.g. their total mass, charge and spin). It then holds that $n_C=n_{C'}$ if and only if $\mathcal{Q}_{C}=\mathcal{Q}_{C'}$. Since the collections that we are concerned with in our class of phenomena - the collections of particles present at various spacelike hypersurfaces in  $\mathcal{C}^{(-)}_X$ and $\mathcal{C}^{(-)}_y$ - necessarily contain particles of the same type, instead of referring to the particle number, we can thus refer to the non-dynamical quantities that characterize the matter present in the phenomenon under consideration.\footnote{See Appendix 1 for details on why the experiments that feature mediating events elicited by non-number-preserving measurements necessarily rely on an insertion of particles of the \textit{same} type.} 

The particle-number-preserving condition can thereby be partially abstracted away, and transformed into a condition that concerns how quantity $P_{S}(\mathcal{Q})$ compares at different hypersurfaces $S$, where $P_{S}(\mathcal{Q})$ is the probability that the non-dynamical quantities that characterize the totality of matter present in spatiotemporal region $S$ are equal to $\mathcal{Q}$. More precisely, the condition will state that \textit{for any} spacelike hypersurface $S_y \subset \mathcal{C}^{(-)}_y$, \textit{there is} a spacelike hypersurface $S_X \subset \mathcal{C}^{(-)}_X$, such that $P_{S_X}(\mathcal{Q})=P_{S_y}(\mathcal{Q})$. The reason we are not imposing equality $P_{S_X}(\mathcal{Q})=P_{S_y}(\mathcal{Q})$ for \textit{any two} spacelike hypersurfaces $S_y \subset \mathcal{C}^{(-)}_y$ and $S_X \subset \mathcal{C}^{(-)}_X$ is related to our remark at the end of Section IV.II, concerning the treatment of mixed states of incoming particles: namely, we want to generally allow for the mediating events to arise out of measurements on the incoming particles \textit{and} on their environment, i.e. on the classical records or purifying degrees of freedom. While the latter need not be present at \textit{every} surface $S_X \subset \mathcal{C}^{(-)}_X$, they are necessarily present on at least \textit{some} surface $S_X \subset \mathcal{C}^{(-)}_X$ (e.g. at the beginning of the experiment).

Note that our appeal to the non-dynamical quantities $\mathcal{Q}$ would definitely require further elaboration here. Indeed, there might be other ways of abstracting away our constraint that there be no change in the number of particles, by referring to other quantities that characterize the matter present in the phenomenon under consideration - and these ways \textit{might} turn out to be more readily applicable universally across phenomena. It might for instance be sufficient to refer to the total \textit{energy} content of the spatiotemporal regions in question. More precisely, it might the case that the particle-number-preserving condition can be transformed into $P_{S_X}(E)=P_{S_y}(E)$, where $P_{S}(E)$ is the probability that the total energy present in spatiotemporal region $S$ is equal to $E$. An adequate exploration of whether this simplification holds or not can however not be undertaken here, which is why we will, for the time being, proceed with our current formulation in terms of quantities $\mathcal{Q}$.\footnote{Here is, briefly, why the energy-strategy, albeit appearing to be promising, demands further work. If the particles present in the experiment are photons, then the non-number-preserving mediating events probabilistically introduce photons of the \textit{same} frequency, thus changing the energy. For instance, in Example 2, the total energy is changed, up to a multiplicative constant, from $\nu$ to $0$ (with probability $\frac{1}{4}$) or to $2\nu$ (with probability $\frac{1}{4}$), where $\nu$ is the frequency of the photon. The photonic case therefore \textit{does} clearly introduce a probabilistic change in the total energy of the system. On the other hand, in case the particle used is an electron (or some other fermion or charged boson), then the mediating events require a deterministic insertion of another electron or positron. The total rest mass of the system thereby increases deterministically. However, while this \textit{does} imply a change in total energy in the \textit{non}-relativistic regime (when the speed of the particles is much smaller than the speed of light), a more careful investigation would be needed to inspect whether this is also the case in the relativistic regime.} With all of these qualifications in place, let us now introduce two auxiliary definitions in terms of which our principle will finally be formulated. First, taking into account the last of the above qualifications, let us introduce a subclass of $\text{GIP}_{\text{m}}$-s defined by those in which the aforementioned quantities of matter are accordingly preserved.

\bigskip

\textbf{Definition 4.} Let $T$ be a $\text{GIP}_{\text{m}}$. $T$ is said to be \textit{closed} if for any spacelike hypersurface $S_y \subset \mathcal{C}^{(-)}_y$, there is a spacelike hypersurface $S_X \subset \mathcal{C}^{(-)}_X$, such that $P_{S_X}(\mathcal{Q})=P_{S_y}(\mathcal{Q})$, where $P_{S}(\mathcal{Q})$ is the probability that the non-dynamical quantities that characterize the totality of matter present in spatiotemporal region $S$ are equal to $\mathcal{Q}$.

\bigskip

A $\text{GIP}_{\text{m}}$ is therefore closed if, roughly speaking, the amount of matter that flows into location $y$ equals the amount of matter that flows towards locations $X$; or, whatever ends up flowing towards $y$ should have previously already flown somewhere within the past of $X$. Now, taking into account the remaining qualifications above, let us proceed by defining \textit{n-local-completions} of a $\text{GIP}_{\text{m}}$, which, stated concisely, are \textit{other} $\text{GIP}_{\text{m}}$-s that contain $n$ additional events that both preserve and mediate the correlation between the events that occur at $X$ and at $y$. 

\bigskip

\textbf{Definition 5.} Let $T$ and $T^*$ be $\text{GIP}_{\text{m}}$-s. $T^*$ is said to be a \textit{n-local-completion} of $T$ if there exists $Y=(y_1...y_n)$, with $y_i \in \mathcal{C}^{(+)}_{X}\cap \mathcal{C}^{(-)}_{y}$ and $y_i \sim y_j$, such that for any PE-model $P$ of $T$, there is a PE-model $P^*$ of $T^*$ that satisfies the following conditions:
\begin{enumerate}
    \item $P^*_Z=P_Z$, for all $Z \subset \mathcal{C}^{(-)}_{X}$
    \item $P^*_y(\omega_y|\omega_X)=P_y(\omega_y|\omega_X)$
    \item $P^*_y(\omega_y|\omega_X)=\sum_{\omega_Y}P^*_y(\omega_y|\omega_Y)P^*_Y(\omega_Y|\omega_X)$
    \item There is a $z \in \mathcal{C}^{(-)}_{y}$, such that $P^*_{y_i}(\omega_{y_i}|\omega_X\omega_z)= P^*_{y_i}(\omega_{y_i}|\omega_{S^{(\omega_z)}_i(X)}\omega_z)$, where $S^{(\omega_z)}_i(X) \subset X$ and $|S^{(\omega_z)}_i(X)|\leq \lceil \frac{m}{n} \rceil$
    \item For any refinement $P'$ of $P^*$, there is \textit{no} $y' \in \mathcal{C}^{(-)}_{y}$ for which $P'_{y'}(\omega_{y'}|\omega_X)=P'_y(\omega_y|\omega_X)$
\end{enumerate}

\bigskip

\begin{figure}[H]
\centering
\begin{subfigure}{.5\textwidth}
  \centering
  \includegraphics[width=0.95\linewidth]{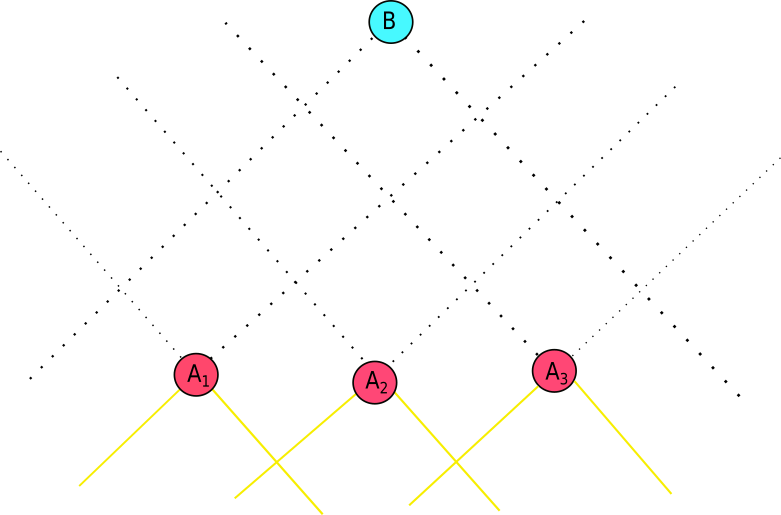}
\end{subfigure}%
\begin{subfigure}{.5\textwidth}
  \centering
  \includegraphics[width=0.95\linewidth]{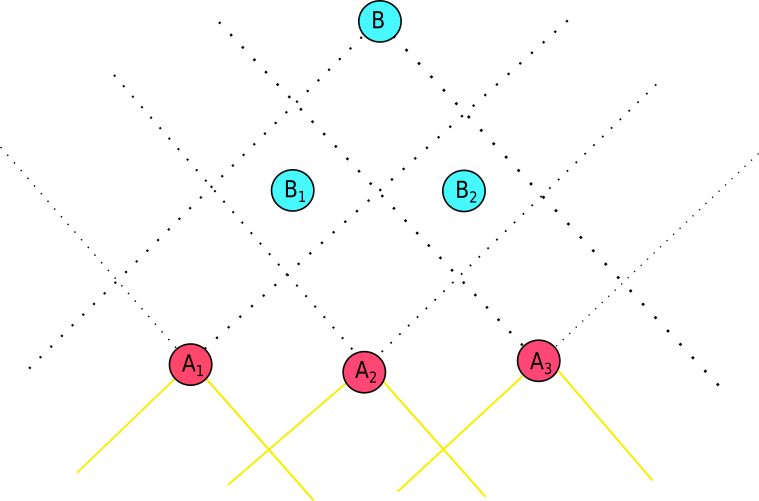}
\end{subfigure}
\caption{A 2-local-completion (right figure) of a $\text{GIP}_{\text{3}}$ (left figure). The 2-local completion contains two additional mutually spacelike separated events $(b_1,b_2)$ that mediate and preserve the correlation between $(a_1,a_2,a_3)$ and $b$. The yellow lines indicate that the events in the causal past of $(a_1,a_2,a_3)$ are equally distributed in both phenomena.}
\label{figlast}
\end{figure}

In the particular class of cases in which $T$ and $T^*$ are such that
$\lceil \frac{m}{n} \rceil=2$, we will say that $T^*$ is a \textit{bi-local completion} of $T$. Let us now tread more carefully through the above definition, in order to see how the aforementioned qualifications are codified therein - see Fig. \ref{figlast}. Note first that $T$ and $T^*$ are assumed, per condition 1, to agree on what happens in the causal past of $X$. Furthermore, $T^*$ is supposed to contain, according to some PE-model $P^*$, $n$ additional mutually spacelike separated events at $Y$, that both preserve (condition 2), and mediate (condition 3), the correlation between the events at $X$ and the event at $y$. Condition 4 in turn secures that each mediating event is limited in its dependence to a subset of at most $\lceil m/n \rceil$ events at $X$: note that we also enabled the possibility of this subset to probabilistically depend on another event at some location $z$.\footnote{Here is the motivation for why we allow for the said probabilistic dependence. Consider a variation of Example 3, in which the $n$ particles are sent in a mixed state $\rho=\frac{1}{2}(\ket{\psi}\bra{\psi} + \ket{\psi'}\bra{\psi'})$, where $\ket{\psi}$ is the state considered in Example 3 (eq. \eqref{random 3}), and $\ket{\psi'}$ is a very similar state, but one in which the first particle is sent through devices $(1,3)$ and the second particle through devices $(2,4)$, whereas the remaining particles are in the same state as in $\ket{\psi}$. Now, we want to legitimize a local completion that consists in letting $b_1$ be elicited by an interferometric measurement on paths $(1,2)$ \textit{if} the incoming state is $\ket{\psi}$, and letting $b_1$ be elicited by an interferometric measurement on paths $(1,3)$ \textit{if} the incoming state is $\ket{\psi'}$. The mediating event $b_1$ thus probabilistically mediates either events $(a_1,a_2)$ \textit{or} events $(a_1,a_3)$.} Finally, condition 5 ensures that there is no event at $y'$, prior to $Y$, that trivializes the mediation by already establishing the required correlation with $X$. 

We have now finally reached the stage at which we are able to state our principle, the final abstraction of the possibility-impossibility structure exhibited in interferometric experiments, the structure that motivated this whole investigation. 

\bigskip

\textbf{Principle.} Every \textit{maximal} $\text{GIP}_{\text{m}}$ has a physically possible \textit{closed bi-local completion}, for any $m$. Conversely, \textit{not} every maximal $\text{GIP}_{\text{m}}$ has a physically possible closed $n$-local completion, for $\lceil \frac{m}{n} \rceil<2$. 

\bigskip

The principle contains two parts that amount to abstractions of, respectively, the \textit{impossibility of 3rd-order} and the \textit{possibility of 2nd-order} particle interference. Indeed, the 3rd-order impossibility has been transformed into the universal existence of closed bi-local completions, whereas the 2nd-order possibility has been morphed into the existence of phenomena that do not have closed completions that are stronger than the bi-local ones. 

In order to avoid misunderstandings, let us now make fully explicit the relationship that holds between the just stated principle on the one hand, and the quantum-mechanical proofs and examples from the previous sections on the other hand. Note that the principle is stated in general ``theory-independent'' terms: it speaks of phenomena that are characterized in terms of events and their statistical correlations, rather than, say, in terms of quantum particles and quantum measurement devices. As such, the validity of the principle is \textit{not implied} by quantum theory, but only \textit{evidentially supported} by it, thereby amounting to a fallible conjecture. More precisely: (i) Corollary 1 and Theorem 2 provide evidence for the validity of the first part of the principle, namely, that ``every maximal $\text{GIP}_{\text{m}}$ has a physically possible closed bi-local completion, for any $m$''; (ii) Examples 2 and 3 provide evidence for the validity of the second part of the principle, namely, that ``not every maximal $\text{GIP}_{\text{m}}$ has a physically possible closed $n$-local completion, for $\lceil \frac{m}{n} \rceil<2$''. These examples and analytic proofs provide the latter evidence in the following way: (a) they provide evidence that quantum mechanics validates our principle; and (b) since quantum mechanics is, as of now, empirically corroborated, the latter supports the claim that our principle \textit{really} holds of real phenomena, and not only according to quantum theory. It might be useful here to once again draw an analogy to the results proved by Bell and Tsirelson. A way of summarizing their contribution is to say that they showed that quantum mechanics validates the following $\text{principle}^*$: ``Bell's inequalities \textit{can} be violated, but \textit{only} up to Tsirelson's bound''. Since quantum mechanics is, as of now, empirically corroborated, this supports the claim that the $\text{principle}^*$ really holds. In sum, both our principle and Bell-Tsirelson's $\text{principle}^*$ are thus not per se implied by quantum theory, but are instead validated by it, and thus evidentially supported by it.\footnote{The Bell-Tsirelson $\text{principle}^*$ speaks of the violation of Bell's inequalities, and thus of the amount by which quantum mechanical predictions - and empirical evidence - depart from the predictions of a certain class of ``locally realistic'' theories. In other words, theories within this latter class do not satisfy Bell-Tsirelson's $\text{principle}^*$. On the other hand, \textit{our} principle seems not to have anything to do with the assumptions that lead to Bell inequalities: thus, ``locally realistic'' theories might as well validate our principle. A detailed examination of the exact relationship between the validity of our principle and ``local realism'' is however to be conducted elsewhere.}

Let us now proceed by reflecting on some properties of our principle. Notice that the principle concerns only \textit{maximal} $\text{GIP}_{\text{m}}$-s, and that it thus, strictly speaking, amounts to an abstraction of properties of interferometric experiments that exhibit maximal interference $I_m=\frac{1}{2}$. We will not address here whether the principle, or some cognate thereof, could be extended to the probabilistic case as well, the case that would generalize the entire \textit{probabilistic} possibility-impossibility structure exhibited by particle-interference phenomena. Furthermore, another dimension across which a generalization might be viable, and that exceeds the bounds of this work, is the consideration of generalized interference phenomena in which the events in question, instead of being representable by bits, can take a higher finite or possibly even (countably or uncountably) infinite number of values. In (Horvat \& Dakić, 2021a) it was shown that a simple modification of semi-general interference experiments, in which the number of possible configurations of each device is some prime number $d\geq 2$, the same possibility-impossibility structure is still retained.\footnote{The interference term is in this case $I_m^{(d)}=\frac{1}{d^m}\sum_{\mathbf{a}} P(s^{(d)}_{\mathbf{a}}|\mathbf{a})$, where $s^{(d)}_{\mathbf{a}}\equiv (\sum_i a_i)$ mod $d$. See Horvat \& Dakić (2021a) for details.} It is thus an open question whether this structure - or even a more general one that is still to be discovered - can also be lifted to the universal level that we have reached here for the binary case.

Note that our investigation has concerned only those phenomena that can be described in terms of PE-models, which are defined by assignments of events to spatiotemporal regions, all of which can be canvassed on a \textit{Minkowski} spacetime. It is however simple to see that everything that has been said here can trivially be transposed to more general spacetimes, making the Minkowskian choice an immaterial pedagogical device. Indeed, the principle is sensitive only to \textit{causal relations} between events - to whether the events in question are spacelike, null-like or timelike separated - and to how these interrelate to their \textit{statistical correlations}. Both the causal relations and the statistical correlations here can obviously be abstracted away and instantiated in other general-relativistic spacetimes, thereby inducing the general-relativistic version of our principle. This further generalization accordingly makes the principle more conjectural, in that quantum interferometric phenomena in the presence of relatively large gravitational fields are still to be empirically probed. 

As announced at the outset, the principle we have reached in our investigation is formulated, \textit{almost exclusively}, in terms of correlations between adequately spatiotemporally localized events, unconcerned with the natural or artificial objects that may or may not play out in them.\footnote{The ``almost exclusively'' here refers to the appearance of quantities $\mathcal{Q}$ (or, more speculatively, of the total energy of the system) in the definition of a closed $\text{GIP}_{\text{m}}$. Note however, that, despite this reference, we have still managed to reach a principle that regulates event-occurrences universally across phenomena, in that the preservation of quantities $\mathcal{Q}$, or of some cognate quantity that characterizes the matter present in the phenomenon (such as the total energy), can \textit{arguably} be applied to any physical phenomenon whatsoever.} We managed to reach this form despite the aforementioned ontological drawback - the lack of clarity concerning event-occurrences in quantum phenomena - by referring to \textit{pairs} of physically possible phenomena, that are such as to agree on what occurs in a certain spatio-temporal region.\footnote{Notice that a crucial step here - the one that climbed up the quantum-mechanical ladder before \textit{dropping} it - was the turning of the equivalence of ($\rho$, $U^{(\mathbf{a})}$) in experiments $T$ and $T'$ into the equivalence of distributions of event-occurrences in spatio-temporal region $\mathcal{C}^{(-)}_X$ in experiments $T$ and $T'$.} Our principle thereby inherently relies on \textit{physical-modal relations}: it speaks of a certain class of phenomena being such that there (do not) exist \textit{other} physically possible phenomena with such and such properties. This makes our principle quite different from Bell-Tsirelson's $\text{principle}^*$, which appears to be phraseable by reference to \textit{individual} phenomena: roughly stated, all phenomena within a wide class are such as not to violate Tsirelson's bound, whereas some phenomena within this same class do violate Bell's bounds. Our principle - with its appeal to properties of \textit{pairs} of adequately related phenomena - thus appears to be of a new form, one that has arguably not as of yet appeared within physical theorizing. This opens up the possibility of there being other such principles, still awaiting to be formulated, possibly implicit in other structural features of quantum phenomena, and perhaps even in structural features of relativistic phenomena.\footnote{A propaedeutic investigation into this possibility is to be carried out in a separate future work.}

\section*{VI. What have we learned?}

In this paper we took hold of quantum interference - a phenomenon that has been part of our natural-scientific home for a while now - and, as announced at the beginning, we observed it from a different angle, interpreting its properties as signifiers of a yet untold universal principle. In particular, we proposed that some structural features of interferometric phenomena suggest the validity of a principle that regulates correlations between event-occurrences in a wide range of physical phenomena, if not in any physical phenomenon whatsoever. Informally stated, the said principle mandates the following, up to various qualifications: a joint influence of $m$ spacelike separated events on another event lying in their common future is such that the same influence \textit{could} have been established via $\lceil \frac{m}{2} \rceil$ mediating events. Or, in other words, any such joint influence can be partially \textit{separated} by $\lceil \frac{m}{2} \rceil$ mediating events. While the possibility of second-order particle-interference thus signifies that events can influence each other in a non-separable fashion, the impossibility of higher-order particle-interference signifies that this non-separability has a certain exactly quantifiable limit, exhibited in its bi-local structure.\footnote{Note that the kind of non-separability that we have been concerned with - the non-separability of joint influences between spatiotemporally localized events - differs from the other kinds of non-separability that are usually discussed in relation to quantum mechanics, such as the non-separability of properties of composite quantum systems (e.g., Ismael \& Schaffer, 2020, and references therein), or, more recently, the supposed ``causal non-separability'' of some quantum phenomena that feature ``indefinite causal order'' (e.g., Oreshkov \& Giarmatzi, 2016).} 

The principle that we have thereby reached concerns the relationship between the causal-spatiotemporal relations between events on the one hand, and their statistical correlations on the other. In these regards, it is of a similar form to the aforementioned $\text{principle}^*$ that can be extracted from Bell's and Tsirelson's possibility and impossibility results, and to the so-called no-signalling principle, which constrains statistical correlations between spacelike separated events. Our principle however also bears significant differences with respect to these, at least in its current formulation. One of the differences, as already mentioned, is that reference is made to pairs of different adequately related statistical phenomena, instead of referring only to single statistical phenomena and to their properties. Another difference is that reference is made, albeit minimally, to certain material features of the phenomena in question (the quantities $\mathcal{Q}$, or more speculatively, the total energy), which can arguably not be reduced to mere structural relations between events. Both of these differences would presumably have a quite different shape if we knew how to speak of event-occurrences in arbitrary regions of quantum phenomena, and not only in those that feature considerable decoherence. However, as stated already multiple times, a principled identification of event-occurrences in quantum phenomena is still out of reach.

If our principle, or some close cognate thereof, holds - and do mind that this is still a considerable \textit{if} - its universality is bound to stay with us, and to persist throughout future developments in our physical theorizing, at least as long as we keep speaking of spatiotemporally localized events and of statistical correlations obtaining among them. Indeed, note that our principle's validity does \textit{not} hinge on the validity of quantum theory, similarly as the validity of Tsirelson's bound does not hinge upon it: while we did use quantum theory to obtain \textit{evidence} for our principle, the latter is phrased in \textit{non}-quantum-mechanical terms, thus being in principle applicable even to those potentially yet-to-be-elicited phenomena to which quantum theory \textit{might not} be applicable.\footnote{To be sure, we have \textit{not} suggested any concrete new phenomena: (i) we have not suggested an experiment in which a deviation from quantum theory is to be expected (but which would still validate our principle), and (ii) we have not suggested any interesting examples of $\text{GIP}_{\text{m}}$-s that cannot be phrased as semi-general interference experiments. Regarding point (ii), we merely alluded that, besides semi-general interference experiments, there might be examples of $\text{GIP}_{\text{m}}$-s that can only be described with the aid of quantum field theory. This is however a mere speculation that should be further explored elsewhere. More generally, what was \textit{not} within the scope of the current article, but which does merit its due attention, is a thorough investigation of novel concrete examples of $\text{GIP}_{\text{m}}$-s. Our point here was rather to propose and to defend the validity of a new physical principle.} That is, even if quantum theory were one day to be superseded by a new theory, the proposed principle - as much as Tsirelson's bound - \textit{could} still turn out to hold.\footnote{Conversely, it is also of course possible that the principle - as much as Tsirelson's bound - turns out \textit{not} to hold after all: we are indeed dealing with fallible conjectures here. The point is that we do, as of now, have sufficient empirically supported reasons to fallibly believe that both the hereby suggested principle, and Tsirelson's bound, really \textit{do} hold - that they do not merely hold according to quantum theory, but that they simply hold \textit{tout court}.} In other words, the quantum was used here as a ladder, one that may as well be dropped after climbing it, after instrumentally using it to reach our desired higher ground. 

Our principle - or, more realistically, a more developed and heavily polished version thereof - is thereby ideally to be added to the inventory of items that we are to hold on to in our future physics, a law that is to the very least to constrain, if not ideally to guide, future constructive theorizing, in Einstein's terms.\footnote{See Flores (1999) and Felline (2011) for a discussion of Einstein's distinction between principle and constructive theories.} If, or when, this mature stage is reached, we will be in a position to relate our principle to other ones within our inventory, and to theorize about their mutual interdependencies. It is perhaps also reasonable to hope that \textit{other} such \textit{causal-statistical} principles can be found - laws that constrain the relationship between the causal and the statistical relations between events - and that a unification thereof within a systematic \textit{causal-statistical theory} can one day be developed. While we are far from having reached the thus glimpsed utopia, I hope that our investigation, all of its imperfections notwithstanding, to the very least points clearly in its direction.\footnote{There have already been several attempts at developing principles that imply the Bell-Tsirelson possibility-impossibility structure (see Scarani, 2019, Chapter 10, and references therein). However, all of these principles have so far been information-theoretic in their nature, in that they speak not of what can or cannot \textit{be the case}, but of what can or cannot \textit{be done} by a certain community of spatiotemporally separated agents. While such principles certainly have a value of their own, they should not be thought of as \textit{replacing} physical principles, the reaching for the latter arguably being one of the primary goals of physics, a reaching that we should not leave buried in the past. The development of a causal-statistical physical theory is thus, if possible at all, still in its beginnings.}

\newpage

\section*{Acknowledgments}
Many thanks to Borivoje Dakić for illuminating discussions and to three anonymous reviewers for their constructive comments, questions and suggestions, which helped improve this paper. This research was funded in whole or in part by the Austrian Science Fund (FWF) [10.55776/P36994] and [10.55776/DOC162]. For open access purposes, the author has applied a CC BY public copyright license to any authoraccepted manuscript version arising from this submission.

\section*{References}

Aharonov, Y., \& Vaidman, L. (2000). Nonlocal aspects of a quantum wave. \textit{Physical Review A, 61}(5), 052108.

\smallskip

Barnum, H., Lee, C. M., Scandolo, C. M., \& Selby, J. H. (2017). Ruling out higher-order interference from purity principles. \textit{Entropy, 19}(6), 253.

\smallskip

Catani, L., Leifer, M., Schmid, D., \& Spekkens, R. W. (2023a). Why interference phenomena do not capture the essence of quantum theory. \textit{Quantum, 7}, 1119.

\smallskip

Catani, L., Leifer, M., Scala, G., Schmid, D., \& Spekkens, R. W. (2023b). Aspects of the phenomenology of interference that are genuinely nonclassical. \textit{Physical Review A, 108}(2), 022207.

\smallskip

Chen, X., Zhang, Y., Winter, A., Lorenz, V. O., \& Chitambar, E. (2024). Information carried by a single particle in quantum multiple-access channels. \textit{Physical Review A, 109}(6), 062420.

\smallskip

Felline, L. (2011). Scientific explanation between principle and constructive theories. \textit{Philosophy of Science, 78}(5), 989-1000.

\smallskip

Feynman, R. P., Leighton, R. B. \& Sands, M. L. (1963). \textit{The Feynman Lectures
on Physics.} Addison-Wesley world
student series. 

\smallskip

Flores, F. (1999). Einstein's theory of theories and types of theoretical explanation. \textit{International Studies in the Philosophy of Science, 13}(2), 123-134.

\smallskip

Freire, O., Bacciagaluppi, G., Darrigol, O., Joas, C., \& Kojevnikov, A. (Eds.). (2022). \textit{The Oxford handbook of the history of quantum interpretations}. Oxford University Press.

\smallskip

Gisin, N. (1989). Stochastic quantum dynamics and relativity. \textit{Helv. Phys. Acta, 62}(4), 363-371.

\smallskip

Goldstein, S. (2025). Bohmian Mechanics, \textit{The Stanford Encyclopedia of Philosophy }(Fall 2025 Edition), Edward N. Zalta \& Uri Nodelman (eds.).

\smallskip

Goyal, P. (2023). The role of reconstruction in the elucidation of quantum theory. In \textit{Phenomenology and QBism} (pp. 338-389). Routledge.

\smallskip

Helstrom, C. W. (1969). Quantum detection and estimation theory. \textit{Journal of Statistical Physics, 1}(2), 231-252.

\smallskip

Horvat, S. (2019). \textit{Quantum superposition as a resource for quantum communication} (Master thesis, University of Zagreb. Faculty of Science. Department of Physics).

\smallskip

Horvat, S., Allard Guérin, P., Apadula, L., \& Del Santo, F. (2020). Probing quantum coherence at a distance and Aharonov-Bohm nonlocality.\textit{ Physical Review A, 102}(6), 062214.

\smallskip

Horvat, S., \& Dakić, B. (2021a). Interference as an information-theoretic game. \textit{Quantum, 5}, 404.

\smallskip

Horvat, S., \& Dakić, B. (2021b). Quantum enhancement to information acquisition speed. \textit{New Journal of Physics, 23}(3), 033008.

\smallskip

Horvat, S., \& Toader, I. D. (2025). Quantum logic and meaning. \textit{Journal of Philosophical Logic, 54}(6), 1323-1343.

\smallskip

Horvat, S. (2025). Notes on a future quantum event-ontology. \textit{arXiv preprint arXiv:2502.08823.}

\smallskip

Hughston, L. P., Jozsa, R., \& Wootters, W. K. (1993). A complete classification of quantum ensembles having a given density matrix. \textit{Physics Letters A, 183}(1), 14-18.

\smallskip

Ismael, J., \& Schaffer, J. (2020). Quantum holism: Nonseparability as common ground. \textit{Synthese, 197}(10), 4131-4160.
\smallskip

Maisriml, J., Horvat, S., \& Dakić, B. (2026). Acquisition of Delocalized Information via Classical and Quantum Carriers. \textit{Physical Review Research 8}, 013102.

\smallskip

Maudlin, T. (2005). The tale of quantum logic. \textit{Hilary Putnam}, 156-187.

\smallskip

Oreshkov, O., \& Giarmatzi, C. (2016). Causal and causally separable processes. \textit{New Journal of Physics, 18}(9), 093020.

\smallskip

Passon, O. (2025). Why isn't every physicist a Bohmian? Common objections and. \textit{Guiding Waves in Quantum Mechanics: One Hundred Years of de Broglie-Bohm Pilot-Wave Theory}, 47.

\smallskip

Putnam, H. (1969). Is logic empirical?. In \textit{Boston Studies in the Philosophy of Science: Proceedings of the Boston Colloquium for the Philosophy of Science 1966/1968} (pp. 216-241). Dordrecht: Springer Netherlands.

\smallskip

Scarani, V. (2019). \textit{Bell nonlocality} (p. 239). Oxford University Press.

\smallskip

Schmid, D., Selby, J. H., \& Spekkens, R. W. (2024). Addressing some common objections to generalized noncontextuality. \textit{Physical Review A, 109}(2), 022228.

\smallskip

Sinha, U., Couteau, C., Jennewein, T., Laflamme, R., \& Weihs, G. (2010). Ruling out multi-order interference in quantum mechanics. \textit{Science, 329}(5990), 418-421.

\smallskip

Tsirelson, B. S. (1980). Quantum generalizations of Bell's inequality. Letters in \textit{Mathematical Physics, 4}(2), 93-100.

\smallskip

Wightman, A. S. (1995). Superselection rules; old and new. \textit{Il Nuovo Cimento B (1971-1996), 110}(5), 751-769.

\smallskip

Zhang, Y., Chen, X., \& Chitambar, E. (2022). Building multiple access channels with a single particle. \textit{Quantum, 6}, 653.

\newpage

\section*{Appendix 0: List of notations}

For readability, here is a list of the main terms that appear throughout the paper:

\begin{itemize}
    \item $x,y,... \in \mathbb{R}^4$: spatiotemporal points
    \item $x \prec y$ . . . there is a future-oriented time-like or null-like curve from $x$ to $y$
    \item $x \sim y$ . . . $x$ and $y$ are spacelike separated
    \item $S,S',... \subset \mathbb{R}^4$ . . . spacelike hypersurfaces
    \item $X,Y,... \subset \mathbb{R}^4$ . . . finite sets of spatiotemporal points 
    \item $\mathcal{C}^{(-)}_X \equiv \left\{y \in \mathbb{R}^4| \exists x \in X: y \prec x \right\}$ . . . union of causal pasts of points in $X$
    \item $\Omega_x$ . . . set of events that can occur at $x$
    \item $\Omega_X \equiv \Pi_{x \in X} \Omega_x$ . . . joint set of events that can occur at points in $X$
    \item $P_X(\omega_X)$ . . . probability that events $\omega_X \in \Omega_X$ occur at points in $X$
    \item $\mathbf{P}=\left\{(\Omega_X,\mathcal{F}_X, P_X) \right\}_{X\in \mathcal{X}}$ . . . ``probabilistic event model'' or ``PE-model'' (see Definition 1)
    \item $I_m\equiv \frac{1}{2^m}\sum_{\omega_X}P_y(\oplus_{i=1}^m\omega_{x_i}|\omega_X)-\frac{1}{2}$ . . . $m$-th order interference term associated to points $(X,y)$
    \item $\text{GIP}_{\text{m}}$ . . . abbreviation for ``general interference phenomenon of order $m$'' (see Definition 3)
    \item ``\textit{maximal} $\text{GIP}_{\text{m}}$'' . . . a $\text{GIP}_{\text{m}}$ that satisfies $I_m=\frac{1}{2}$ (it generates maximal interference)
    \item $T=\left(\rho, U^{(\mathbf{a})}, \Pi \right)^{n}_d$ . . . characterization of a semi-general interference experiment involving $n$ particles whose internal degrees of freedom have dimensionality $d$. The particles are initially in state $\rho$; they undergo transformation $U^{(\mathbf{a})}$, for some fixed $\mathbf{a}$; they finally undergo measurement $\Pi$.
    \item $I_m(T)$ . . . the $m$-th order interference term generated in a semi-general interference experiment characterized by $T$
    \item $\mathcal{Q}$ . . . the non-dynamical quantities that characterize a collection of particles of the same type (total mass, charge, spin, etc.)
    \item $P_S(\mathcal{Q})$ . . . probability that the non-dynamical quantities that characterize the matter present on spacelike hypersurface $S$ is equal to $\mathcal{Q}$
    \item \textit{closed} $\text{GIP}_{\text{m}}$ . . . a $\text{GIP}_{\text{m}}$ in which quantities $\mathcal{Q}$ are preserved (see Definition 4)
    \item $n$-local completion of a $\text{GIP}_{\text{m}}$ . . . another $\text{GIP}_{\text{m}}$ that contains $n$ additional events that mediate and preserve the interferometric correlation exhibited by the original $\text{GIP}_{\text{m}}$ (see Definition 5)
    \item the ``principle'': every \textit{maximal} $\text{GIP}_{\text{m}}$ has a physically possible \textit{closed bi-local completion}, for any $m$. Conversely, \textit{not} every maximal $\text{GIP}_{\text{m}}$ has a physically possible closed $n$-local completion, for $\lceil \frac{m}{n} \rceil<2$.
\end{itemize}

\section*{Appendix 1: Discussion of Example 2}

Here we will analyze some properties of the experiments surveyed in Example 2. We are in particular concerned with the existence of additional events that preserve and mediate the correlation between $(a_1,a_2)$ and $b$. In particular, the additional events are supposed to be elicited by measurements on state 

\begin{equation}\label{a1}
    \ket\psi_{a_1a_2}=\frac{1}{\sqrt{2}} \left( e^{i\pi a_1}\ket{1} + e^{i\pi a_2}\ket{2}\right),
\end{equation}
where $\ket{i}$ is the state of the particle being localized at $\text{Alice}_i$'s location. Note that the quantum state can be rewritten, up to a global phase, in the second quantization notation, as 
\begin{equation}
    \ket\psi_{a_1a_2}=\frac{1}{\sqrt{2}} \left(c_1^{\dagger}+(-1)^{a_1+a_2}c_2^{\dagger} \right) \ket{0},
\end{equation}
where $\ket{0}$ is the vacuum state, and $c_i^{\dagger}$ is the creation operator that ``creates'' one particle at $\text{Alice}_i$'s location. 

Let us first suppose that the particle in question is an uncharged boson (e.g. a photon), and that thus no superselection rule applies. Let $\Pi=\left\{ \Pi^{(1)}_{b_1} \otimes \Pi^{(2)}_{b_2} \right\}_{b_i=0,1}$ be a projective measurement, where $\Pi^{(i)}_{x}=\ket{x}_i\bra{x}$, with

\begin{equation}
    \ket{x}_i=\frac{1}{\sqrt{2}}\left(\mathds{1}+(-1)^{x} c_i^{\dagger} \right)\ket0.
\end{equation}

The outcomes of a measurement of $\Pi$
on state $\ket\psi_{a_1a_2}$ are distributed as
\begin{equation}
    P(b_1b_2|a_1a_2)=\delta_{b_1 \oplus b_2, a_1 \oplus a_2},
\end{equation}
which obviously preserves and mediates the maximal correlation between $b$ and $a_1 \oplus a_2$. However, as noted in the main text, the mediating measurements are not number-preserving. Indeed, if the measurement outcomes are $(b_1,b_2)$, the output state is

\begin{equation}
    \ket\psi_{b_1b_2}=\frac{1}{2} \left(\mathds{1} + (-1)^{b_1} c_1^{\dagger} + (-1)^{b_2} c_2^{\dagger} + (-1)^{b_1+b_2} c_1^{\dagger}c_2^{\dagger} \right)\ket0,
\end{equation}
which has probability $\frac{1}{4}$ of containing two particles and probability $\frac{1}{4}$ of containing no particles. While the measurement therefore preserves the mean value of the particle number, it does not preserve its other moments. Or, in other words, it introduces a probabilistic change in the particle number.

Let us now suppose that the particle is charged or fermionic (e.g. an electron), making the above measurement prohibited by superselection rules. As mentioned in the main text, the correlation between $(a_1,a_2)$ and $b$ can nevertheless still be mediated and preserved, if an additional ancillary particle is used. Indeed, suppose that $\text{Bob}_1$ and $\text{Bob}_2$, at the moment in which they are to implement the intermediate measurements, share another particle of the same type (e.g. an electron), such that the two particles are in joint state 

\begin{equation}
    \ket\psi_{a_1a_2}=\frac{1}{2} \left( c_{11}^{\dagger} + c_{21}^{\dagger}\right) \left( c_{12}^{\dagger} + (-1)^{a_1+a_2}c_{22}^{\dagger}\right) \ket0,
\end{equation}
where $c_{ij}^{\dagger}$ creates one particle at location `$ij$', where `$i1$' and `$i2$' are locations that are, roughly speaking, localized close to each other at $\text{Bob}_i$'s location. Let $\Pi=\left\{ \Pi_1, \Pi_2\right\}$ be a POVM, where $\Pi_1$ is a projector on the subspace spanned by $\left\{ c_{11}^{\dagger} c_{22}^{\dagger}\ket0, c_{21}^{\dagger} c_{12}^{\dagger}\ket0 \right\}$ and $\Pi_2$ is a projector on the subspace spanned by $\left\{ c_{11}^{\dagger} c_{12}^{\dagger}\ket0, c_{21}^{\dagger} c_{22}^{\dagger}\ket0 \right\}$. $\Pi_1$ thus projects on the subspace in which exactly one particle is present at each $\text{Bob}_i$'s location, whereas $\Pi_2$ projects on the subspace in which the two particles are both either at $\text{Bob}_1$'s or at $\text{Bob}_2$'s location. 

It follows that a measurement of $\Pi$ on state $\ket\psi_{a_1a_2}$ yields either of the two outcomes with equal probability $\frac{1}{2}$. Let us suppose that the outcome of the measurement is the one corresponding to projector $\Pi_1$, and that the post-measurement state is thus 

\begin{equation}
    \ket\psi^{(1)}_{a_1a_2}=\frac{1}{\sqrt{2}} \left( c_{21}^{\dagger} c_{12}^{\dagger} + (-1)^{a_1+a_2} c_{11}^{\dagger} c_{22}^{\dagger}\right) \ket0.
\end{equation}

Let $\Tilde{\Pi}_1=\left\{ \Pi^{(11)}_{b_1} \otimes \Pi^{(12)}_{b_2} \right\}_{b_i=0,1}$ be a projective measurement, where $\Pi^{(1i)}_{x}=\ket{x}^1_i\bra{x}$, with 
\begin{equation}
    \ket{x}^1_i=\frac{1}{\sqrt{2}}\left(c_{i1}^{\dagger}+(-1)^{x} c_{i2}^{\dagger} \right)\ket0.
\end{equation}

It is easy to check that the outcomes of measurement $\Tilde{\Pi}_1$ on state $\ket\psi^{(1)}_{a_1a_2}$ are distributed as 

\begin{equation}
    P(b_1b_2|a_1a_2)=\delta_{b_1 \oplus b_2, a_1 \oplus a_2},
\end{equation}
which once again mediates and preserves the maximal correlation between $b$ and $a_1 \oplus a_2$. 

Conversely, if the outcome of measurement $\Pi$ is the one corresponding to $\Pi_2$, then the same distribution of outcomes $(b_1,b_2)$ is obtained for measurement
$\Tilde{\Pi}_2=\left\{ \Pi^{(21)}_{b_1} \otimes \Pi^{(22)}_{b_2} \right\}_{b_i=0,1}$, where $\Pi^{(2i)}_{x}=\ket{x}^2_i\bra{x}$ and 

\begin{equation}
    \ket{x}^2_i=\frac{1}{\sqrt{2}}\left(c_{1i}^{\dagger}+(-1)^{x} c_{2i}^{\dagger} \right)\ket0. 
\end{equation}

Events that mediate and preserve the correlation between $b$ and $(a_1,a_2)$ can therefore be identified in an experiment that consists of: (i) a measurement of $\Pi$, which, if it results in outcome corresponding to $\Pi_i$, is followed by (ii) a measurement of $\Tilde{\Pi}_i$. Note that, while both of the involved operations are number-preserving - and thereby not prohibited by superselection rules - the overall processes nevertheless increases the particle number, in that it relies on an insertion of an additional ancillary particle. Indeed, the particle number is in this case doubled, and deterministically so.

To summarize, regardless of the nature of the involved particle, there is a possible experiment in which intermediate events are established that mediate and preserve the correlation between $b$ and $(a_1,a_2)$. However, these experiments are not particle-number-preserving: the case of uncharged bosons introduces a probabilistic change in the number of particles, whereas the case of charged particles or fermions introduces a deterministic doubling in the number of particles. Note finally that the inserted particle is in both cases of the same type as the one originally present in the experiment. In particular, in the electronic case, if the additional particle were not of the same type, then the interferometric measurement would be prohibited by the charge and parity superselection rules.\footnote{In the electronic case, one could alternatively insert a \textit{positron}, instead of a second electron, and mediate the correlation via a similar method, which instead of interfering the two electrons, lets the particle and its antiparticle annihilate into two photons. In this case, the number of electrons (and the total charge) is reduced deterministically from 1 to 0. For details, see Aharonov \& Vaidman (2000) and Horvat \textit{et al}. (2020).}

\section*{Appendix 2: Proof of Lemma 1}
Let a semi-general interference experiment of order $2n$ be characterized by triple $T=\left(\rho, U^{(\mathbf{a})}, \Pi \right)_d^{n}$, where $\rho=\sum_{\psi} p_{\psi}\ket{\psi}\bra{\psi}$. We want to show that, if $I_{2n}(T)=\frac{1}{2}$, then $U^{(\mathbf{a})}\ket\psi$ has the particular form stated in Lemma 1, for each $\ket{\psi}$. 

\bigskip

The proof will contain the following enumerated steps:
\begin{enumerate}
    \item Simplification of the form of $U^{(\mathbf{a})}$, without loss of generality.
    \item Application of Helstrom's bound that leads to a consequence of $I_{2n}(T)=\frac{1}{2}$, formulated as a condition on the \textit{trace norm} of a certain \textit{operator}.
    \item Remarks that simplify the form of the said operator.
    \item Calculation of the trace norm of the said operator.
    \item Consequences of the trace norm condition obtained in Step 2 on the form of $U^{(\mathbf{a})}\ket\psi$. End of proof.

\end{enumerate}

Let us now proceed step by step.

\subsection*{Step 1: Simplification of $U^{(\mathbf{a})}$}

In the first step we will simplify the form of $U^{(\mathbf{a})}$, without loss of generality. But before doing so, we will first introduce a notation that will be useful in what follows. For any string $\mathbf{j}=(j_1...j_N)$, we will denote with $D_{\mathbf{j}}$ the number of different values appearing in $\mathbf{j}$: for example, if $\mathbf{j}=(1,1,...,1)$ then $D_{\mathbf{j}}=1$; if $\mathbf{j}=(1,2,...,N)$ then $D_{\mathbf{j}}=N$. Before proceeding, let us recall the definitions and constraints that we are hereby to abide by. The incoming state $\rho \in \mathcal{L}(\mathbb{C}^{2n}\otimes \mathbb{C}^d)^{\otimes n}$ is an arbitrary state of $n$ particles, each of which has an associated $2n$-dimensional spatial state space and a $d$-dimensional internal state space. The unitaries $U^{(\mathbf{a)}}$ have the general form 

\begin{equation}
    U^{(\mathbf{a)}}=\sum_{\mathbf{i}}\ket{\mathbf{i}}\bra{\mathbf{i}} \otimes U^{(\mathbf{a_i})},
\end{equation}
where $\mathbf{i}\equiv(i_1...i_n)$ and $\mathbf{a_i}\equiv (a_{i_1}...a_{i_n})$. In particular, for those strings $\mathbf{i}$ that satisfy $D_{\mathbf{i}}=n$, it holds that $U^{(\mathbf{a_i})} = \bigotimes_{l=1}^n U_{i_l}^{(a_{i_l})}$. The measurement $\Pi = \left\{ \Pi_0, \Pi_1 \right\}$ is an arbitrary binary POVM on $\mathcal{L}(\mathbb{C}^{2n}\otimes \mathbb{C}^d)^{\otimes n}$. Finally, the interference term is defined as 
\begin{equation}
    I_{2n}(T)\equiv \frac{1}{2^{2n}}\sum_ {\mathbf{a}} P_T(\oplus_i a_i | \mathbf{a}) - \frac{1}{2},
\end{equation}
where $P_T(b | \mathbf{a})=\Tr (\Pi_b U^{(\mathbf{a)}} \rho U^{(\mathbf{a)}\dagger})$.

Let us now start by noting that $P_T(b | \mathbf{a}) = P_{\tilde{T}}(b | \mathbf{a})$, for any $\tilde{T}=(\rho, \tilde{U}^{(\mathbf{a})}, \tilde\Pi)$, for which $\tilde{U}^{(\mathbf{a})}=G^{\dagger}U^{(\mathbf{a)}}$ and $\tilde\Pi=G^{\dagger}\Pi G$, for some unitary $G$. Consider in particular
\begin{equation}\label{G}
    G=\sum_{\mathbf{i}} \ket{\mathbf{i}}\bra{\mathbf{i}} \bigotimes_{l=1}^n U_{i_l}^{(0)}.
\end{equation}
This choice of $G$ implies that $\tilde{U}^{(\mathbf{a})}=\sum_{\mathbf{i}} \ket{\mathbf{i}}\bra{\mathbf{i}}\otimes\tilde{U}^{(\mathbf{a_i})}$, where $\tilde{U}^{(\mathbf{a_i})}$ are such that, for all strings $\mathbf{i}$ that satisfy $D_{\mathbf{i}}=n$:

\begin{equation}
    \tilde{U}^{(\mathbf{a_i})} = \bigotimes_{l=1}^n (\tilde{U}_{i_l})^{a_{i_l}},
\end{equation}
where $\tilde{U}_{i_l}=U_{i_l}^{(0)\dagger}U_{i_l}^{(1)}$. In other words, $\tilde{U}^{(\mathbf{a})}$ is such that, for states in which the particles have no spatial overlap (i.e. $D_{\mathbf{i}}=n$), each local unitary applies the identity operator $\mathds{1}$ in case its pertaining configuration is set to 0, and a non-trivial unitary operator in case its configuration is set to 1. Since $P_T(b | \mathbf{a}) = P_{\tilde{T}}(b | \mathbf{a})$, it is also the case that $I_{2n}(T)=\frac{1}{2}$ if and only if $I_{2n}(\tilde{T})=\frac{1}{2}$. In what follows, we will prove that $\tilde{U}^{(\mathbf{a})} \ket\psi$ has a certain general form. This will then trivially also imply the general form of $U^{(\mathbf{a})}\ket\psi$, in that, $U^{(\mathbf{a})}\ket\psi=G\tilde{U}^{(\mathbf{a})} \ket\psi$. Let us henceforth, for simplicity, suppress the ``tildes'', while keeping in mind that our end result will be valid up to the global unitary $G$. More precisely, we will assume, for $D_{\mathbf{i}}=n$, that $U^{(\mathbf{a_i})} = \bigotimes_{l=1}^n (U_{i_l})^{a_{i_l}}$ and exhibit the general form of $U\ket \psi$, recognizing that this form will be valid up to a unitary $G$ that has the form stated in Eq. \eqref{G}.

\subsection*{Step 2: Application of Helstrom's bound}

In this step we will note a simple consequence that the assumed condition $I_{2n}(T)=\frac{1}{2}$ imposes on the trace norm of a certain operator that depends both on $U^{(\mathbf{a})}$ and on $\ket\psi$. Namely, note that the interference term amounts to
\begin{equation}
    I_{2n}(T)=\frac{1}{2} \left(\Tr \left(\Pi_0 \rho^{(0)} \right) + \Tr \left(\Pi_1 \rho^{(1)} \right)\right) - \frac{1}{2}, 
\end{equation}
where 

\begin{equation}
    \rho^{(s)}=\frac{1}{2^{2n-1}}\sum_{\oplus_ia_i=s} U^{(\mathbf{a)}} \rho U^{(\mathbf{a)}\dagger}.
\end{equation}
For fixed $(\rho^{(0)},\rho^{(1)})$, the maximum value of $I_{2n}(T)$ is given by Helstrom (1969)'s bound:
\begin{equation}
    \text{max}_{\Pi}I_{2n}(T)=\frac{1}{4}||\rho^{(1)}-\rho^{(0)}||,
\end{equation}
where $||A||$ is the trace norm of operator $A$. Therefore, a necessary condition for $I_{2n}(T)=\frac{1}{2}$ to hold is 
\begin{equation}\label{=2}
    ||\rho^{(1)}-\rho^{(0)}||=2.
\end{equation}
Now, note that $\rho^{(1)}-\rho^{(0)}=\sum_{\psi}p_{\psi} \left(\rho^{(1)}_{\psi}-\rho^{(0)}_{\psi} \right)$, where 
\begin{equation}
    \rho^{(s)}_{\psi}=\frac{1}{2^{2n-1}}\sum_{\oplus_ia_i=s} U^{(\mathbf{a)}} \ket\psi \bra\psi U^{(\mathbf{a)}\dagger}.
\end{equation}
Since, per the triangle inequality, $||\sum_i A_i|| \leq \sum_i|| A_i||$, Eq. \eqref{=2} also implies, for all $\psi$:
\begin{equation}\label{=22}
    || \rho^{(1)}_{\psi}-\rho^{(0)}_{\psi}||=2.
\end{equation}

\subsection*{Step 3: Simplification of $\rho^{(1)}_{\psi}-\rho^{(0)}_{\psi}$}

Our goal is to inspect what the condition $|| \rho^{(1)}_{\psi}-\rho^{(0)}_{\psi}||=2$ implies about $U^{(\mathbf{a)}}$ and  $\ket\psi$. In order to do that we will first note here certain properties of $\rho^{(1)}_{\psi}-\rho^{(0)}_{\psi}$ that will simplify its form, enabling an easier subsequent calculation of its trace norm.

The most general form that each state $\ket{\psi}$ can take is given by 
\begin{equation}
    \ket\psi=\sum_{\mathbf{i}\mathbf{k}}e^{i\phi_{\mathbf{i}\mathbf{k}}}\sqrt{p_{\mathbf{i}\mathbf{k}}} \ket{\mathbf{i}}\ket{\mathbf{k}},
\end{equation}
where $\left\{\ket{\mathbf{i}}\right\}$ and $\left\{\ket{\mathbf{k}}\right\}$ form orthonormal bases of $(\mathbb{C}^{2n})^{\otimes n}$ and $(\mathbb{C}^{d})^{\otimes n}$ respectively, the probabilities $p_{\mathbf{i}\mathbf{k}}$ are normalized, and the phases $\phi_{\mathbf{i}\mathbf{k}}$ are arbitrary. The latter form is trivially seen to be equivalent to 
\begin{equation}
    \ket\psi=\sum_{\mathbf{i}}\sqrt{p_{\mathbf{i}}} \ket{\mathbf{i}}\ket{\phi^{(\mathbf{i})}},
\end{equation}
where $\ket{\phi^{(\mathbf{i})}}\equiv \frac{1}{\sqrt{p_{\mathbf{i}}}} \sum_{\mathbf{k}}e^{i\phi_{\mathbf{i}\mathbf{k}}}\sqrt{p_{\mathbf{i}\mathbf{k}}} \ket{\mathbf{k}}$ are normalized and possibly non-mutually orthogonal vectors, and $p_{\mathbf{i}}=\sum_{\mathbf{k}}p_{\mathbf{i}\mathbf{k}}$.

Letting $\ket{\phi^{(\mathbf{i})}_{\mathbf{a_i}}} \equiv U^{(\mathbf{a_i})}\ket{\phi^{(\mathbf{i})}}$, it follows that
\begin{equation}
    \rho^{(s)}_\psi=\frac{1}{2^{2n-1}}\sum_{\mathbf{i},\mathbf{j}}\sqrt{p_{\mathbf{i}}p_{\mathbf{j}}}\ket{\mathbf{i}}\bra{\mathbf{j}} \otimes \sum_{\oplus_ia_i=s} \ket{\phi^{(\mathbf{i})}_{\mathbf{a_i}}}\bra{\phi^{(\mathbf{j})}_{\mathbf{a_j}}}.
\end{equation}
It is simple to notice that the latter implies (up to an irrelevant minus-sign):
\begin{equation}
    \rho^{(1)}_{\psi}-\rho^{(0)}_{\psi}=\frac{1}{2^{2n-1}}\sum_{\mathbf{i},\mathbf{j}}\sqrt{p_{\mathbf{i}}p_{\mathbf{j}}}\ket{\mathbf{i}}\bra{\mathbf{j}} \otimes E_{\mathbf{i}\mathbf{j}},
\end{equation}
where 
\begin{equation}
    E_{\mathbf{i}\mathbf{j}} \equiv \sum_{\mathbf{a}} (-1)^{\sum_ia_i } \ket{\phi^{(\mathbf{i})}_{\mathbf{a_i}}}\bra{\phi^{(\mathbf{j})}_{\mathbf{a_j}}}.
\end{equation}
A moment of inspection reveals that most elements of $E_{\mathbf{i}\mathbf{j}}$ vanish: indeed, if the number of different values in the string obtained by appending $\mathbf{i}$ and $\mathbf{j}$ is less than $2n$ - i.e. if $D_{\mathbf{i}\mathbf{j}}<2n$ - then there is at least one index $l$, such that $E_{\mathbf{i}\mathbf{j}}=\sum_{a_l=0}^1 (-1)^{a_l} K_{\mathbf{i}\mathbf{j}}=0$, where $K_{\mathbf{i}\mathbf{j}}$ is some quantity that does not depend on $a_l$. Therefore:
\begin{equation}
     D_{\mathbf{i}\mathbf{j}}<2n \quad \longrightarrow \quad E_{\mathbf{i}\mathbf{j}} = 0,
\end{equation}
and
\begin{equation}\label{r1-r0}
    \rho^{(1)}_{\psi}-\rho^{(0)}_{\psi}=\frac{1}{2^{2n-1}}\sum_{D_{\mathbf{i}\mathbf{j}}=2n}\sqrt{p_{\mathbf{i}}p_{\mathbf{j}}}\ket{\mathbf{i}}\bra{\mathbf{j}} \otimes E_{\mathbf{i}\mathbf{j}}.
\end{equation}
Since $D_{\mathbf{i}\mathbf{j}}=2n$ holds only if all values in string $(\mathbf{i}\mathbf{j})$ differ from each other, it follows that, for $D_{\mathbf{i}\mathbf{j}}=2n$:
\begin{equation}
\begin{split}
    E_{\mathbf{i}\mathbf{j}}&=\sum_{\mathbf{a}}(-1)^{\sum_i a_i} (U_{i_1})^{a_{i_1}} \otimes. . . \otimes (U_{i_n})^{a_{i_n}} \ket{\phi^{(\mathbf{i})}} \bra{\phi^{(\mathbf{j})}} (U^{\dagger}_{j_1})^{a_{j_1}} \otimes. . . \otimes (U^{\dagger}_{j_n})^{a_{j_n}}\\
    &=\ket{\alpha^{(\mathbf{i})}}\bra{\alpha^{(\mathbf{j})}},
\end{split}
\end{equation}
where $\ket{\alpha^{(\mathbf{i})}} \equiv \bigotimes_{l=1}^{n}(\mathds{1}-U_{i_l})\ket{\phi^{(\mathbf{i})}}$. We have accordingly obtained the following simplified form:

\begin{equation}\label{M1}
    \rho^{(1)}_{\psi}-\rho^{(0)}_{\psi}=\frac{1}{2^{2n-1}}\sum_{D_{\mathbf{i}\mathbf{j}}=2n}\sqrt{p_{\mathbf{i}}p_{\mathbf{j}}}\ket{\mathbf{i}}\bra{\mathbf{j}} \otimes \ket{\alpha^{(\mathbf{i})}}\bra{\alpha^{(\mathbf{j})}}.
\end{equation}

The sum over strings $\mathbf{i},\mathbf{j}$, constrained by $D_{\mathbf{i}\mathbf{j}}=2n$, can be simplified by introducing the following re-parametrization. Let $\mathcal{B}_{2n}$ be the set of all bipartitions $B: \left\{1,...,2n \right\} \rightarrow \left\{0,1\right\}$ that satisfy $|B^{-1}(0)|=|B^{-1}(1)|=n$. Let $\mathcal{S}_n$ be the symmetric group on elements $\left\{1,...,n \right\}$. The said reparametrization amounts to the following: to each vector $\ket{\mathbf{i}}$ we associate vector $\ket{Sx}_B$, for some $S \in \mathcal{S}_n, B \in \mathcal{B}_{2n}$ and $x \in \left\{0,1 \right\}$ , if and only if $S(B^{-1}(x))=(i_1...i_n)$. The intuition here is that any string $\mathbf{i}$ taking $n$ different values among $2n$ possible values can be specified by (i) choosing the subset of the $2n$ values that figure in $\mathbf{i}$, this subset here being determined by $B$ and $x$, and (ii) choosing the ordering by which the $n$ chosen values appear in $\mathbf{i}$, here determined by a permutation on $n$ elements. We will accordingly map the $\mathbf{i}$-dependent objects $p_{\mathbf{i}}$ and $\ket{\alpha^{(\mathbf{i})}}$ to their correspodning re-parametrized $p_{B}^{(Sx)}$ and $\ket{\alpha^{(Sx)}}_B$.

The advantage of the new parametrization is that - as clear after a moment of inspection - Eq. \eqref{M1} turns out to reduce to a direct sum over bipartitions $B$:

\begin{equation}\label{M2}
    \rho^{(1)}_{\psi}-\rho^{(0)}_{\psi}=\frac{1}{2^{2n-1}} \bigoplus_{B \in \mathcal{B}_{2n}} M_B,
\end{equation}
where 
\begin{equation}
    M_B=\sum_{S,S' \in \mathcal{S}_n}  \left [ \sqrt{p_{B}^{(S0)}p_{B}^{(S'1)}} \ket{S0}_B \bra{S'1} \otimes \ket{\alpha^{(S0)}_B}\bra{\alpha^{(S'1)}} + \sqrt{p_{B}^{(S'0)}p_{B}^{(S1)}} \ket{S1}_B \bra{S'0} \otimes \ket{\alpha^{(S1)}_B}\bra{\alpha^{(S'0)}}  \right].
\end{equation}

\subsection*{Step 4: Calculation of the trace norm}

The trace norm that we are after thus reduces to 
\begin{equation}\label{M2}
    ||\rho^{(1)}_{\psi}-\rho^{(0)}_{\psi}||=\frac{1}{2^{2n-1}} \sum_{B \in \mathcal{B}_{2n}} ||M_B||.
\end{equation}
Now, each operator $M_B$ is further reducible to the following simple form:
\begin{equation}
    M_B=\ket{0}_B \bra{1} + \ket{1}_B \bra{0},
\end{equation}
where $\ket{x}_B \equiv \sum_{S \in \mathcal{S}_n} \sqrt{p_{B}^{(Sx)}} \ket{Sx}_B \otimes \ket{\alpha^{(Sx)}_B} $. Since each $M_B$ is a Hermitian operator, its trace norm is given by $M_B=\sum_i |\lambda^{B}_i|$, where $\lambda_i$ are the eigenvalues of $M_B$. Now, since $\ket{0}_B$ and $\ket{1}_B$ are orthogonal, the eigenvalues of $M_B$ are given by $\pm \sqrt{{}_{B}\!\left\langle 0 \middle| 0 \right\rangle_{B} {}_{B}\!\left\langle 1 \middle| 1 \right\rangle_{B}}$. We therefore obtain

\begin{equation}
    ||M_B||=2 \sqrt{{}_{B}\!\left\langle 0 \middle| 0 \right\rangle_{B} {}_{B}\!\left\langle 1 \middle| 1 \right\rangle_{B}}
\end{equation}

\subsection*{Step 5: Consequences of $||\rho_1-\rho_0 ||=2$}

Now that we obtained a simple expression for our desired trace norm, we are finally ready to inspect the consequences of the condition $||\rho_1-\rho_0 ||=2$, which was derived in Step 2. The final consequence will result in the form of $U^{(\mathbf{a})}\ket\psi$ that was stated in the Lemma, thus finalizing the proof.

First, note that 
\begin{equation}
\begin{split}
    {}_{B}\!\left\langle x \middle| x \right\rangle_{B}&=\sum_s p_B^{(Sx)} \braket{\alpha^{(Sx)}_B|\alpha^{(Sx)}_B}\\
    &= 2^n \sum_s p_B^{(Sx)} \braket{\phi_B^{(Sx)}| \bigotimes_{l=1}^n \left[\mathds{1} - \frac{1}{2}\left(U_{S(l)}^{(Bx)} + U_{S(l)}^{(Bx)\dagger} \right) \right] | \phi_B^{(Sx)}},
\end{split}
\end{equation}
where $U_{S(l)}^{(Bx)}$ is the re-parametrized notation of $U_{i_l}$, and $\ket{\phi_B^{(Sx)}}$ is the re-parametrized notation of $\ket{\phi^{(\mathbf{i})}}$. All of the above then implies:
\begin{equation}
\begin{split}
    ||M_B||^2=2^{2n+2} &\left\{ \sum_s p_B^{(S0)} \braket{\phi_B^{(S0)}| \bigotimes_{l=1}^n \left[\mathds{1} - \frac{1}{2}\left(U_{S(l)}^{(B0)} + U_{S(l)}^{(B0)\dagger} \right) \right] | \phi_B^{(S0)}} \right\}\\
    &\left\{ \sum_s p_B^{(S1)} \braket{\phi_B^{(S1)}| \bigotimes_{l=1}^n \left[\mathds{1} - \frac{1}{2}\left(U_{S(l)}^{(B1)} + U_{S(l)}^{(B1)\dagger} \right) \right] | \phi_B^{(S1)}} \right\}.
    \end{split}
\end{equation}

Next, note that for any list $(U_1,...U_n)$ of unitary operators and any state $\ket\phi$:
\begin{equation}
    \braket{\phi| \bigotimes_{l=1}^n \left[\mathds{1} - \frac{1}{2}\left(U_l + U_l^{\dagger} \right) \right] | \phi}=\sum_{x_1...x_n=0}^1 (-1)^{\sum_i x_i} \braket{\phi| \bigotimes_{l=1}^n \left[\frac{1}{2}\left(U_l + U_l^{\dagger} \right) \right]^{x_l} | \phi} \leq 2^n.
\end{equation}
In particular, note that the maximal value of the latter expression is obtained for
\begin{equation}
    \braket{\phi| \bigotimes_{l=1}^n \left[\frac{1}{2}\left(U_l + U_l^{\dagger} \right) \right]^{x_l} | \phi}=(-1)^{\sum_i x_i},
\end{equation}

which implies 
\begin{equation}
    \bigotimes_{l\neq j} \mathds{1}_l \otimes U_j \ket\phi=-\ket\phi,
\end{equation}
for all $j=1,...,n$.\footnote{The expression $\mathds{1}_l \otimes U_j$ in the above equation is shorthand for $\bigotimes_{l=1}^{j-1} \mathds{1}_l \otimes U_j \otimes \bigotimes_{k=j+1}^{n}\mathds{1}_k$, i.e. the operator that acts with $U_j$ on the internal state space of the $j$-th particle, and trivially on the internal state spaces of the remaining particles. We will keep using this notation throughout the rest of the Appendices.} Applying this observation to our case of interest, the maximal value of $||M_B||$ is obtained if
\begin{equation}\label{units}
    \bigotimes_{l\neq j} \mathds{1}_l \otimes U^{(Bx)}_{S(j)} \ket{\phi_B^{(Sx)}}=-\ket{\phi_B^{(Sx)}},
\end{equation}
and amounts to
\begin{equation}
    \text{max}_{\ket\phi, U} ||M_B||^2=2^{2n+2} 2^{2n} (\sum_S p_B^{(S0)}) (\sum_S p_B^{(S1)}).
\end{equation}
Given that $0 \leq\sum_S (p_B^{(S0)}+ p_B^{(S0)}) \equiv q_B \leq 1$, a simple application of the method of Lagrange multipliers yields that the maximal value is obtained for 
\begin{equation}\label{probs}
    \sum_S p_B^{(S0)} = \sum_S p_B^{(S1)} = \frac{q_B}{2},
\end{equation}
and amounts to 
\begin{equation}
    \text{max} ||M_B||=2^{2n}q_B,
\end{equation}
which accordingly implies 
\begin{equation}
    \text{max}||\rho^{(1)}_{\psi}-\rho^{(0)}_{\psi}||=\frac{1}{2^{2n-1}} \sum_{B \in \mathcal{B}} || \text{max}M_B||=2\sum_Bq_B.
\end{equation}
The latter quantity is obviously equal to 2 only if $\sum_Bq_B=1$. In other words, the weights $p_{\mathbf{i}}$ - pertaining to the initial quantum state $\ket\psi$ - need to be non-zero only for those strings $\mathbf{i}$ that satisfy $D_{\mathbf{i}}=n$. We therefore found out that $||\rho^{(1)}_{\psi}-\rho^{(0)}_{\psi}||=2$ - and therefore that $I_{2n}(T)=\frac{1}{2}$ - only if: (i) $U^{(\mathbf{a})}$ and $\ket\psi$ are such that conditions \eqref{units} and \eqref{probs} hold, and if (ii) $p_{\mathbf{i}}=0$ for all those $\mathbf{i}$ for which $D_{\mathbf{i}}<n$. 

Let us now inspect what the general form of $U^{(\mathbf{a})}\ket\psi$ accordingly looks like. An arbitrary state $\ket\psi$ that satisfies condition (ii) can be expressed in our re-parametrized form as:

\begin{equation}
    \ket\psi=\sum_{B,S,x} \sqrt{q_B} \sqrt{\tilde{p}_B^{(Sx)}}\ket{Sx}_B \ket{\phi_B^{(Sx)}},
\end{equation}
where we introduced $\tilde{p}_B^{(Sx)} \equiv \frac{p_B^{(Sx)}}{q_B}$. It then follows that 

\begin{equation}
    U^{(\mathbf{a})}\ket\psi= \sum_{B,S,x} \sqrt{q_B} \sqrt{\tilde{p}_B^{(Sx)}} \ket{Sx}_B\bigotimes_{l=1}^n \left( U_{S(l)}^{(Bx)} \right)^{a_{S(l)}^{(Bx)}} \ket{\phi_B^{(Sx)}},
\end{equation}
where we used the analogous notation $a_{S(l)}^{(Bx)}$ for $a_{i_l}$. Using condition \eqref{units}, we obtain
\begin{equation}
\begin{split}
    U^{(\mathbf{a})}\ket\psi&=\sum_B \sqrt{q_B} \left[ (-1)^{\sum_l a_{l}^{(B0)}} \sum_S \sqrt{\tilde{p}_B^{(S0)}}\ket{S0}_B \ket{\phi_B^{(S0)}} + (-1)^{\sum_l a_{l}^{(B1)}} \sum_S \sqrt{\tilde{p}_B^{(S1)}}\ket{S1}_B \ket{\phi_B^{(S1)}}  \right]\\
    &=\sum_B \sqrt{q_B} (-1)^{\sum_l a_{l}^{(B0)}} \left[\sum_S \sqrt{\tilde{p}_B^{(S0)}}\ket{S0}_B \ket{\phi_B^{(S0)}} + (-1)^{\sum_l a_l} \sum_S \sqrt{\tilde{p}_B^{(S1)}}\ket{S1}_B \ket{\phi_B^{(S1)}} \right],
\end{split}
\end{equation}
where in the first equality we used $\sum_l a_{S(l)}^{(Bx)}=\sum_l a_{l}^{(Bx)}$ (since the sum is permutation invariant) and in the second equality we used $\sum_l(a_{l}^{(B0)}+a_{l}^{(B1)})=\sum_l a_l$ (since $x=0$ and $x=1$ together cover both partitions of any bipartition $B$). Our expression can further be simplified as follows:
\begin{equation}\label{final1}
    U^{(\mathbf{a})}\ket{\psi}=\sum_{B\in\mathcal{B}_{2n}}e^{i\phi_B}\sqrt{q_B}\left[\frac{1} {\sqrt{2}} \left(\ket{\beta^{(0)}}_B + (-1)^{\sum_la_l}\ket{\beta^{(1)}}_B \right) \right],
\end{equation}
where $\ket{\beta^{(x)}}_B\equiv \sum_{S\in\mathcal{S}_n}\sqrt{2\tilde{p}_S^{(Bx)}} \ket{Sx}_B\otimes \ket{\phi_S^{(Bx)}}$ and $\phi_B=\sum_l a_{l}^{(B0)}$. It is easily checked that $\ket{\beta^{(0)}}_B$ and $\ket{\beta^{(1)}}_B$ are orthogonal and normalized to 1. Recall finally that the most general form of $U^{(\mathbf{a})}\ket\psi$ that concords with our assumptions allows for a further global unitary $G$ of the form stated in \eqref{G} to be appended to the unitaries $U^{(\mathbf{a})}$, and thus appended to Eq. \eqref{final1}. We have thereby proved the statement of Lemma 1, i.e. that for each $\ket\psi$, the following holds:

\begin{equation}
    U^{(\mathbf{a})}\ket{\psi}=G\sum_{B\in\mathcal{B}_{2n}}e^{i\phi_B}\sqrt{q_B}\left[\frac{1} {\sqrt{2}} \left(\ket{\beta^{(0)}}_B + (-1)^{\sum_la_l}\ket{\beta^{(1)}}_B \right) \right].
\end{equation}

\section*{Appendix 3: Proof of Theorem 1}
Here we are going to prove Theorem 1, which states that for any semi-general interference experiment of order $2n$ characterized by triple $T=\left(\rho, U^{(\mathbf{a})}, \Pi \right)_d^{n}$, and such that $I_{2n}(T)=\frac{1}{2}$, there exists another semi-general interference experiment of order $2n$ that is characterized by triple $T'=\left(\rho, U^{(\mathbf{a})}, \Pi' \right)_d^{n}$, for which $I_{2n}(T')=\frac{1}{2}$, and such that $\Pi'$ has the specific form stated in Theorem 1. 

Let us first consider the case in which $\rho$ is a pure state $\rho=\ket{\psi} \bra{\psi}$. Lemma 1 states that $U^{(\mathbf{a})}\ket{\psi}$ takes the following form:
\begin{equation}
    U^{(\mathbf{a})}\ket{\psi}=G\sum_{B\in\mathcal{B}_{2n}}e^{i\phi_B}\sqrt{q_B}\left[\frac{1} {\sqrt{2}} \left(\ket{\beta^{(0)}}_B + (-1)^{\sum_la_l}\ket{\beta^{(1)}}_B \right) \right].
\end{equation}
Let $T'$ be an experiment in which the state $U^{(\mathbf{a})}\ket{\psi}$ undergoes a series of several transformations:
\begin{enumerate}
    \item First the unitary $G^{\dagger}$ is applied resulting in state 
    \begin{equation}
    \sum_{B\in\mathcal{B}_{2n}}e^{i\phi_B}\sqrt{q_B}\left[\frac{1} {\sqrt{2}} \left(\ket{\beta^{(0)}}_B + (-1)^{\sum_la_l}\ket{\beta^{(1)}}_B \right) \right].
\end{equation}
    \item The latter state then undergoes a joint projective measurement $\left\{E_B \right\}$, where $E_B\equiv \sum_{Sx}\ket{Sx}_B\bra{Sx}_B \otimes \mathds{1}$. Outcome $B$ obtains with probability $q_B$ and results in output state 
      \begin{equation}
    \frac{1} {\sqrt{2}} \left(\ket{\beta^{(0)}}_B + (-1)^{\sum_la_l}\ket{\beta^{(1)}}_B \right). 
\end{equation}
    
    \item Then a unitary operator $H_B$ is applied, conditioned on the outcome of the previous measurement. Each $H_B$ acts as
    \begin{equation}\label{hb}
    H_B \ket{\beta^{(x)}}_B=\ket{Sx}_B \otimes \ket\phi_B,
\end{equation}
for some $S \in \mathcal{S}_n$ and $\ket{\phi}_B \in (\mathbb{C}^d)^{\otimes n}$. For each $B$, the output state is thus transformed into
\begin{equation}
     \frac{1} {\sqrt{2}} \left(\ket{S0}_B + (-1)^{\sum_la_l}\ket{S1}_B \right) \otimes \ket\phi_B.
\end{equation}
The state can accordingly be re-parametrized using the initial notation as
\begin{equation}
    \frac{1} {\sqrt{2}} \left(\ket{\mathbf{i}^{(B)}} + (-1)^{\sum_la_l}\ket{\mathbf{j}^{(B)}} \right) \otimes \ket\phi_B,
\end{equation}
where $\mathbf{i}^{(B)}$ and $\mathbf{j}^{(B)}$ are two strings that satisfy $D_{\mathbf{i}^{(B)}\mathbf{j}^{(B)}}=2n$. 

\item The latter state then undergoes another projective measurement $\left\{ M^{(B,1)}_{b_1} \otimes . . . \otimes M^{(B,n)}_{b_n} \right\}_{b_i =0,1}$, where 
\begin{equation}
    M^{(B,l)}_{b_l}=\frac{1}{2} \left( \ket{i^{(B)}_l} + (-1)^{b_l} \ket{j^{(B)}_l} \right) \left( \bra{i^{(B)}_l} + (-1)^{b_l} \bra{j^{(B)}_l} \right).
\end{equation}
It is simple to verify that the outcomes are distributed as 
\begin{equation}
    P_{T'}(b_1...b_n|\mathbf{a}B)= \frac{1}{2^{n-1}} \delta_{\oplus_i b_i, \oplus_j a_j}.
\end{equation}

\item Finally, a binary measurement $\Pi^{(B)} =\left\{ \Pi^{(B)}_0, \Pi^{(B)}_1\right\}$ is applied, where 
\begin{equation}
    \Pi^{(B)}_b= \sum_{\oplus_i b_i = b} M^{(B,1)}_{b_1} \otimes . . . \otimes M^{(B,n)}_{b_n},
\end{equation}
whose outcomes are thus distributed as 
\begin{equation}
    P_{T'}(b|b_1...b_n) =\delta_{b, \oplus_i b_i}.
\end{equation}

\end{enumerate}

The effective correlation between the final outcome $b$ and inputs $\mathbf{a}$ is therefore:

\begin{equation}
\begin{split}
    P_{T'}(b|\mathbf{a}) &= \sum_{B,b_1,...b_n} P(b|b_1...b_n) P(b_1...b_n|\mathbf{a}B) P(B|\mathbf{a})\\
    &= \frac{1}{2^{n-1}} \sum_{B,b_1,...b_n} \delta_{b, \oplus_i b_i}  \delta_{\oplus_i b_i, \oplus_j a_j} q_B \\
    &= \delta_{b,\oplus_i a_i},
\end{split}
\end{equation}
which accordingly induces maximal interference $I_{2n}=\frac{1}{2}$. We have thereby constructed experiment $T'$, which features the same input state $\ket{\psi}$ and unitaries $U^{(\mathbf{a})}$ as the original experiment $T$, and which also generates maximal interference, \textit{but} which features intermediate operations of a specific structure. The latter operations can in particular be summarized in form of the binary POVM $\Pi'= \left \{\Pi_0', \Pi_1' \right\}$, with elements
\begin{equation}
    \Pi'_b=\sum_{b_1...b_n,B}G E_B H_B^{\dagger} \left(M^{(B,1)}_{b_1} \otimes . . . \otimes M^{(B,n)}_{b_n} \right) \Pi^{(B)}_b \left(M^{(B,1)}_{b_1} \otimes . . . \otimes M^{(B,n)}_{b_n} \right) H_B E_B G^{\dagger},
\end{equation}
which concords with the corresponding statement in Theorem 1. 

In case the input state $\rho = \sum_{\psi} p_{\psi} \ket\psi \bra{\psi}$ is mixed, all of the above steps can remain the same, \textit{except }for step 3. Namely, note that this step invokes a unitary operator $H_B$ which is supposed to map - per Eq. \eqref{hb} - a pair of orthogonal states $(\ket{\beta^{(0)}}_B,\ket{\beta^{(1)}}_B)$ into another specifically chosen pair of orthogonal states $(\ket{S0}_B\ket{\phi}_B,\ket{S1}_B\ket{\phi}_B)$. Since the states $\ket{\beta^{(x)}}_B$ generally depend on the initial state $\ket\psi$, the choice of $H_B$ that does the job will also generally depend on $\ket\psi$. Now, according to the GHJW theorem, for any quantum-mechanical system $S$ in mixed state $\rho_S = \sum_{\psi} p_{\psi} \ket\psi_S \bra{\psi}$, there is another system $E$, such that their joint state is $\rho_{SE}$, and such that the following holds: there is a measurement $\left\{\Pi_{\psi}\right\}_{\psi}$ on system $E$, which is such that, if outcome $\psi$ is obtained, the post-measurement state of $S$ is $\ket\psi$ (Gisin, 1989; Hughston \textit{et al}., 1993). This holds both if $\rho_{SE}$ is mixed - i.e. if $S$ is a classically prepared mixture - and if $\rho_{SE}$ is pure - i.e. if the mixture arises due to entanglement. Experiment $T'$ can therefore be modified simply by turning the unitary in step 3 into 
\begin{equation}
    H_B=\sum_{\psi}H_B^{(\psi)} \otimes \Pi_{\psi},
\end{equation}
where $\Pi_{\psi}$ acts on the state of the additional system $E$, and $H_B^{(\psi)}$ is attuned to satisfy Eq. \eqref{hb}, for input state $\ket\psi$. The other steps remain the same as before, involving operations that act trivially (via the identity operator) on the additional system $E$. Given that, as shown above, for each $\ket\psi$ the correlation between $b$ and $\mathbf{a}$ is such as to generate $I_{2n}=\frac{1}{2}$, maximal interference is thus also generated for the mixed state $\rho$. We have therefore shown that for any experiment $T$ with maximal interference, there is a $T'$ with maximal interference, whose final measurement $\Pi'$ has the form stated in Theorem 1.

\section*{Appendix 4: Proof of Theorem 2}
In this appendix we will present a proof analogous to the one presented in Appendices 2 and 3, but that applies to the case of odd $m=2n-1$. As mentioned in the main text, the statement hereby proved will however be less general, in that it will apply only to those $\rho=\sum_p \ket{\psi} \bra{\psi}$, whose states $\ket{\psi}$ each have support on exactly two vectors $\ket{\mathbf{i}},\ket{\mathbf{j}} \in (\mathbb{C}^{2n-1})^{n}$.\footnote{The reader will notice that the proof can trivially be extended to a larger class of states, having overlap on more than two different spatial configurations, but whereby these different configurations need to stand in a certain relation. We will however not delve here into a systematization of this further generalization.} Let us first note that the steps we took in Appendix 2 - in the case of even $m=2n$ - can be trivially transposed to the odd case, all up to Eq. \eqref{r1-r0}, by a simple replacement of $2n$ by $2n-1$. We therefore know that the condition for interference to be maximal is $||\rho^{(1)}_{\psi}-\rho^{(0)}_{\psi}||=2$, where

\begin{equation}\label{r1-r0 odd}
    \rho^{(1)}_{\psi}-\rho^{(0)}_{\psi}=\frac{1}{2^{2n-2}}\sum_{D_{\mathbf{i}\mathbf{j}}=2n-1}\sqrt{p_{\mathbf{i}}p_{\mathbf{j}}}\ket{\mathbf{i}}\bra{\mathbf{j}} \otimes E_{\mathbf{i}\mathbf{j}},
\end{equation}
and 
\begin{equation}
    E_{\mathbf{i}\mathbf{j}} \equiv \sum_{\mathbf{a}} (-1)^{\sum_ia_i } \ket{\phi^{(\mathbf{i})}_{\mathbf{a_i}}}\bra{\phi^{(\mathbf{j})}_{\mathbf{a_j}}}.
\end{equation}

The intuitive reason we cannot transpose the rest of the even proof to the odd case is that now the non-zero components $E_{\mathbf{i}\mathbf{j}}$ are given by strings $\mathbf{i}$ and $\mathbf{j}$, each of length $n$, that satisfy $D_{\mathbf{i}\mathbf{j}}=2n-1$. This implies that the joint string $(\mathbf{i}\mathbf{j})$ contains one value that repeats itself twice, which was not the case in the even case, and which happens to block a simple reproduction of the previous proof.  We will thus focus on the aforementioned restricted case for which 
\begin{equation}
    p_{\mathbf{i}}\neq 0 \longleftrightarrow \mathbf{i} \in \left\{\mathbf{k},\mathbf{l} \right\}.
\end{equation}
In particular, given that the sum in \eqref{r1-r0 odd} ranges only over strings that satisfy $D_{\mathbf{i}\mathbf{j}}=2n-1$, we can assume without loss of generality that our two chosen strings $\mathbf{i} \in \left\{\mathbf{k},\mathbf{l} \right\}$ satisfy $D_{\mathbf{k}\mathbf{l}}=2n-1$. This can either be the case if (i) there is a repetition within string $\mathbf{k}$ but no repetition in string $\mathbf{l}$ (or viceversa), or if (ii) there is no repetition neither within $\mathbf{k}$ nor within $\mathbf{l}$, but only within the joint string $(\mathbf{k},\mathbf{l})$. We will need to address these two cases separately.

\bigskip

\textbf{Case (i).} Let us address the first case, and thus assume that $D_{\mathbf{l}}=n$ and $D_{\mathbf{k}}=n-1$. This means that there exist $r \neq s$, such that $k_r=k_s$, whereas all other values within $\mathbf{k}$ and $\mathbf{l}$ differ from each other. Let us for simplicity, and without loss of generality, assume that $r=1$ and $s=2$, and thus $k_1=k_2$. It follows that
\begin{equation}
    E_{\mathbf{k}\mathbf{l}} = (\mathds{1}_1 \otimes \mathds{1}_2 - U_{k_1} \otimes U_{k_1} ) \bigotimes_{l\neq 1,2} (\mathds{1}_l-U_{k_l})\ket{\phi^{(\mathbf{k})}}\bra{\phi^{(\mathbf{l})}} \bigotimes_{j} (\mathds{1}_j-U^{\dagger}_{l_j}).
\end{equation}
This implies that 
\begin{equation}
    \rho^{(1)}_{\psi}-\rho^{(0)}_{\psi}=\frac{1}{2^{2n-2}}\left(\ket{\theta^{(\mathbf{k})}}\bra{\theta^{(\mathbf{l})}} + \ket{\theta^{(\mathbf{l})}}\bra{\theta^{(\mathbf{k})}} \right),
\end{equation}
where 
\begin{equation}
\begin{split}
    &\ket{\theta^{(\mathbf{k})}}\equiv (\mathds{1}_1 \otimes \mathds{1}_2 - U_{k_1} \otimes U_{k_1} ) \bigotimes_{l\neq 1,2} (\mathds{1}_l-U_{k_l})\ket{\phi^{(\mathbf{k})}}\\
    &\ket{\theta^{(\mathbf{l})}}\equiv  \bigotimes_{l} (\mathds{1}_l-U_{k_l})\ket{\phi^{(\mathbf{l})}}.
\end{split}
\end{equation}
The form that we have here obtained is similar to the one we had obtained in Appendix 2. An application of the same techniques accordingly implies that $||\rho^{(1)}_{\psi}-\rho^{(0)}_{\psi}||=2$ only if the following conditions are satisfied:

\begin{equation}
\begin{split}
     &\bigotimes_{j\neq i} \mathds{1}_j \otimes U_{k_i} \ket{\phi^{(\mathbf{k})}}=-\ket{\phi^{(\mathbf{k})}}, \quad \forall i\neq 1,2\\
     &\left(U_{k_1}\otimes U_{k_1}\right)\otimes \bigotimes_{j\neq 1,2} \mathds{1}_j  \ket{\phi^{(\mathbf{k})}}=-\ket{\phi^{(\mathbf{k})}}\\
     &\bigotimes_{j\neq i} \mathds{1}_j \otimes U_{l_i} \ket{\phi^{(\mathbf{l})}}=-\ket{\phi^{(\mathbf{l})}}, \quad \forall i\\
     & p_{\mathbf{k}}=p_{\mathbf{l}}=\frac{1}{2}.
\end{split}
\end{equation}
Again, analogously to the even case, the latter conditions imply (up to a global phase):
\begin{equation}
    U^{(\mathbf{a})}\ket{\psi}=G\frac{1}{\sqrt{2}} \left( \ket{\mathbf{k}}\ket{\phi^{(\mathbf{k})}}+ (-1)^{\sum_i a_i} \ket{\mathbf{l}}\ket{\phi^{(\mathbf{l})}}\right),
\end{equation}
where we reintroduced the global unitary $G$ of the form \eqref{G}, analogously to the even case. Following the techniques from Appendix 3, it is now obvious that the latter form allows for maximal interference to be generated in an experiment $T'$, that consists of the following steps:
\begin{enumerate}
    \item The unitary operator $G^{\dagger}$ is applied.
    \item A unitary operator $H$ is applied which acts as 
    \begin{equation}
        \begin{split}
        &H\ket{\mathbf{k}}\ket{\phi^{(\mathbf{k})}}=\ket{\mathbf{k}}\ket{\phi}\\  &H\ket{\mathbf{l}}\ket{\phi^{(\mathbf{l})}}=\ket{\mathbf{l}}\ket{\phi},
        \end{split}
    \end{equation}
for some $\ket\phi \in (\mathbb{C}^d)^{n}$.

\item Projective measurement $\left\{ M^{(1)}_{b_1} \otimes . . . \otimes M^{(n)}_{b_n} \right\}_{b_i =0,1}$ is applied, where 
\begin{equation}
    M^{(i)}_{b_i}=\frac{1}{2} \left( \ket{k_i} + (-1)^{b_i} \ket{l_i} \right) \left( \bra{k_i} + (-1)^{b_i} \bra{l_i} \right).
\end{equation}
\item Finally, binary measurement $\Pi=\left\{\Pi_0,\Pi_1 \right\}$ is applied, where $\Pi_b=\sum_{b=\oplus_ib_i} M^{(1)}_{b_1} \otimes . . . \otimes M^{(n)}_{b_n}$.
\end{enumerate}

It follows that experiment $T'$ generates maximal interference $I_{2n-1}=\frac{1}{2}$ and that it features $n$ additional possibly spacelike separated measurements with outcomes $(b_1,...b_n)$, that satisfy 
\begin{equation}
     P_{T'}(b|\mathbf{a})=\sum_{b_1...b_n} P_{T'}(b|b_1...b_n)P_{T'}(b_1...b_n|\mathbf{a}),
\end{equation}
where $P_{T'}(b|b_1...b_n)=\delta_{b,\oplus_i b_i}$
and $P_{T'}(b_1...b_n|\mathbf{a})=\delta_{\oplus_ib_i,\oplus_ja_j}$.

\bigskip

\textbf{Case (ii)}. Let us now address case (ii), and thus assume that $D_{\mathbf{l}}=n$ and $D_{\mathbf{k}}=n$, but $D_{\mathbf{k}\mathbf{l}}=2n-1$. This means that there exist $r,s$, such that $k_r=l_s$, whereas all other values within $\mathbf{k}$ and $\mathbf{l}$ differ from each other. Let us for simplicity, and without loss of generality, assume that $r=1$ and $s=1$, and thus $k_1=l_1$.

It follows that
\begin{equation}
\begin{split}
    E_{\mathbf{k}\mathbf{l}} &= \mathds{1}_1\otimes \bigotimes_{i\neq 1} 
    \left(\mathds{1}_i-U_{k_i} \right)\ket{\phi^{(\mathbf{k})}}\bra{\phi^{(\mathbf{l})}}\mathds{1}_1\otimes \bigotimes_{j\neq 1} 
    \left(\mathds{1}_j-U^{\dagger}_{l_j} \right)\\
    &- U_{k_1}\otimes \bigotimes_{i\neq 1} 
    \left(\mathds{1}_i-U_{k_i} \right)\ket{\phi^{(\mathbf{k})}}\bra{\phi^{(\mathbf{l})}}U_{k_1}\otimes \bigotimes_{j\neq 1} 
    \left(\mathds{1}_j-U^{\dagger}_{l_j} \right).
\end{split}
\end{equation}
This implies that 
\begin{equation}
    \rho^{(1)}_{\psi}-\rho^{(0)}_{\psi}=\frac{1}{2^{2n-2}}\left(\ket{A_0}\bra{B_0}-\ket{A_1}\bra{B_1} + \ket{B_0}\bra{A_0}-\ket{B_1}\bra{A_1} \right),
\end{equation}
where 
\begin{equation}
\begin{split}
    &\ket{A_x}\equiv \sqrt{p_{\mathbf{k}}} \ket{\mathbf{k}} \otimes (U_{k_1})^x\otimes \bigotimes_{i\neq 1} 
    \left(\mathds{1}_i-U_{k_i} \right)\ket{\phi^{(\mathbf{k})}}\\ 
    &\ket{B_x}\equiv \sqrt{p_{\mathbf{l}}} \ket{\mathbf{l}} \otimes (U_{k_1})^x\otimes \bigotimes_{i\neq 1} 
    \left(\mathds{1}_i-U_{l_i} \right)\ket{\phi^{(\mathbf{l})}}.
\end{split}
\end{equation}
Note that $\rho^{(1)}_{\psi}-\rho^{(0)}_{\psi}=\frac{1}{2^{2n-2}}(C+C^{\dagger})$, where $C\equiv \ket{A_0}\bra{B_0}-\ket{A_1}\bra{B_1}$. Since $C$ and $C^{\dagger}$ are orthogonal - i.e. $C(C^{\dagger})^{\dagger}=C^{\dagger}C^{\dagger}=0$ - it follows that $||C+C^{\dagger}||=||C|| + ||C^{\dagger}||$. Therefore, since $||C||= ||C^{\dagger}||$, we obtain 
\begin{equation}\label{rho fin}
    ||\rho^{(1)}_{\psi}-\rho^{(0)}_{\psi}||=\frac{1}{2^{2n-3}}||C||.
\end{equation}
Next, let $C=UDV^{\dagger}$ be the singular value decomposition of $C$, where $U$ and $V$ are unitary, $D=\text{diag}(c_1,c_2)$ and $c_1,c_2$ are the two singular values of $C$. Now, note that 
\begin{equation}
\begin{split}
    &|\text{det}(C)|=|\text{det}(U)\text{det}(D)\text{det}(V^{\dagger})|=|\text{det}(D)|=c_1c_2,\\
    &\Tr (C^{\dagger}C)=\Tr (D^{\dagger}D)=c_1^2+c_2^2.
\end{split}    
\end{equation}
Therefore
\begin{equation}\label{C}
\begin{split}
    ||C||& = c_1+c_2 =\sqrt{(c_1+c_2)^2}=\sqrt{\Tr (C^{\dagger}C) + 2|\text{det}(C)| }\\
    &=\sqrt{\Tr (C^{\dagger}C) + 2\sqrt{\text{det}(C^{\dagger}C)} }.
\end{split}
\end{equation}
Now, let $\left\{\ket{0}, \ket{1}\right\}$ be an orthonormal basis that spans the space spanned by $\left\{\ket{B_0}, \ket{B_1}\right\}$ and let $\ket{B_x}=\sum_{k=0}^1 b^{(x)}_k \ket{k}$. A few lines of algebra show that 
\begin{equation}
    C^{\dagger}C=BAB^{\dagger},
\end{equation}
where 
\begin{equation}
\begin{split}
    &A=\sum_{kl=0}^1 (-1)^{k+l}\braket{A_k|A_l}\ket{k}\bra{l},\\
    &B=\sum_{kl=0}^1 b^{(l)}_k \ket{k}\bra{l}.
\end{split}
\end{equation}
In particular, it holds that $(B^{\dagger}B)_{kl}=\braket{B_k|B_l}$. Using the cyclic property of the trace and of the determinant, one can then easily show that Eqs. \eqref{rho fin} and \eqref{C} imply

\begin{equation}
    ||\rho^{(1)}_{\psi}-\rho^{(0)}_{\psi}||=\frac{1}{2^{2n-3}} \sqrt{a_{00}b_{00} + a_{11}b_{11}-2\text{Re}(a_{10}b_{01})+2\sqrt{(a_{00}a_{11}-|a_{01}|^2)(b_{00}b_{11}-|b_{01}|^2)}},
\end{equation}
where $a_{xy}\equiv \braket{A_x|A_y}$ and $b_{xy}\equiv \braket{B_x|B_y}$. It is now simple to verify that the maximum $||\rho^{(1)}_{\psi}-\rho^{(0)}_{\psi}||=2$ is attained if and only if $a_{xx}$ and $b_{yy}$ both attain maximal values and if $a_{01}=-b_{10}$. In other words, the following conditions need to hold
\begin{equation}
\begin{split}
    &\bigotimes_{i\neq j} 
     \mathds{1}_i \otimes U_{k_j} \ket{\phi^{(\mathbf{k})}}=-\ket{\phi^{(\mathbf{k})}} , \quad \forall j\neq 1,\\
     &\bigotimes_{i\neq j} 
     \mathds{1}_i \otimes U_{l_j} \ket{\phi^{(\mathbf{l})}}=-\ket{\phi^{(\mathbf{l})}} , \quad \forall j\neq 1,\\
     & \bra{\phi^{(\mathbf{k})}}U_{k_1} \otimes \bigotimes_{i\neq 1} 
     \mathds{1}_i  \ket{\phi^{(\mathbf{k})}}=-\bra{\phi^{(\mathbf{l})}}U_{k_1} \otimes \bigotimes_{i\neq 1} 
     \mathds{1}_i  \ket{\phi^{(\mathbf{l})}},\\
     &p_{\mathbf{k}}=p_{\mathbf{l}}=\frac{1}{2}.
\end{split}
\end{equation}

Taking into account the global unitary $G$, of form \eqref{G}, the latter conditions imply (up to a global phase):
\begin{equation}
    U^{(\mathbf{a})}\ket{\psi}=G \frac{1}{\sqrt{2}}\left( \ket{\mathbf{k}}\otimes \ket{\phi_{a_{k_1}}^{(\mathbf{k})}} + (-1)^{\sum_{i\neq k_1}a_i} \ket{\mathbf{l}}\otimes \ket{\phi_{a_{k_1}}^{(\mathbf{l})}} \right),
\end{equation}
where $\ket{\phi_{x}^{(\mathbf{k})}}\equiv(U_{k_1})^x \otimes \bigotimes_{i\neq 1} 
     \mathds{1}_i\ket{\phi^{(\mathbf{k})}}$ and $\ket{\phi_{x}^{(\mathbf{l})}}\equiv(U_{k_1})^x \otimes \bigotimes_{i\neq 1} 
     \mathds{1}_i\ket{\phi^{(\mathbf{l})}}$, whereby the previously concluded conditions imply $\braket{\phi_{0}^{(\mathbf{k})}|\phi_{1}^{(\mathbf{k})}}=-\braket{\phi_{0}^{(\mathbf{l})}|\phi_{1}^{(\mathbf{l})}}$.

Now we are again going to construct an experiment $T'$ that will achieve maximal interference, while featuring $n$ mediating measurements.\footnote{The experiment hereby constructed will for simplicity apply for pure input states $\rho=\ket{\psi}\bra{\psi}$. It is however obvious that the mixed case can be trivially covered by the use of the same technique employed at the end of Appendix 3.} For simplicity, we will introduce the notation: $\ket{\phi_{0}^{(\mathbf{k})}}=\ket{0}$, and $\ket{\phi_{1}^{(\mathbf{k})}}= \alpha \ket{0} + \beta \ket{1}$, where $\braket{1|0}=0$, and $\alpha \equiv \braket{\phi_{0}^{(\mathbf{k})}|\phi_{1}^{(\mathbf{k})}}$. Experiment $T'$ then consists of the following steps:
\begin{enumerate}
    \item  Unitary $G^{\dagger}$ is applied.
    \item  Unitary $M\equiv \ket{\mathbf{k}}\bra{\mathbf{k}}\otimes \mathds{1} + \ket{\mathbf{l}}\bra{\mathbf{l}}\otimes M_{\mathbf{l}}$ is applied, where 

\begin{equation}
    \begin{split}
        &M_{\mathbf{l}} \ket{\phi_{0}^{(\mathbf{l})}}=\ket{0}\\
        &M_{\mathbf{l}} \ket{\phi_{1}^{(\mathbf{l})}}=-\alpha\ket{0} + \beta \ket{1},
    \end{split}
\end{equation}
where the second line ensures the condition that $\braket{\phi_{0}^{(\mathbf{k})}|\phi_{1}^{(\mathbf{k})}}=-\braket{\phi_{0}^{(\mathbf{l})}|\phi_{1}^{(\mathbf{l})}}$. The state is thereby transformed into
\begin{equation}
\begin{split}
    &\frac{1}{\sqrt{2}}\left( \ket{\mathbf{k}} + (-1)^{\sum_{i \neq k_1}a_i}\ket{\mathbf{l}} \right)\ket{0}, \quad  \quad \text{if} \quad a_{k_1}=0\\
    &\frac{1}{\sqrt{2}}\left[ \alpha \left(\ket{\mathbf{k}} - (-1)^{\sum_{i \neq k_1}a_i}\ket{\mathbf{l}}\right)\ket{0} + \beta \left(\ket{\mathbf{k}} + (-1)^{\sum_{i \neq k_1}a_i}\ket{\mathbf{l}}\right)\ket{1} \right], \quad  \quad \text{if} \quad a_{k_1}=1
\end{split}
\end{equation}

\item A further unitary is applied which retains all components of the quantum state invariant except for $\ket{\mathbf{l}}\ket{1} \rightarrow - \ket{\mathbf{l}}\ket{1}$, thus transforming the overall state into 

\begin{equation}
    \frac{1}{\sqrt{2}}\left( \ket{\mathbf{k}} + (-1)^{\sum_{j=1}^{2n-1}a_j}\ket{\mathbf{l}} \right)\ket{\phi_{a_{k_1}}^{(\mathbf{k})}}.
\end{equation}

Note that the inputs $\mathbf{a}$ are now all encoded in the phase $(-1)^{\sum_{j=1}^{2n-1}a_j}$, as in the previous cases. 

\item In order to proceed with the interferometric measurements, we need to first ensure that each pair $(k_i,l_i)$ is such that $k_i \neq l_i$. We thus apply a further unitary that shifts $\ket{l_1l_2...l_n} \rightarrow \ket{l_2l_2...l_n}$, ensuring that $k_i \neq l_i$, for all $i$. Then the standard interferometric measurement $\left\{ M^{(1)}_{b_1} \otimes . . . \otimes M^{(n)}_{b_n} \right\}_{b_i =0,1}$ is applied, where 
\begin{equation}
    M^{(i)}_{b_i}=\frac{1}{2} \left( \ket{k_i} + (-1)^{b_i} \ket{l_i} \right) \left( \bra{k_i} + (-1)^{b_i} \bra{l_i} \right).
\end{equation}
\item Finally, binary measurement $\Pi=\left\{\Pi_0,\Pi_1 \right\}$ is applied, where $\Pi_b=\sum_{b=\oplus_ib_i} M^{(1)}_{b_1} \otimes . . . \otimes M^{(n)}_{b_n}$.

\end{enumerate}

It once again follows that experiment $T'$ generates maximal interference $I_{2n-1}=\frac{1}{2}$, while featuring $n$ additional possibly spacelike separated mediating events $(b_1,...b_n)$.

\section*{Appendix 5: Further observations on the case of odd $m$}
In Appendix 4 we analyzed only cases in which the input state is a mixture of states that have each support on exactly two different spatial configurations $\mathbf{k}$ and $\mathbf{l}$. Let us now consider an example where the input state has support on \textit{three} spatial configurations. We will here for simplicity focus on the case $n=2$, but it is clear that the example can be generalized to a larger class of states, for higher $n$. Suppose accordingly that the input state is given as 
\begin{equation}
    \ket\psi=\sqrt{p_{11}}\ket{11}\ket{\phi^{(11)}}+\sqrt{p_{23}}\ket{23}\ket{\phi^{(23)}}+\sqrt{p_{12}}\ket{12}\ket{\phi^{(12)}},
\end{equation}
and thus has support on three spatial configurations. The transformed state is then given by
\begin{equation}\label{3 supp}
    \ket\psi_{\mathbf{a}}\equiv U^{(\mathbf{a})}\ket\psi=\sqrt{p_{11}}\ket{11}U_{11}^{(a_1)}\ket{\phi^{(11)}}+\sqrt{p_{23}}\ket{23}U_{2}^{(a_2)}\otimes U_{3}^{(a_3)}\ket{\phi^{(23)}}+\sqrt{p_{12}}\ket{12}U_{1}^{(a_1)}\otimes U_{2}^{(a_2)}\ket{\phi^{(12)}}.
\end{equation}
A necessary condition for $I_3 = \frac{1}{2}$ is that the following orthogonality conditions hold: 

\begin{equation}
    \braket{\psi_{0a_2a_3}|\psi_{1a_2a_3}}=\braket{\psi_{a_10a_3}|\psi_{a_11a_3}}=\braket{\psi_{a_1a_20}|\psi_{a_1a_21}}=0,
\end{equation}
which implies the following equations
\begin{equation}
    \begin{split}
        &p_{11}+\phi_1p_{23}+\phi_2p_{12}=0\\
        &\phi_3p_{11}+p_{23}+\phi_4p_{12}=0\\
        &p_{11}+\phi_5p_{23}+p_{12}=0
    \end{split}
\end{equation}
where we introduced the abbreviations $\phi_1 \equiv \braket{\phi^{(23)}|U_2 \otimes \mathds{1}|\phi^{(23)}}$, $\phi_2 \equiv \braket{\phi^{(12)}|\mathds{1} \otimes U_2|\phi^{(12)}}$, $\phi_3 \equiv \braket{\phi^{(11)}|U_{11}|\phi^{(11)}}$, $\phi_4 \equiv \braket{\phi^{(12)}|U_1 \otimes \mathds{1}|\phi^{(12)}}$, $\phi_5 \equiv \braket{\phi^{(23)}|\mathds{1} \otimes U_3|\phi^{(23)}}$. 

Now, it is simple to note that - since each $\phi_i \in [-1,1]$ and $\sum_{ij}p_{ij}=1$ - the third equation implies that $p_{23} \geq \frac{1}{2}$. Furthermore, the second equation implies $|p_{23}+\phi_4 p_{12}|\leq p_{11}$, and thus $\phi_4 \leq \frac{p_{11}-p_{23}}{p_{12}}$, which, since $\phi_4 \geq -1$, implies that $p_{23}\leq \frac{1}{2}$. The two results thus entail $p_{23} = \frac{1}{2}$ and accordingly $\phi_5=-1$. Similarly, taking into account the latter result, equation two further implies that $\phi_3=-\frac{1+\phi_4(1-2p_{11})}{2p_{11}}$. Since $p_{11}< \frac{1}{2}$ and $|\phi_3|\leq 1$, it follows that $1+\phi_4(1-2p_{11})\leq 2p_{11}$. But, since $\phi_4\geq -1$, it also holds that $1+\phi_4(1-2p_{11})\geq 1-(1-2p_{11})=2p_{11}$. The two latter results thus imply that $1+\phi_4(1-2p_{11})=2p_{11}$, which in turn implies $\phi_4=-1$ and $\phi_3=-1$. Taking stock:
\begin{equation}
    \begin{split}
        &p_{23}=\frac{1}{2}=p_{11}+p_{12}\\
        &U_{11}\ket{\phi^{(11)}}=-\ket{\phi^{(11)}}\\
        &U_{1} \otimes \mathds{1}\ket{\phi^{(12)}}=-\ket{\phi^{(12)}}\\
        &\mathds{1} \otimes U_3\ket{\phi^{(23)}}=-\ket{\phi^{(23)}}.
    \end{split}
\end{equation}

Now, using the same techniques as in the previous Appendices, it is easily seen that the quantity of interest takes the following form:
\begin{equation}
    ||\rho^{(1)}-\rho^{(0)}||=\sqrt{2} ||C||,
\end{equation}
where
\begin{equation}
    C=\sqrt{p_{11}} \ket{A}\bra{B_0} -\sqrt{p_{11}} \ket{A}\bra{B_1} +\sqrt{p_{12}} \ket{A_0}\bra{B_0} -\sqrt{p_{12}} \ket{A_1}\bra{B_1},
\end{equation}
and

\begin{equation}
\begin{split}
    &\ket{A}\equiv \ket{11}\ket{\phi^{(11)}}\\
    &\ket{A_x}\equiv \ket{12}\left(\mathds{1}\otimes U_2 \right)^{x}\ket{\phi^{(12)}}\\
    &\ket{B_y}\equiv \ket{23}\left(U_2\otimes \mathds{1} \right)^{y}\ket{\phi^{(23)}}\\
\end{split}
\end{equation}

The operator $C$ has support on a two-dimensional subspace, which enables us to apply the same techniques used in Appendix 4, leading to 

\begin{equation}
\begin{split}
    ||C||&=\sqrt{\Tr (C^{\dagger}C) + 2|\text{det}(C)|}\\
    &= \sqrt{1-2p_{11} \text{Re}(b_{10}) - 2p_{12} \text{Re}(a_{01}b_{10}) + 2\sqrt{\left(1-|b_{01}|^2 \right) \left(\frac{1}{4}-|p_{11}+p_{12}a_{01}|^2\right)} },
\end{split}
\end{equation}
where $a_{xy}=\braket{A_x|A_y}$ and $b_{xy}=\braket{B_x|B_y}$. A few lines of calculations show that the maximum $||C||=\sqrt{2}$ (i.e. $||\rho^{(1)}-\rho^{(0)}||=2$) is attained if and only if 
\begin{equation}
    b_{01}=2p_{12}(1-a_{01})-1,
\end{equation}
or, in other words:
\begin{equation}\label{sup 3 2}
    \braket{\phi^{(23)}|U_2 \otimes \mathds{1}|\phi^{(23)}}=2p_{12}(1-\braket{\phi^{(12)}|\mathds{1} \otimes U_2|\phi^{(12)}})-1.
\end{equation}

Taking all of the above into account, Eq. \eqref{3 supp} amounts, up to the global unitary $G$, to 

\begin{equation}
    \ket{\psi_{\mathbf{a}}}=G\frac{1}{\sqrt{2}} \left( (-1)^{a_1+a_3} \ket{23} \ket{\phi^{(23)}_{a_{2}}} + \ket{\theta_{a_2}} \right),
\end{equation}
where $\ket{\phi^{(23)}_{x}} \equiv (U_2\otimes \mathds{1})^x \ket{\phi^{(23)}}$ and $\ket{\theta_x}\equiv \sqrt{2p_{11}}\ket{11}\ket{\phi^{(11)}}+\sqrt{2p_{12}}\ket{12}(\mathds{1}\otimes U_2)^x \ket{\phi^{(12)}}$. In particular, notice that Eq. \eqref{sup 3 2} implies 
\begin{equation}
    \braket{\phi^{(23)}_{0}|\phi^{(23)}_{1}}=-\braket{\theta_0|\theta_1}.
\end{equation}
The latter relation enables the construction of an experiment $T'$ analogous to the one we presented in Appendix 4. Namely, its steps are listed as follows.

\begin{enumerate}
    \item Unitary operator $G^{\dagger}$ is applied
    \item Unitary $M\equiv \ket{23}\bra{23}\otimes \mathds{1} + \left(\ket{11}\bra{11}+\ket{12}\bra{12}\right)\otimes \tilde{M}$ is applied, where 

\begin{equation}
    \begin{split}
        &\tilde{M} \ket{\theta_0}=\ket{12}\otimes\ket{0}\\
        &\tilde{M} \ket{\theta_1}=\ket{12}\otimes(-\alpha\ket{0} + \beta \ket{1}),
    \end{split}
\end{equation}
where $\ket0\equiv \ket{\phi^{(23)}_{0}}$, $\ket{1}$ is a vector orthogonal to $\ket{0}$, and $\ket{\phi^{(23)}_{1}}=\alpha \ket{0} + \beta \ket{1}$. The resulting state is 

\begin{equation}
\begin{split}
    &\frac{1}{\sqrt{2}}\left( (-1)^{a_1+a_3}\ket{23} + \ket{12} \right)\ket{0}, \quad  \quad \text{if} \quad a_{2}=0\\
    &\frac{1}{\sqrt{2}}\left[ \alpha \left((-1)^{a_1+a_3}\ket{23} - \ket{12}\right)\ket{0} + \beta \left((-1)^{a_1+a_3} \ket{23} + \ket{12}\right)\ket{1} \right], \quad  \quad \text{if} \quad a_{2}=1
\end{split}
\end{equation}

\item A further unitary is applied that acts as $\ket{12}\ket{1} \rightarrow -\ket{12}\ket{1}$, thus turning the state, up to a global phase, into

\begin{equation}
    \frac{1}{\sqrt{2}}\left( \ket{23} + (-1)^{a_1+a_2+a_3}\ket{12}\right).
\end{equation}

\item Interferometric measurement $\left\{ M^{(1)}_{b_1} \otimes M^{(2)}_{b_2} \right\}_{b_i =0,1}$ is then applied, where 
\begin{equation}
\begin{split}
    &M^{(1)}_{b_1}=\frac{1}{2} \left( \ket{2} + (-1)^{b_1} \ket{1} \right) \left( \bra{2} + (-1)^{b_1} \bra{1} \right)\\
    &M^{(2)}_{b_2}=\frac{1}{2} \left( \ket{3} + (-1)^{b_2} \ket{2} \right) \left( \bra{3} + (-1)^{b_2} \bra{2} \right)
\end{split}
\end{equation}

\item Finally, binary measurement $\Pi=\left\{\Pi_0,\Pi_1 \right\}$ is applied, where $\Pi_b=\sum_{b=b_1 \oplus b_2} M^{(1)}_{b_1} \otimes M^{(2)}_{b_2}$.

\end{enumerate}

We have thereby again constructed an experiment $T'$ that generates maximal interference $I_3 = \frac{1}{2}$, while featuring $n=2$ possibly spacelike separated mediating events $(b_1,b_2)$. It is furthermore clear that the same proof goes through for a larger class of states for higher $n$. We will however leave a systematic study of this larger class for another occasion.

\section*{Appendix 6: Remark on (im)proper mixed states}

As stated at the end of Section IV.II., Theorems 1 and 2 apply directly to those semi-general interference experiments in which the final measurement $\Pi$ acts solely on the $n$ particles that are initially prepared in state $\rho$. However, the proofs of the theorems generally rely on the accessibility of environmental degrees of freedom that are either classically correlated or entangled to the $n$ particles. Prompted by this observation, one may envisage a broader class of semi-general interference experiments in which the final measurement $\Pi$ acts both on the $n$ particles and on their environment, and wonder whether our theorems apply to such experiments as well. We will now first show that, \textit{if} $\rho$ is a proper mixture, our theorems \textit{do} apply also to such more general cases. Then we will indicate why the case of improper mixtures exceeds our current proofs.

Let $S$ be the abbreviation for the system of $n$ particles and $E$ be the abbreviation for the environment. Let us assume that the joint system $S+E$ is initially in the following \textit{classically correlated} mixed state

\begin{equation}
    \rho_{SE}=\sum_{\psi}p_{\psi} \ket{\psi}\bra{\psi} \otimes \ket{E_{\psi}}\bra{E_{\psi}}.
\end{equation}

The reduced state of $S$ is thus accordingly the \textit{proper} mixture $\rho=\sum_{\psi}p_{\psi} \ket{\psi}\bra{\psi}$. The joint system then undergoes transformation $U^{(\mathbf{a})} \otimes \mathds{1}$, which, by assumption, acts trivially on $E$. Finally, the joint system undergoes a \textit{joint} measurement $\Pi$. It is simple to notice that the steps in the proof of Lemma 1 conducted in Appendix 2 can be reiterated up to Eq. \eqref{=2}, thus resulting in the condition 

\begin{equation}
    || \rho^{(1)}-\rho^{(0)}||=2,
\end{equation}
with the sole difference being that $\rho^{(s)}$ is now defined as follows:

\begin{equation}
    \rho^{(s)}=\frac{1}{2^{2n-1}} \sum_{\oplus_ia_i=s} (U^{(\mathbf{a})} \otimes \mathds{1}) \rho_{SE} (U^{(\mathbf{a})\dagger} \otimes \mathds{1})
\end{equation}

The triangle inequality then implies:

\begin{equation}
    || \rho^{(1)}-\rho^{(0)}|| \leq \sum_{\psi} p_{\psi} || \ket{E_{\psi}}\bra{E_{\psi}} \otimes (\rho^{(1)}_{\psi}- \rho^{(0)}_{\psi} )||= \sum_{\psi} p_{\psi} ||\rho^{(1)}_{\psi}- \rho^{(0)}_{\psi} ||
\end{equation}

Therefore, for each $\psi$, we have:
\begin{equation}
    ||\rho^{(1)}_{\psi}- \rho^{(0)}_{\psi} ||=2,
\end{equation}
resulting in the same condition \eqref{=22} that we arrived at in the proof of Lemma 1. The proof can then follow the same subsequent steps, thereby reaching the same conclusions asserted in Lemma 1 and in Theorem 1. The analogous generalization holds also for Theorem 2.

Let us now point out why the same proof does \textit{not} go through if the mixture $\rho$ is \textit{improper}. Let us accordingly assume that the joint system $S+E$ is initially in the following pure entangled state 

\begin{equation}
    \ket{\phi}=\sum_{\psi} \sqrt{p_{\psi}} \ket{\psi} \otimes \ket{E_{\psi}}
\end{equation}

The reduced state of $S$ is now the improper mixture $\rho=\sum_{\psi} p_{\psi} \ket\psi\bra\psi$. Following the same steps as above then leads us to the following condition:

\begin{equation}
    2=|| \rho^{(1)}-\rho^{(0)}|| \leq \sum_{\psi,\psi'} \sqrt{p_{\psi}p_{\psi'}} || \ket{E_{\psi}}\bra{E_{\psi'}} \otimes (\rho^{(1)}_{\psi\psi'}- \rho^{(0)}_{\psi\psi'} )||= \sum_{\psi,\psi'} \sqrt{p_{\psi}p_{\psi'}} ||\rho^{(1)}_{\psi\psi'}- \rho^{(0)}_{\psi\psi'} ||,
\end{equation}

where 

\begin{equation}
    \rho^{(s)}_{\psi\psi'}=\frac{1}{2^{2n-1}} \sum_{\oplus_ia_i=s} U^{(\mathbf{a})} \ket\psi\bra{\psi'} U^{(\mathbf{a})\dagger}.
\end{equation}

The above inequality does not anymore imply in a straightforward way that $||\rho^{(1)}_{\psi}- \rho^{(0)}_{\psi} ||=2$ needs to hold, for each $\psi$. This case thus demands further separate investigation.

\end{document}